\documentclass[aps,twocolumn,footinbib,floatfix]{revtex4-1}

\usepackage{CJK}
\usepackage{float}
\usepackage{amsmath}
\usepackage{amsfonts}
\usepackage{soul}
\usepackage[caption=false,subrefformat=parens,labelformat=parens]{subfig}
\usepackage{color} 
\usepackage[table,dvipsnames]{xcolor}
\usepackage[colorlinks,linktocpage,bookmarks=false]{hyperref}

\usepackage{graphicx}
\graphicspath{{FIG/}}
\usepackage{multirow}
\usepackage{braket}

\newtheorem{property}{Property}
\newtheorem{definition}{Definition}

\newcommand{\kb}{k_\text{B}}

\newcommand{\tb}{}

\newcommand\myshade{85}
\definecolor{myrulecolor}{RGB}{150,20,0}
\colorlet{mylinkcolor}{violet}
\colorlet{mycitecolor}{YellowOrange}
\colorlet{myurlcolor}{Aquamarine}
\hypersetup{
	linkcolor  = myurlcolor!\myshade!black,
	citecolor  = mycitecolor!\myshade!black,
	urlcolor   =  myrulecolor!\myshade!black,
}

\makeindex

\begin{document} 
\begin{CJK*}{UTF8}{gbsn} 
\title{Hyperbolic Fracton Model, Subsystem Symmetry, and Holography}

\author{Han Yan (闫寒)}

\email{han.yan@oist.jp}
\affiliation{Okinawa Institute of Science and Technology Graduate University, Onna-son, Okinawa 904-0412, Japan}
\date{\today}
\begin{abstract}
We propose that the {fracton models with subsystem symmetry can be}
a class of toy models for {the holographic principle}.
The discovery of the anti-de Sitter/conformal field theory   correspondence as a concrete construction of holography
and the subsequent developments including the subregion duality and
Ryu-Takayanagi formula of entanglement entropy
have revolutionized our understanding of quantum gravity 
and provided powerful tool {sets} for solving various
strongly-coupled quantum field theory problems.
To resolve many mysteries of holography,
toy models can be very helpful.
One example is the holographic tensor networks 
which illuminate the quantum error correcting properties of gravity in the anti-de Sitter space.
In this work we discuss a classical toy model 
{featuring subsystem symmetries and immobile fracton excitations}.
We show that such a model defined on the hyperbolic
lattice satisfies some key properties of the holographic correspondence.
{
The correct subregion duality
and Ryu-Takayanagi formula for mutual information
are established for a connected boundary region.
}
A naively defined black hole's entropy scales as its horizon area.
{We also present discussions on corrections
for more complicated boundary subregions,
the possible generalizations of the model,
and a comparison with the holographic tensor networks.}
\end{abstract}
\maketitle
\end{CJK*}

\section{Introduction}
The holographic principle \cite{Hooft1974,Susskind1995} and anti-de Sitter/conformal field theory (AdS/CFT) correspondence \cite{Maldacena1999,Witten1998}
have profoundly improved our understanding
of quantum gravity.
AdS/CFT is a duality between quantum gravity in $(d+1)$-dimensional asymptotically AdS \tb{spacetime} 
and a $d$-dimensional CFT on its boundary.
It proposes a striking \tb{conjecture} that a gravitational system
is equivalent to a strongly coupled quantum field theory without gravity.
Besides unveiling some of the deepest mysteries of quantum gravity in
its subsequent developments \cite{gubser1998gauge,Aharony2000,Aharony2008,Klebanov2002,Hawking2005,Guica2009},
the AdS/CFT correspondence also serves as a powerful tool
for studying strongly coupled quantum field theories including many-body systems \cite{Hartnoll}.

Another remarkable development in AdS/CFT is the realization of the
intimate relation between the geometry of spacetime
and quantum entanglement.
Ryu and Takayanagi conjectured that the entanglement entropy of
a boundary segment is measured by the area of certain extremal covering surface 
in the AdS geometry \cite{Ryu2006,Ryu2006a}.
Their seminal idea, now known as the Ryu-Takayanagi (RT) formula,
has sparked a series of insightful works {along this direction} (for example, see review Ref.~\cite{Bousso2002}).
%

AdS/CFT has
deep connections with 
various condensed matter theory problems. 
One example is the multiscale entanglement renormalization ansatz (MERA) tensor networks.
Their structure bears considerable similarity with the 
renormalization scale represented by the 
radial direction of AdS space.
Such insight by Swingle \cite{Swingle2012} 
leads to a fruitful field of
building toy models of AdS/CFT with tensor networks \cite{Pastawski2015,Almheiri2015,Hayden2016,Qi2018},
which in return demystify some intriguing properties of holography.
For instance, the perfect tensor networks \cite{Pastawski2015,Almheiri2015} incorporate the 
quantum error correction feature of AdS/CFT 
and help  to clarify the conundrum
of subregion duality.
%

Since conformally invariant or strongly coupled systems
are common themes in many-body physics,
the condensed matter systems often sit on the CFT side
when AdS/CFT is applicable \cite{Hartnoll}.
Examples of many-body systems on the bulk side are rare \cite{Qi2013,Gu2016,Lee2016}.
Therefore it is desirable to 
seek   many-body systems that,
instead of being described by some CFT,
mimic  the behavior of gravity 
and sit  on the AdS side of holography.
%
\tb{Studying such systems 
not only is of interest to the condensed matter community,
but also may provide us insights in understanding gravity.
}

This work aims to show that 
the recently discovered \tb{fracton models} \cite{Haah2011,Vijay2015}
mimic  gravity
and can sit on the AdS side
{as a toy model of holography}.
The fracton phases
cover several types of exotic states in many-body systems,
and have attracted much attention
in the condensed matter community recently.
For example, gapped fracton topological orders 
have
intriguing subextensive ground-state degeneracy 
and (partially) immobile excitations 
\cite{Vijay2016,Halasz2017,Bulmash2018,Shirley2017,Shirley2018,Ma2018,Schmitz2018,Ma2017,Ma2018a} (also see review Ref.~\cite{Nandkishore2018}).
The gapless versions of them are 
described by the rank-2 U(1) gauge theories \cite{Pretko2018,Pretko2017,Pretko2017a,Pretko2017b}.
The fracton topological orders 
can also be obtained by gauging the subsystem symmetries
of the model \cite{Shirly2018-2,Vijay2016},
which inspired study of 
fracton models protected by subsystem symmetries as well \cite{DevakulPhysRevB2018,YouPhysRevB2018}. %
\tb{In this work,} we study a classical fracton model \tb{with subsystem symmetry }on the hyperbolic disk,
or a spatial slice of AdS$_3$ spacetime.
We show that such a system satisfies the major properties of AdS/CFT,
in a manner similar to the holographic tensor networks.
These properties include the AdS-Rindler reconstruction and subregion duality,
and the RT formula for mutual information as the classical analog of entanglement entropy.
\tb{They are satisfied exactly for a connected boundary subregion up to lattice discretization.
The corrections for more complicated boundary subregions are also discussed.
}
The hyperbolic fracton model gives the proper entropy for a naively defined black hole
as well.

The paper is arranged as follows: 
Sec.~\ref{SEC_II_Summary} provides a concise summary of the major results;
Sec.~\ref{SEC_EU_fracton} introduces the fracton model on the Euclidean lattice,
and discusses various hints implying that it could be holographic;
Sec.~\ref{SEC_AdSCFT} presents some essential knowledge of AdS/CFT relevant to our work, mainly for readers not familiar with this discipline;
Sec.~\ref{SEC_Hyper_Fracton} introduces the fracton model on the hyperbolic lattice;
Secs.~\ref{SEC_Rindler},\ref{SEC_RT_formula} and \ref{SEC_BH}
are the major results of this work. 
These sections show that the model satisfies some major properties of AdS/CFT,
and discuss some possible deviations;
\tb{
Sec.~\ref{SEC_generalization}
discusses how to generalize the classical model 
to three dimension and to a quantum version;
Sec.~\ref{SEC_comparison} 
presents a comparison of the hyperbolic fracton model
and the holographic tensor networks
to make clear 
what holographic properties are still beyond the scope of current construction;
finally, Sec.~\ref{SEC_summary_discussion} gives an outlook on the implications
and future problems related to this work.
}

\section{Summary of the Holographic Properties of the Hyperbolic Fracton Model} \label{SEC_II_Summary}
%
%
In this paper
we will demonstrate that 
the \textit{hyperbolic fracton model},
a classical fracton model defined on a hyperbolic disk (a spatial slice of AdS$_3$),
satisfies several 
key properties of AdS/CFT correspondence.
The main results are summarized here,
with detailed proofs and discussions presented subsequently.\\

\noindent
{\bf Rindler reconstruction ---} 
In the hyperbolic fracton model defined by Eq.~\eqref{C4_eqn_hamiltonian}, given the state, or spin configuration on a \tb{connected boundary subregion},
the bulk states within the {\it minimal convex wedge} of the boundary can be reconstructed. 
The minimal convex wedge is essentially the entanglement wedge on a discrete lattice, which
approximates the continuous case.\\

%

\noindent 
{\bf  Ryu-Takayanagi formula for mutual information:}
For a bipartition of the boundary
\tb{into two individually connected subregions}
denoted $A$ and $A^c$,
their mutual information in the  classical model, as the classical analog of entanglement entropy, 
obeys the geometric RT formula:
\begin{equation}
\I(A,A^c) =\kb\log2\times |\gamma_A| \;.
\end{equation}
where $|\gamma_A|$ is the area of the minimal covering surface, 
or in this case the geodesic on the hyperbolic disk.\\

\noindent 
{\bf Black hole entropy:}
A naively defined black hole in the model
has entropy proportional to the area of the black hole horizon.
Also with the presence of black hole, 
the available lowest energy boundary states increase  as expected.

\section{Fracton Model on the  Euclidean Lattice} \label{SEC_EU_fracton}
\subsection{The  Model}
We start with a discussion of the  {fracton model with subsystem symmetry} on 
the Euclidean square lattice,
as an introduction of the major features shown in various  {fracton} models.

%
%
\begin{figure}[t]
	\centering
	\includegraphics[width=0.35\textwidth]{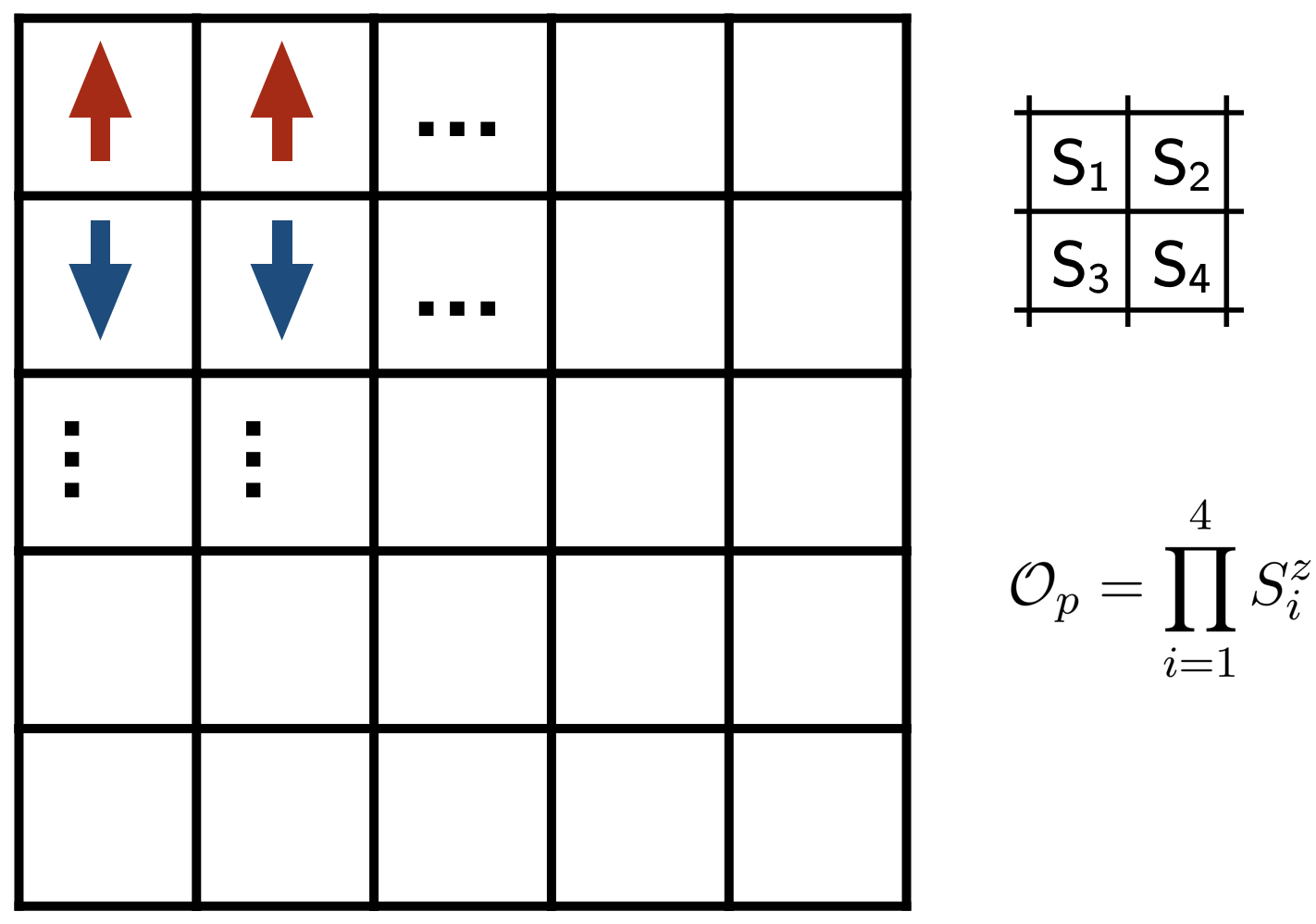}
	\caption{The fracton model on the Euclidean lattice defined by Eq.~\eqref{eqn_hamiltonian}. 
		On \tb{each center of the unit square sits an Ising spin}.
		The right panel shows how operator $O_p$ in Eq.~\eqref{EQN_Op_Def} 
		is defined.
		\label{FIG_EU_Fracton}
	}
\end{figure}
Consider the square lattice with an Ising spin sitting on {the \textit{center} of each square} 
as shown in Fig.~\ref{FIG_EU_Fracton}.
For the four  {spins sitting on squares sharing the same corner, we define} an operator
\begin{equation}\label{EQN_Op_Def}
\mathcal{O}_p=\prod_{i=1}^{4} S^z_i			,
\end{equation}
where $S^z_i=\pm1$ are the Ising spins,
and $i$ runs over its four spins, 

The Hamiltonian of this \tb{classical} Fracton model is defined as the negative sum of such operators on all four-spin clusters,
\begin{equation}\label{eqn_hamiltonian}
\mathcal{H}_\textsf{cl}=-\sum_{p}\mathcal{O}_p			 .
\end{equation}

This model has a rich context in various disciplines of physics.
It is essentially a two-dimensional version of the ``plaquette model'' 
discussed in Ref.~\cite{Vijay2016} . 
It is also a self-dual model with subsystem {symmetries} discussed in Refs.~\cite{Shirly2018-2,YouPhysRevB2018,DevakulPhysRevB2018}.
It is dual to an exactly solvable square-lattice eight-vertex model \cite{baxter2016exactly}, whose implication will be discussed in a future work.
The classical model has also been studied as a spin glass statistical physics problem \cite{Jack05,GarrahanPhysRevE},
and  proposed
as a string regularization known as  the gonihedric Ising model \cite{Savvidy1994,Savvidy1996,Pietig1998}.

\tb{In this work we will focus on the features of this classical model,
	and briefly discuss its quantum version in Sec.~\ref{SEC_generalization}.
}

\subsection{Features of \tb{the Fracton Model}}\label{SEC_Eu_FTO}
{The fracton models exhibit several exotic features, regarding their ground states, entanglement, and excitations.}

{Before elaborating these properties, it should be emphasized that the \textit{subsystem symmetries}  play a crucial role. 
	The same statement is also true for the holographic properties of the hyperbolic fracton model.}\\

\noindent\textbf{Feature one: Sub-extensive ground state degeneracy --- }

%
%
\begin{figure}[t]
	\centering
	\captionsetup[subfigure]{justification=centering}
	\subfloat[Ground state\label{FIG_EU_GS_degeneracy_sub1}]{\includegraphics[width=0.22\textwidth]{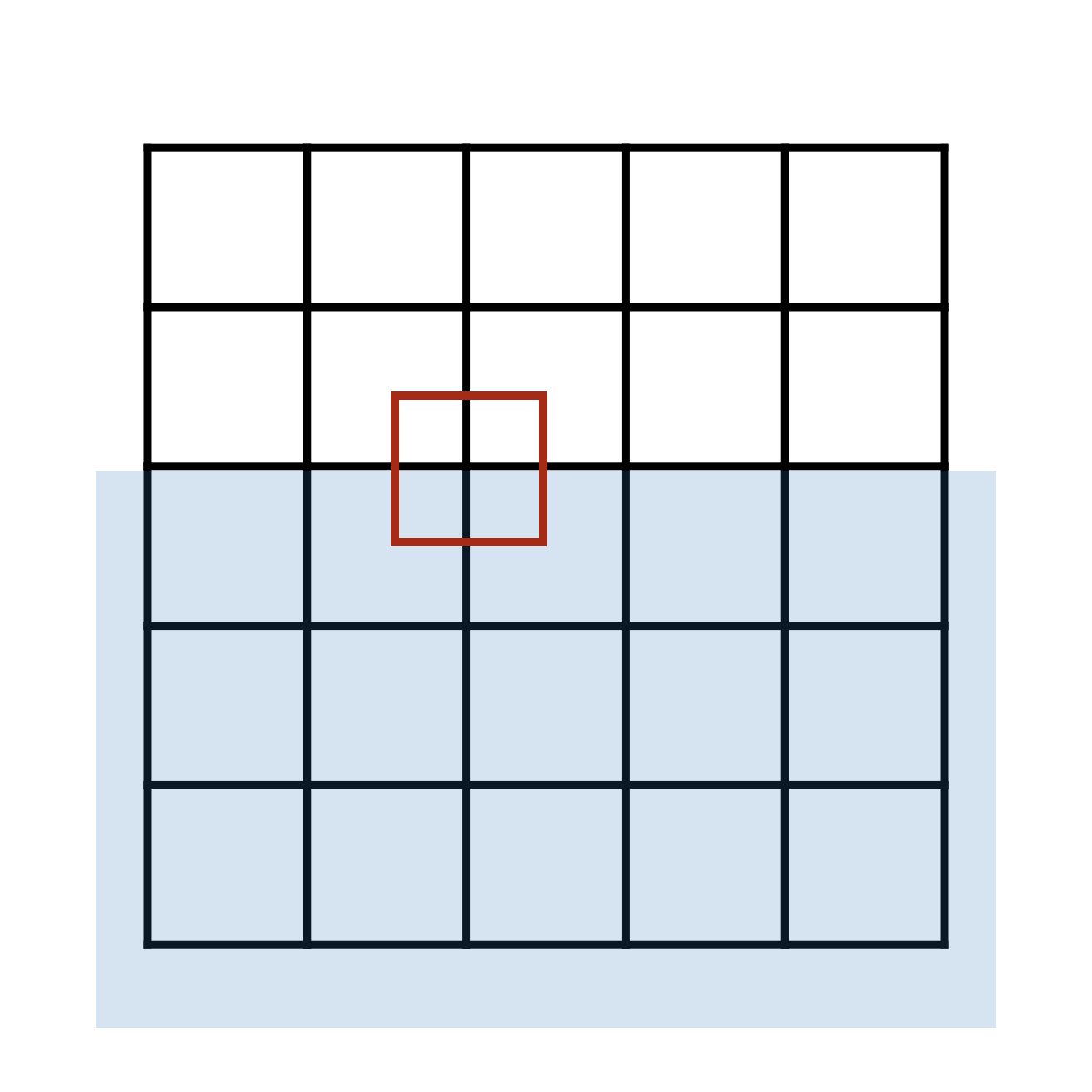}}
	\subfloat[Another ground state\label{FIG_EU_GS_degeneracy_sub2}]{\includegraphics[width=0.22\textwidth]{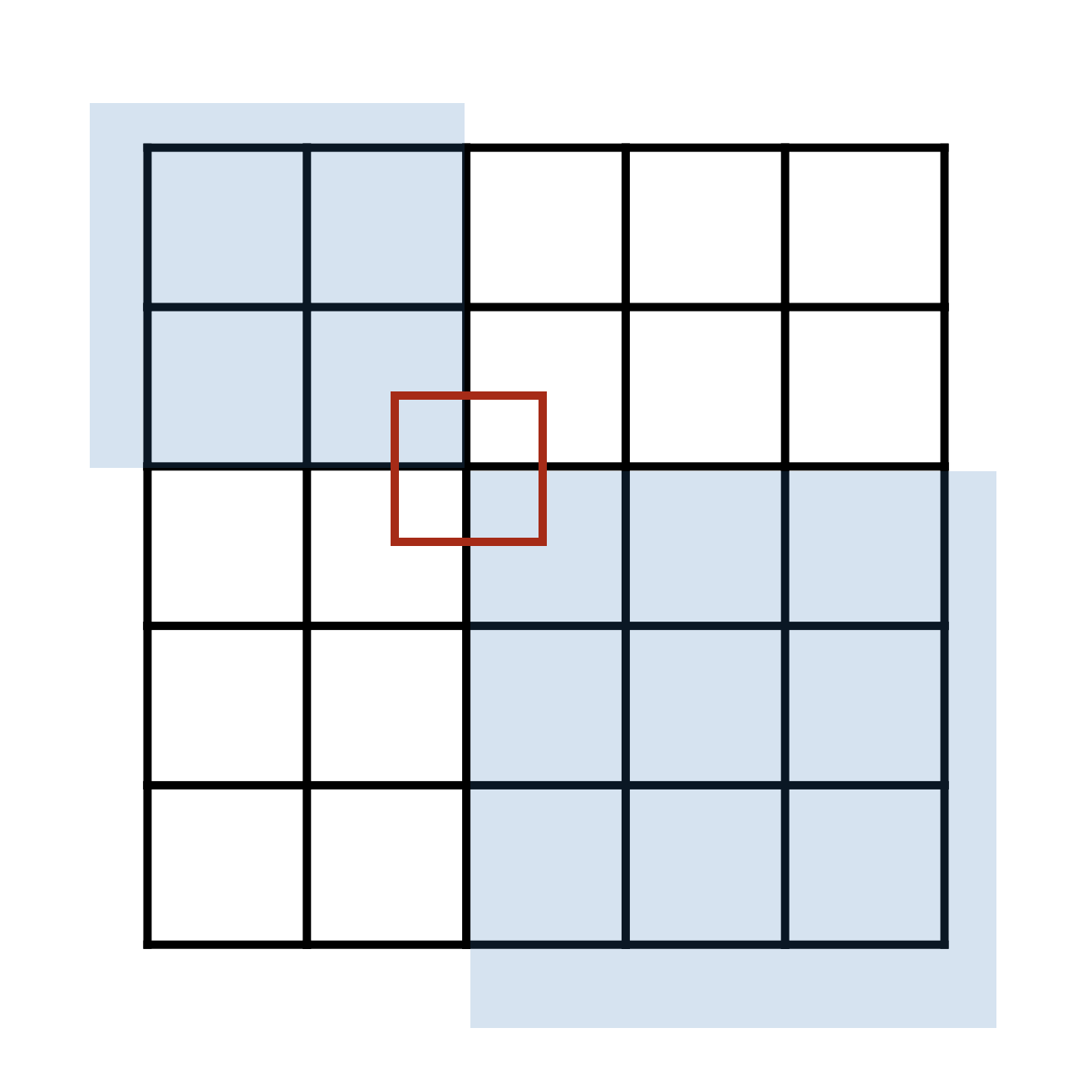}}
	\caption{Ground states for the fracton model on the Euclidean lattice.
		(a) Starting from the ground state  of all spins up,
		a new ground state can be constructed by flipping 
		all spins on one side of a vertical or horizontal line of the lattice.
		In this figure flipping all spins in the blue region will create another ground state.
		(b) By repeating the procedure described in (a) for different lines consecutively,
		any ground state can be constructed. 
		This figure shows a ground state constructed
		by two such flipping operations.
		\label{FIG_EU_GS_degeneracy}
	}
\end{figure}

The classical ground states are the spin configurations satisfying 
\begin{equation}
\mathcal{O}_p=1
\end{equation}
on all four-spin clusters.
Under open boundary condition
the ground state degeneracy and entropy are respectively
\begin{eqnarray} 
\Omega & = & 2^{L_x+L_y -1 } \;,\\
S & = & \kb\log \Omega = \kb\log2\times(L_x+L_y -1 ) \nonumber\\
& \sim & \kb\log2\times \text{(Boundary area)} \;. 
\label{eqn_flat_gs}
\end{eqnarray}
where $L_x,\ L_y$ are the sizes of the boundary in the $x-$ and $y-$direction respectively.

\tb{To construct these ground states}, we start from the obvious ground state of
the all-spin-up configuration, 
then perform the operation of flipping all spins
on one side of a straight line in the $x-$ or $y-$direction,
as shown in Fig.\ref{FIG_EU_GS_degeneracy_sub1}.
Since every \tb{four-spin cluster}  has either zero, two, or four spins flipped,
the values of all operators $\mathcal{O}_p$ remain invariant.
The system stays at its lowest energy,
and another ground state is constructed.

By repeating such operations for different straight lines 
as shown in Fig.\ref{FIG_EU_GS_degeneracy_sub2},
all ground states can be constructed explicitly.
The Shannon entropy scales with the number of straight lines,
which is the size of the boundary, 
hence comes Eq.\eqref{eqn_flat_gs}.
Actually, this
is already a hint of certain similarity between fracton models
and gravity, 
as we shall elaborate later.

{The subextensive ground state degeneracy should not
	be confused with the sub-extensive ground state degeneracy
	of three-dimensional fracton topological orders, even though they are intimately related.
	Here the subextensive ground state degeneracy is a direct reflection
	of the number of independent subsystem symmetries that scales with the boundary
	size.}\\
%
%

\noindent\textbf{Feature two: Immoblie fracton excitations --- }

\begin{figure}[t]
	\centering
	\captionsetup[subfigure]{justification=centering}
	\subfloat[One fracton state\label{FIG_EU_fracton_excitation_sub1}]
	{\includegraphics[width=0.22\textwidth]{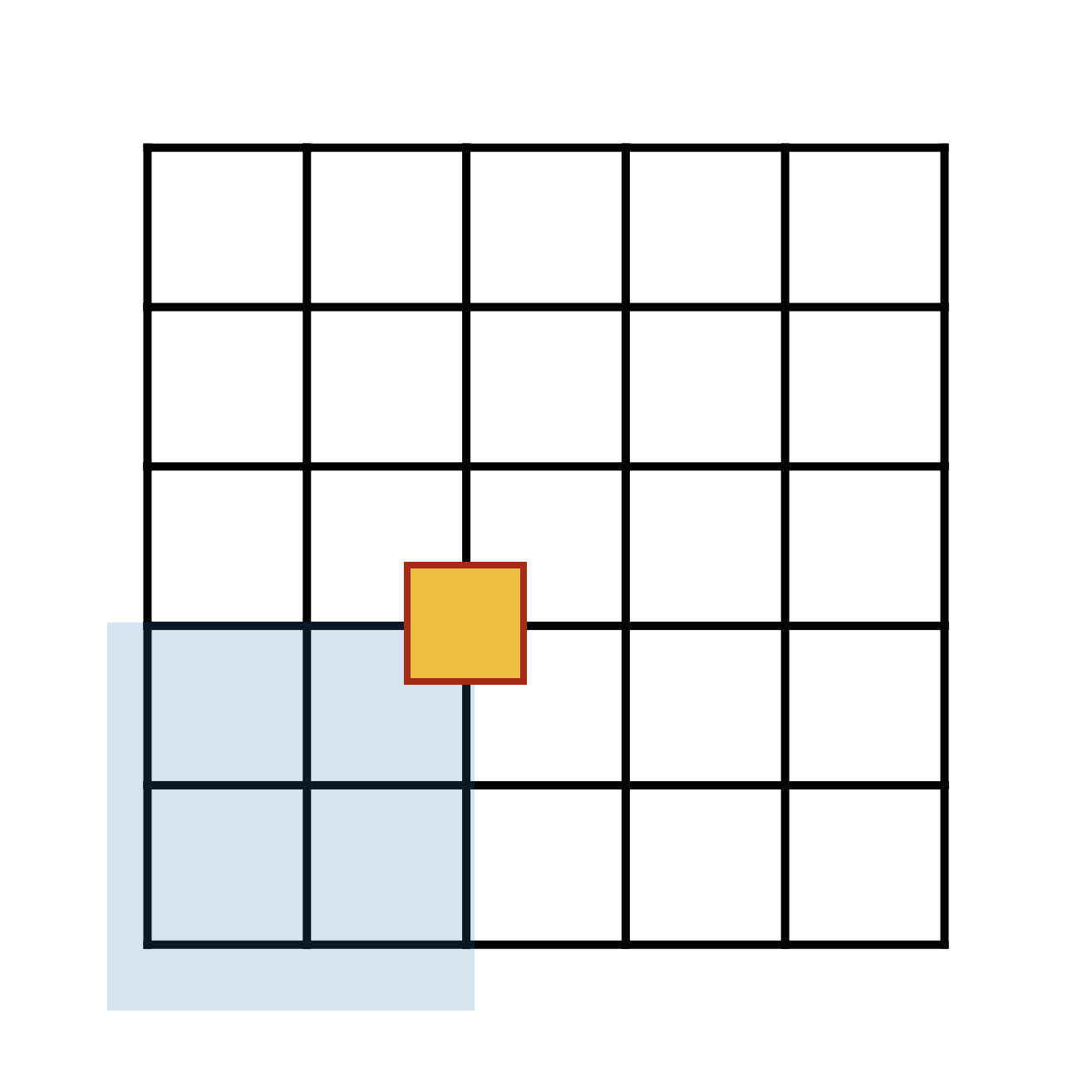}}
	\subfloat[Two-fracton bound state\label{FIG_EU_fracton_excitation_sub2}]
	{\includegraphics[width=0.22\textwidth]{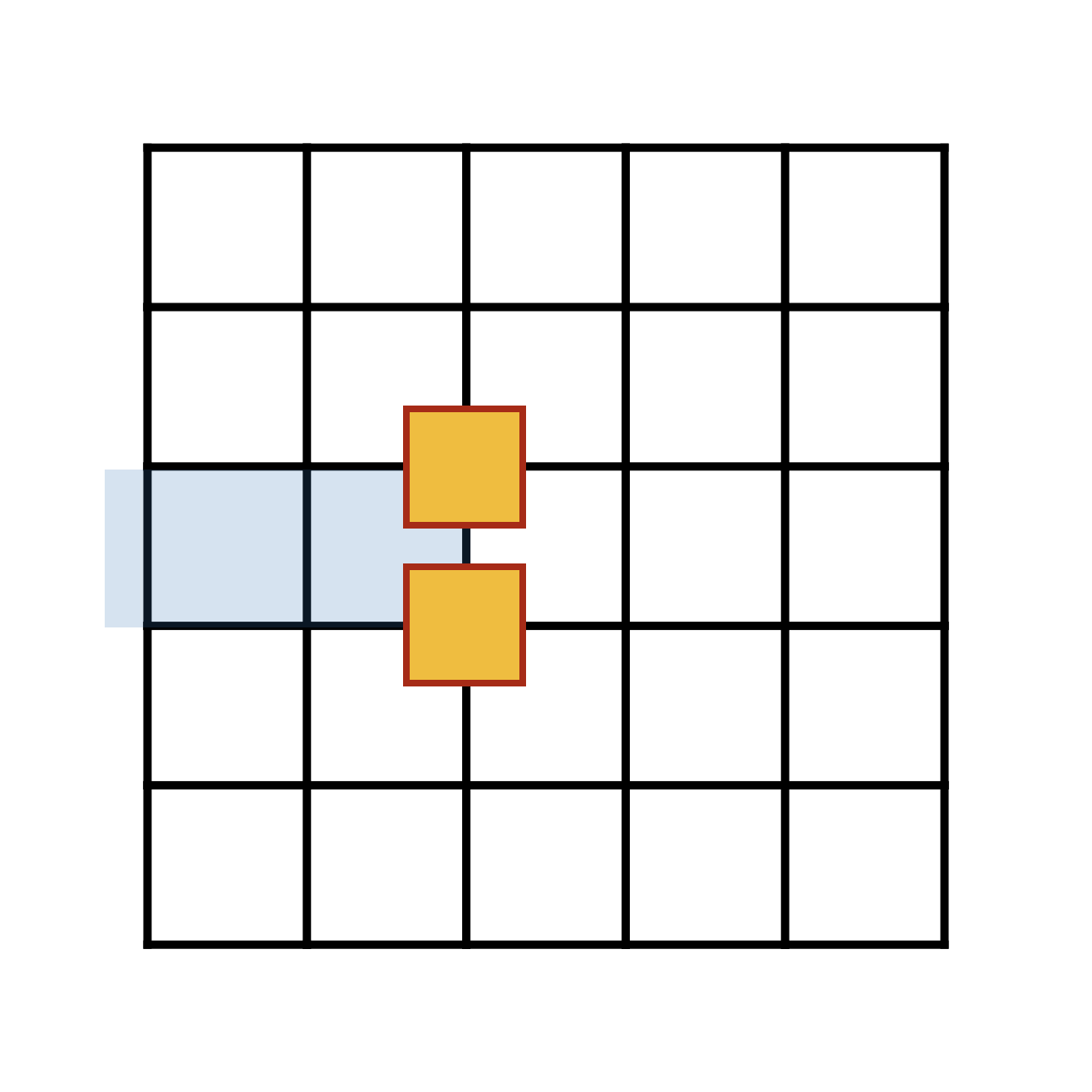}}\\
	\subfloat[Four-fracton bound state\label{FIG_EU_fracton_excitation_sub3}] 
	{\includegraphics[width=0.22\textwidth]{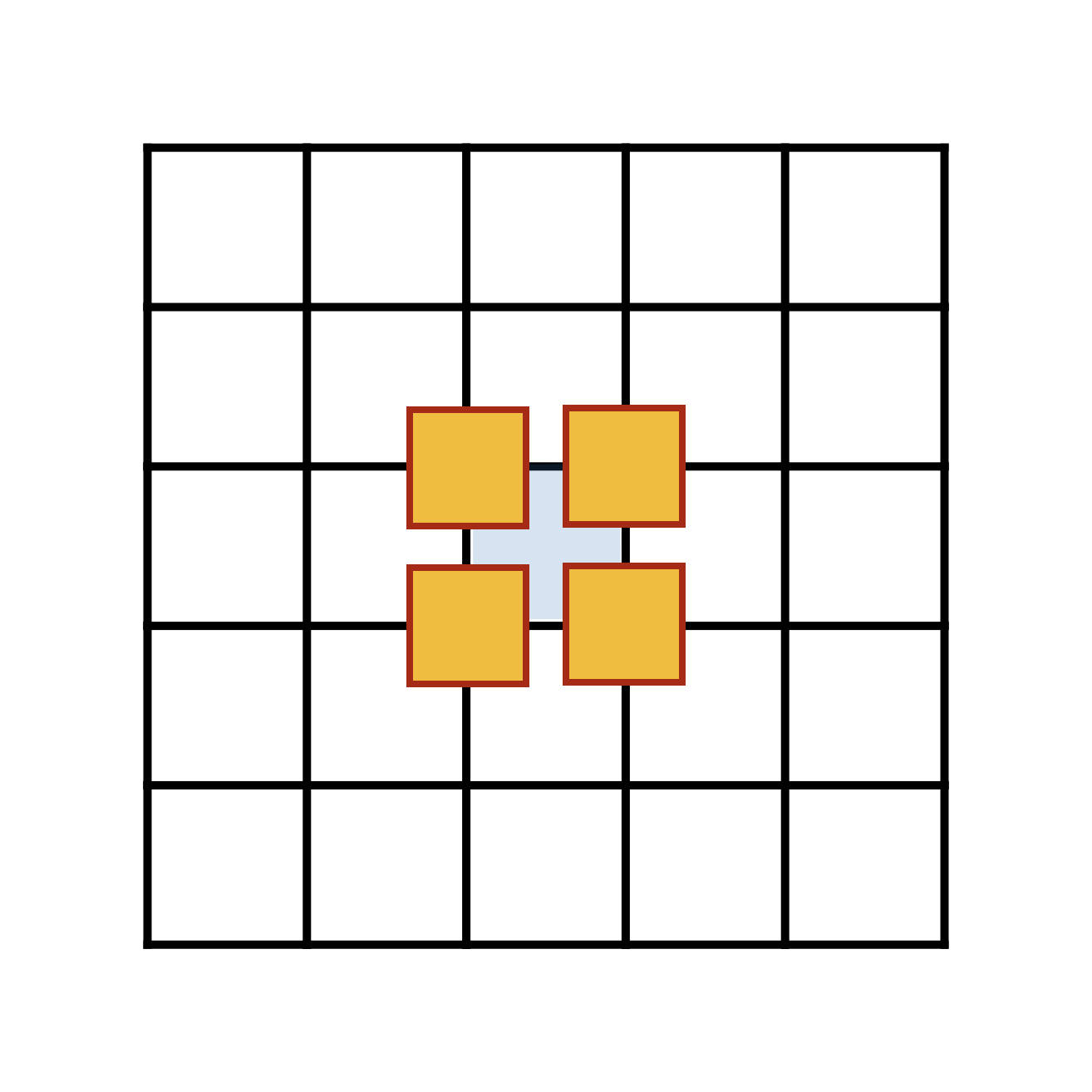}}
	\caption{Fracton excitations and their bound states.
		(a) A single fracton excitation can be created by flipping spins in the blue region bounded by two perpendicular cuts.
		It is a ``topological'' excitation, and not movable by any local, finite number of spin flips without costing the system more energy.
		(b) A two-fracton bound state can be created by flipping a semi-infinite line of spins in the blue region. 
		It is also a ``topological'' excitation. By local operations it can move horizontally but not vertically.
		(c) A four-fracton bound state can be created by a single spin flip.
		It is a local excitation, and can move freely on the lattice by local spin flip.
		\label{FIG_EU_fracton_excitation}
	}
\end{figure}

The first excited state of the model is created by
flipping the sign of only one operator $\mathcal{O}_p$,
while keeping the others invariant.
Its construction is shown in Fig.~\ref{FIG_EU_fracton_excitation_sub1}:
From any ground state one can choose two intersecting lines in the $x-$ and $y-$
directions, which split the lattice into four parts.
Then flip the spins in one quadrant of the lattice.
A fracton will be created at the intersection.
In the limit of infinite lattice, 
such operations become 
``topological'' in the sense that they involve 
infinitely many spins.
It is easy for the readers to convince themselves 
that any local operation, i.e., flipping finitely many spins in the bulk,
will create more than one fracton in the system.

Furthermore, the fracton excitation is immobile
in the sense that it is impossible for a local operation
to move it without creating new fractons and costing more energy.
To move the fracton, a  nonlocal operation of flipping a semi-infinite line of  spins next to the fracton is necessary.\\

\noindent\textbf{Feature three: Fracton bound states with enhanced mobility --- }
Now let us consider the bound state of two fractons,
created by a non-local operation
of flipping a semi-infinite line of spins,
as shown in Fig.~\ref{FIG_EU_fracton_excitation_sub2}.
The bound state can move in a one-dimensional submanifold of the system:
by local operations of extending or shrinking the 
semi-infinite line of flipped spins, the bound state can move along the direction of the line,
but it cannot move perpendicularly.

Finally, a four-fracton bound state can be created by a local single spin flip as shown in Fig.~\ref{FIG_EU_fracton_excitation_sub3},
and is obviously free to move in any direction.

The three features above are common among \tb{many fracton models with subsystem symmetries.
	The behaviors of fractons are highly generic in other types of fracton models
	including the gapped fracton topological orders and gapless rank-2 U(1) theories.
}

\subsection{Hints of Holography}\label{Sec_Hints_of_Holography}
\begin{figure*}
	\centering
	\includegraphics[width=0.8\textwidth]{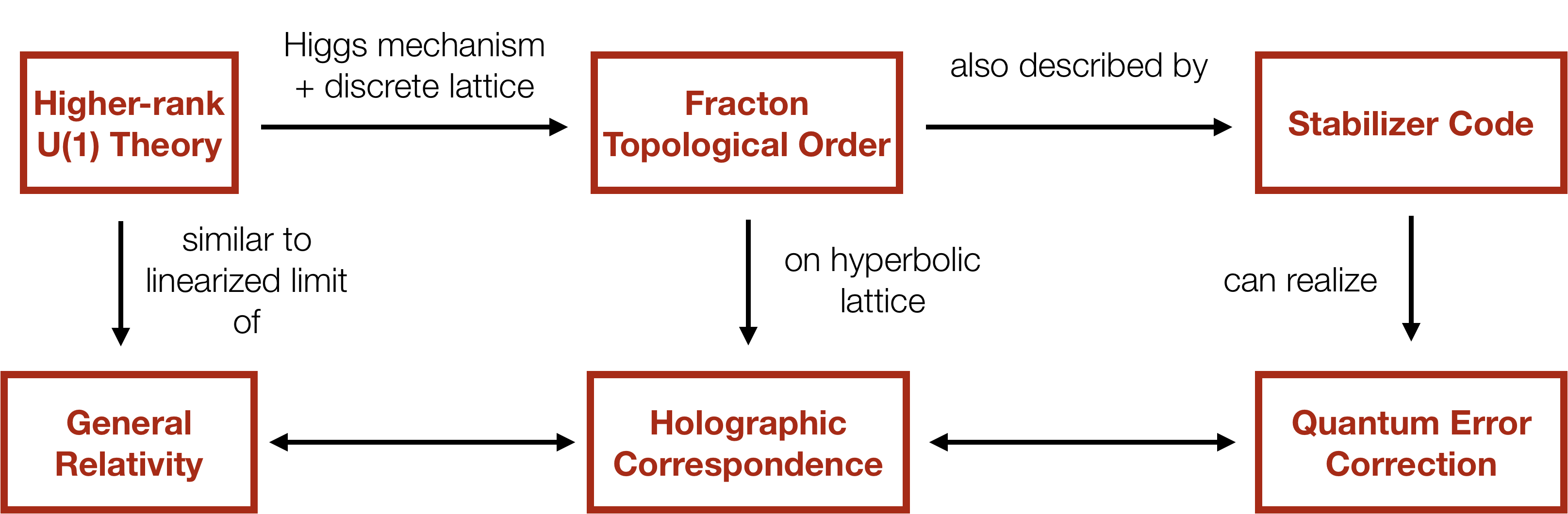}
	\caption{Speculated connections between different theories, where hints of the fracton holography hide.
		Certain rank-two gauge theories have a gauge structure similar to that of linearized general relativity.
		These theories on a lattice with a proper Higgs mechanism yield various \tb{gapped fracton models}.
		Since gravity is holographic, this implies that a fracton model on a hyperbolic lattice may also show features of holography, which is established in this work.
		The \tb{gapped fracton models can also be described by stabilizer code
			tensor networks, which provides another potential venue 
			to study them utilizing methods developed in 
			holographic tensor networks}
		\label{FIG_Big_pic}
	}
\end{figure*}

Though the model has some exotic features, 
it is not obvious how it could be holographic.
Here we 
reveal some hints
suggested by 
properties 
of certain fracton models in two- and three-dimensional Euclidean space.
The overall speculated big picture of these connections is illustrated in Fig.~\ref{FIG_Big_pic}.\\
%
%

\noindent{\bf Sub-extensive ground state degeneracy ---}
The first hint is the entropy of the system's ground state being proportional to 
its boundary area, as already demonstrated in Eq.~\eqref{eqn_flat_gs}.
Readers familiar with quantum gravity and holography will recall 
the same rule for a black hole, 
and gravitational systems in general \cite{Hooft1993,Susskind1995,Bousso2002}.
Indeed, in later sections the construction of degenerate ground states,
or the structure of subsystem symmetries,
is closely related to their holographic properties.\\

\noindent{\bf Similarity between linearized gravity and rank-two U(1) gauge theory ---}
\tb{The second hint lies in the effective theories of the gapless rank-2 U(1) 
	symmetric tensor gauge theories \cite{Pretko2017b,Pretko2017a,Pretko2017,Ma2018,Bulmash2018}.
	The gapped fracton models can usually be obtained by Higgsing the gapless gauge 
	freedoms.}
Here we do not intend to give a self-contained account of these theories,
but will only mention the key results.
More rigorous, detailed analysis can be found in the references cited.

First let us review some facts of linearized Einstein gravity.
The metric of spacetime $g^{\mu\nu}$ is assumed to have only small perturbation
away from the flat spacetime,

\begin{equation}
g^{\mu\nu}  = \eta^{\mu\nu} + h^{\mu\nu} \;,
\end{equation}

where $\eta^{\mu\nu}$ is the Minkowski metric, 
and $ h^{\mu\nu}$ is the small perturbation.
The gauge symmetry is a subset of the differmorphism invariance \cite{leclerc2006faddeev, Pretko2017}

\begin{equation}
h^{\mu\nu}  \rightarrow  h^{\mu\nu} +\partial^\mu \xi^\nu +\partial^\nu \xi^\mu \;.
\end{equation}

It turns out that $h^{00}$ and $h^{i0}$ serve 
as Lagrangian multipliers in the Lagrangian.
The physical degrees of freedom are $h^{ij},\ (i,j=1,2,3)$ , whose canonical conjugates
we denote as $\pi^{ij}$.
We can write down the gauge constraints
and gauge transformations for them. 
For the convenience of comparing to rank-two U(1) theories,
we write them in two groups:
\begin{eqnarray} \label{EQN_LinGR_1}
\partial_i\pi^{ij} & = & T^{0j} \;,\nonumber\\
h_{ij}  & \rightarrow & h^{ij} +\partial^i \xi^j +\partial^j \xi^i \;,
\end{eqnarray}
and
\begin{eqnarray} \label{EQN_LinGR_2}
\partial_i\partial_j h^{ij}-\partial^2 h^i_i  & =&  T^{00} \;,\nonumber\\
\pi^{ij} & \rightarrow & \pi^{ij}+ \partial^i\partial^j\alpha-\delta^{ij}\partial^2\alpha \;.
\end{eqnarray}

Now we turn to the rank-two $U(1)$ theories. 
One version of them has a symmetric tensorial electric field
\begin{equation}
E^{ij}  =  E^{ji} \;,
\end{equation}
with associated vector charge defined as 
\begin{equation} \label{EQN_R2_1_1}
\partial_i E^{ij} = \rho^j \;.
\end{equation}
As a result, the corresponding gauge field has symmetry  \cite{rasmussen2016stable}
\begin{equation} \label{EQN_R2_1_2}
A^{ij} \rightarrow A^{ij} +\partial^i \lambda^j +\partial^j \lambda^i \;.
\end{equation}
If  $E^{ij}$ is identified as the conjugate momentum of $A^{ij}$, 
Eq.~(\ref{EQN_R2_1_1},\ref{EQN_R2_1_2}) are equivalent to
Eq.~\eqref{EQN_LinGR_1}.

Since $h^{ij}$ and $\pi^{ij}$ are conjugate with each other,
we can also treat $\pi^{ij}$ as the gauge field and 
$h^{ij}$ as the momentum.
This is partially captured by another version of the rank-2 $U(1)$ theory,
which has a symmetric, traceless tensorial electric field, and associated scalar charge, defined by
\begin{eqnarray} \label{EQN_R2_2_1}
E^i_{\ i}=0,\; \partial_i E^{ij} = \rho^j \;.
\end{eqnarray}
Its gauge freedom is
\begin{equation} \label{EQN_R2_2_2}
A^{ij} \rightarrow A^{ij} +\partial^i \partial^j \lambda \;.
\end{equation}
%
%
%
\tb{
	As Ref.\cite{Pretko2017} pointed out,
	the similarity between 
	Eq.~(\ref{EQN_R2_2_1},\ref{EQN_R2_2_2}) and
	Eq.~\eqref{EQN_LinGR_2}
	implies some shared properties between gravity and rank-2 U(1)
	theories.}

\tb{
	Other studies have also shown connections
	between fracton models and gravity.
	For example, Ref.\cite{RasmussenPhysRevB2018}
	shows that linearized gravity harbors gapless topological order.
	More recently, the foliated field theory for fracton models 
	has been proposed and 
	found to correspond to a singular limit of tetradic Palatini gravity \cite{Slagle2018arXiv},
	indicating connections between fracton topological order and 
	soft hairs in gravity.
}

%

\section{Brief Review of the AdS/CFT Correspondence}\label{SEC_AdSCFT}
The holographic principle states that 
a gravitational theory describing a region of space is equivalent
to a non-gravitational theory living  on its boundary.
For readers  unfamiliar with holography, 
we present a brief summary of the essential results relevant to this work. 
More thorough introductions can be found in Ref.\cite{Polchinski2010,nastase2007introduction,klebanov2001tasi,maldacena2003tasi}.

\subsection{Black Hole Information Paradox}
This profound principle was firstly motivated by the black hole entropy.
As a pure classical, exact solution to Einstein's equations of general relativity,
a black hole should have zero entropy.
However, this violates the second law of thermodynamics,
since we lose information on whatever objects pass the horizon \tb{when falling into the black hole}.
This is partially resolved by the Bekenstein-Hawking black hole entropy \cite{Bekenstein1973,Hawking},
which states that a black hole actually 
has entropy proportional to the area of its horizon
\begin{equation} \label{EQN_BH_Entropy}
S_{BH}=\frac{A}{4G_N}\;,
\end{equation}
where $A$ is the horizon area, and $G_N$ is the Newtonian constant.
The entropy can be interpreted as counting the microstates of a black hole.
Hence Eq.~\eqref{EQN_BH_Entropy} indicates that the number of degrees of freedom for 
a black hole is proportional to its horizon area,
instead of its volume like conventional quantum field theories.
This echoes the holographic principle, which states that the degrees 
of freedom are living on the boundary instead of in the bulk.

\subsection{AdS/CFT Correspondence}
The AdS/CFT correspondence is a more concrete realization of holography.
It is a duality between a  gravitational theory in $d+1-$dimensional AdS space 
and $d-$dimensional CFT on its boundary. 

\begin{figure}[t]
	\centering
	\includegraphics[width=0.45\textwidth]{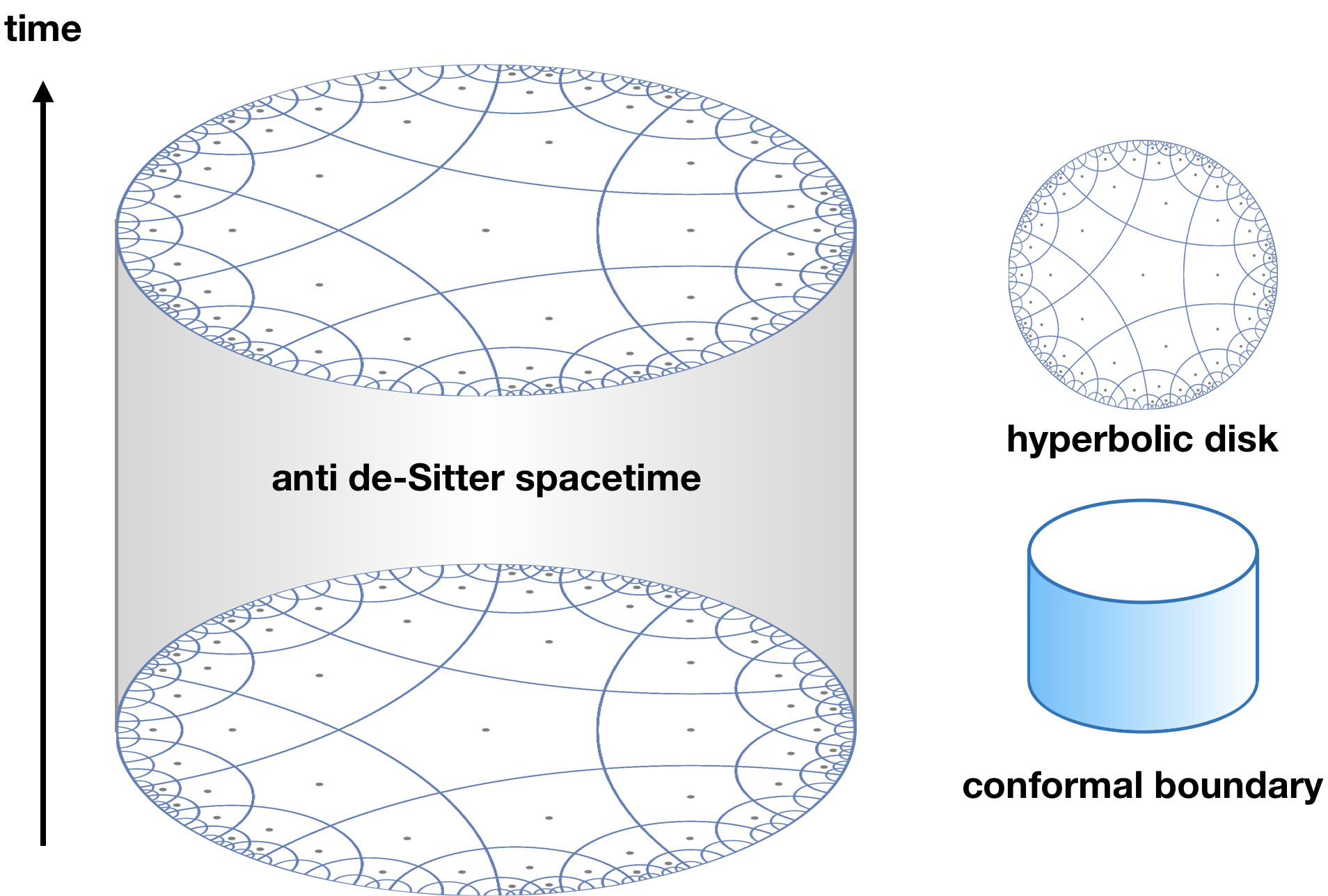}
	\caption{Anatomy of anti-de Sitter (AdS) spacetime. 
		It can be viewed as a stack of constant-negative-curvature spatial slices in temporal direction. 
		Each slice here is a hyperbolic disk.
		\tb{Note that the temporal direction is not simply straight upward.}
		The boundary of the AdS spacetime, as shown on the bottom right panel, is where a conformal field theory (CFT) lives.
		\label{FIG_ADS_Space}
	}
\end{figure}

An AdS space has constant negative curvature,
equipped with the metric
\begin{equation}
ds^2=\frac{R^2}{u^2}(-dt^2+d\vec{x}^2+du^2) \;,
\end{equation}
which can be seen as AdS spatial slices stacked in the temporal direction.
In Fig.~\ref{FIG_ADS_Space}, an AdS$_3$ space is illustrated as a stack of hyperbolic disks.

The first example of AdS/CFT proposed by Maddalena 
is the duality between type-IIB superstring theory in the bulk of AdS$_5\times$S$_5$
and large-$N$ $\mathcal{N}=4$ super-Yang-Mills theory on the boundary \cite{Maldacena1999}.
It suggests that there should be no information loss with black holes in a gravitational system,
since it is equivalent to some non-gravitational quantum physics \tb{in which} information is preserved.

\subsection{Ryu-Takayanagi Formula}
\begin{figure}[t]
	\centering
	\subfloat[RT formula\label{FIG_RT_Rindler_sub_1}]
	{\includegraphics[width=0.22\textwidth]{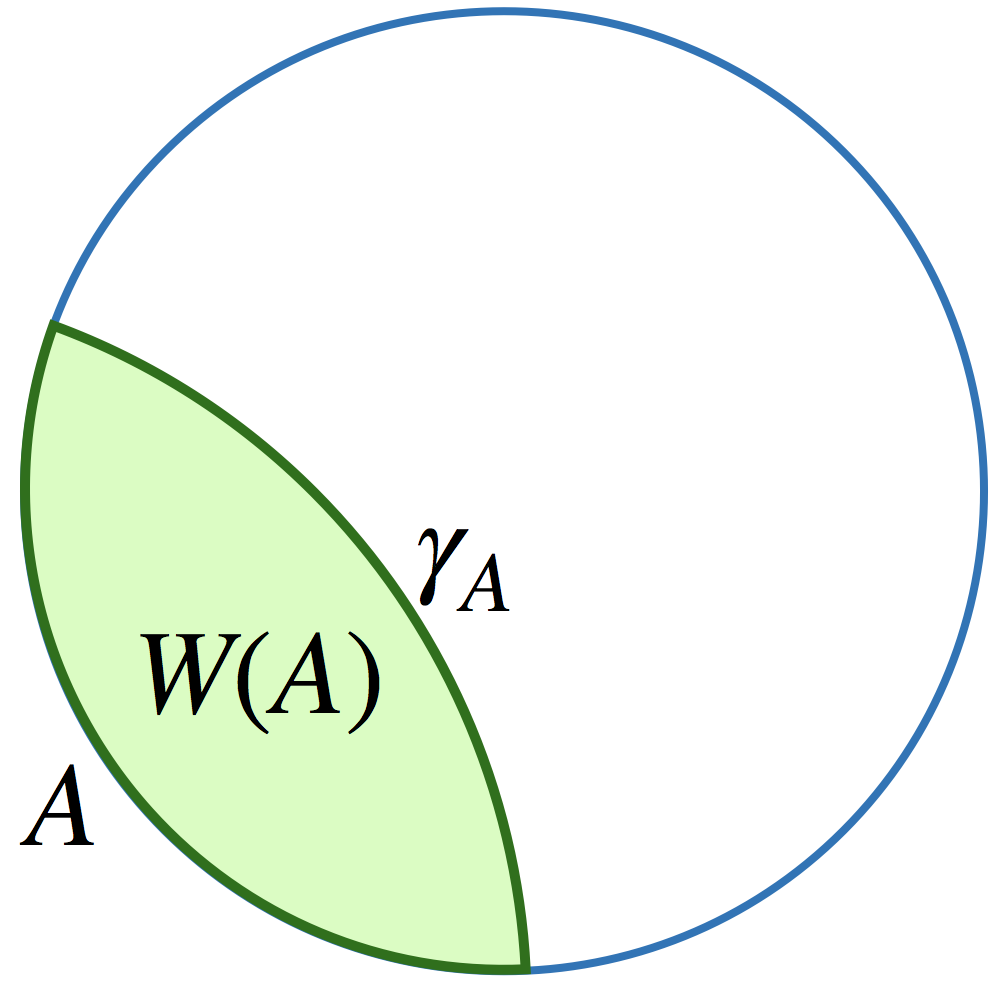}}
	\subfloat[Rindler reconstruction\label{FIG_RT_Rindler_sub_2}]
	{\includegraphics[width=0.22\textwidth]{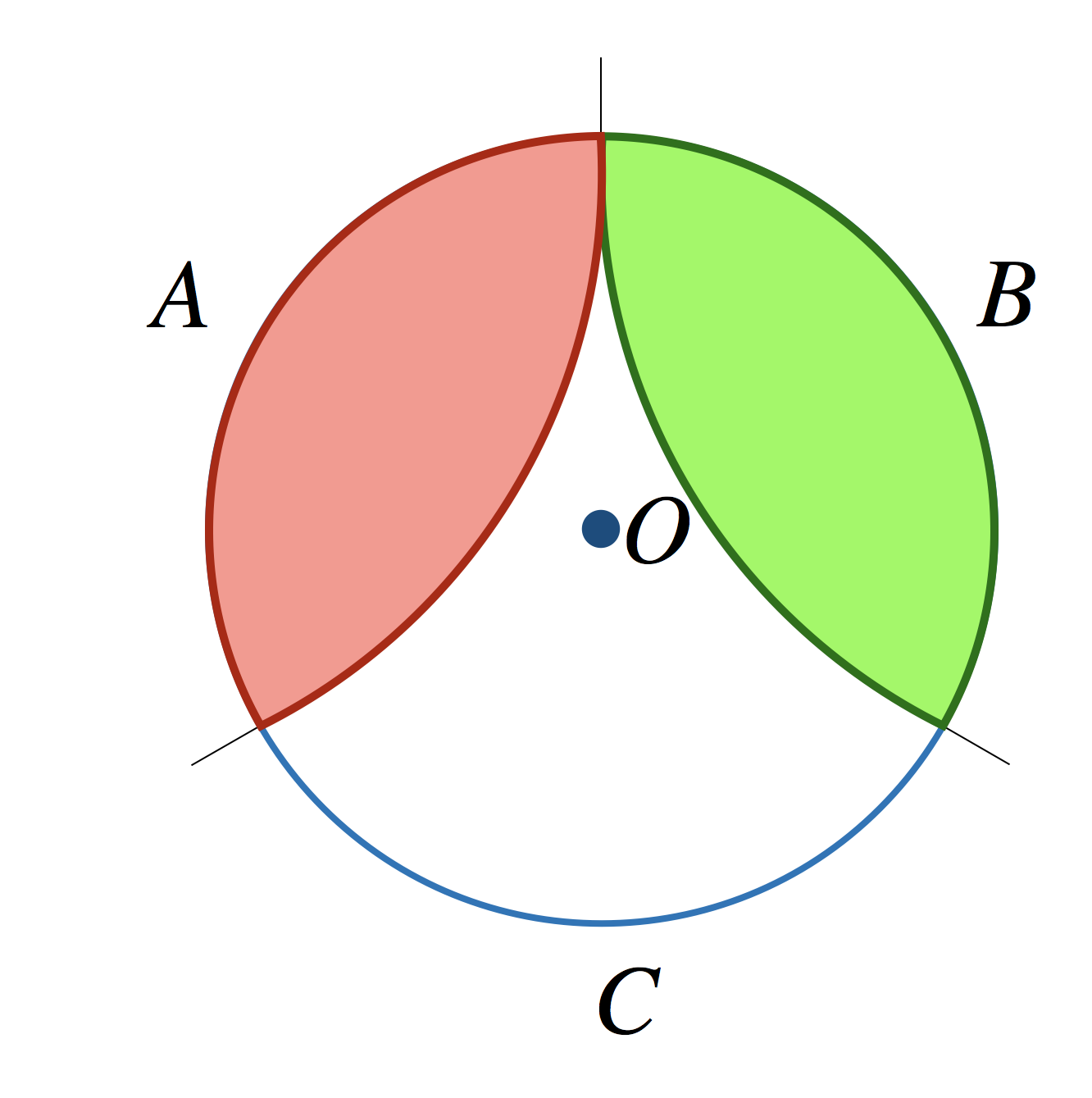}}\\
	\subfloat[Rindler reconstruction\label{FIG_RT_Rindler_sub_3}]
	{\includegraphics[width=0.22\textwidth]{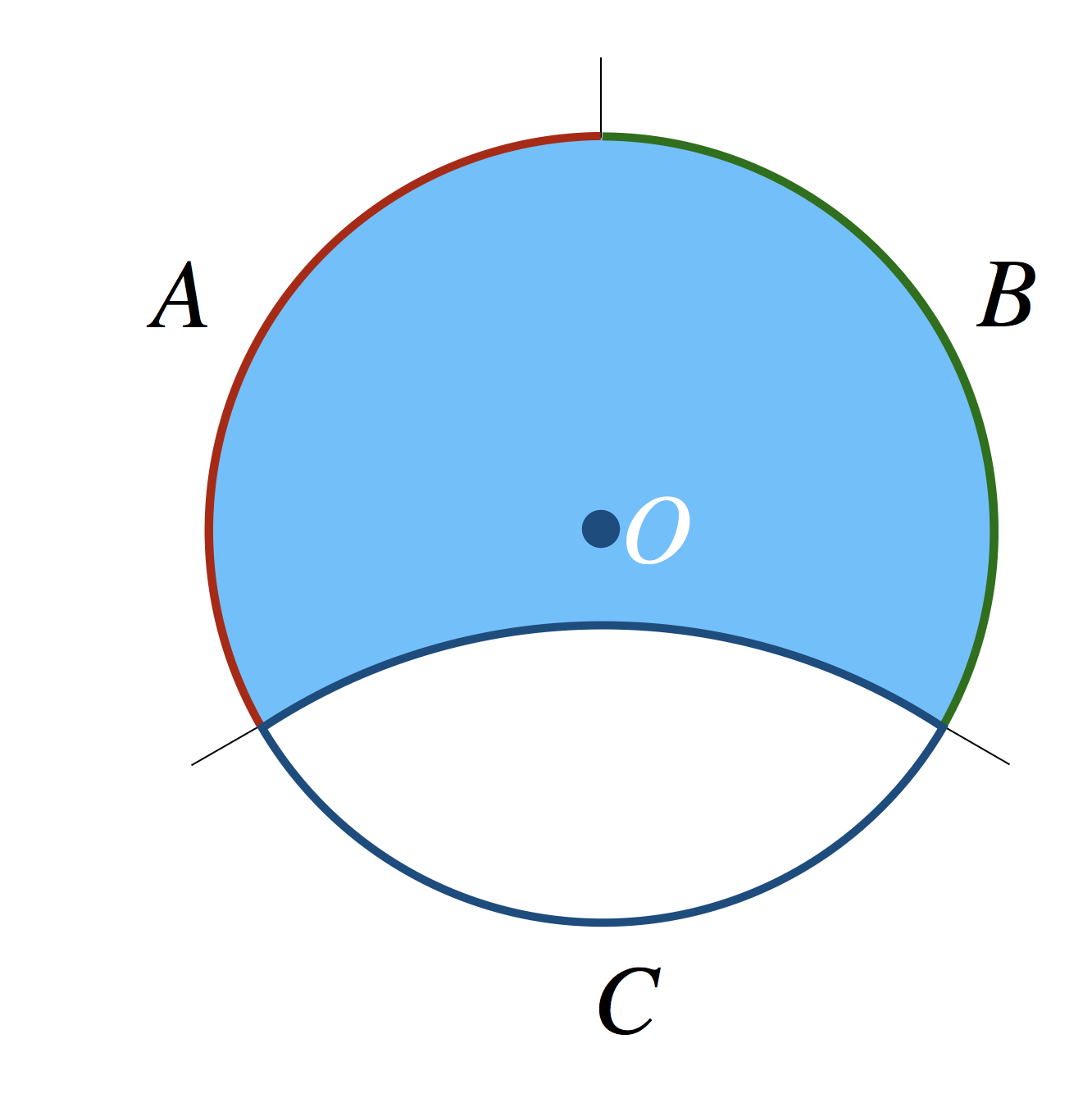}}
	\caption{Ryu-Takayanagi (RT) formula of entanglement entropy and Rindler reconstruction. 
		(a) RT formula for entanglement entropy.
		The boundary subregion A and its complement $A_c$'s entanglement entropy
		is proportional to the area of $\gamma_A$, the minimal surface in the bulk covering $A$.
		Given boundary states on $A$, bulk operators in the entanglement wedge $W(A)$ (shaded volume) can be reconstructed.
		(b),(c) An example of Rindler reconstruction. 
		The bulk operator $\mathcal{O}$ cannot be reconstructed by boundary region $A$, $B$,
		or $C$ individually as it lies outside  
		each individual entanglement wedge (shaded volumes).
		However, it is included in the entanglement wedge of $A\bigcup B$, and can be reconstructed
		when the boundary states on both $A$ and $B$ are known. 
		\label{FIG_RT_Rindler}
	}
\end{figure}

The Ryu-Takayanagi formula reveals the deep connection between the geometry
of the AdS \tb{spacetime} and the entanglement of the boundary CFT states.
Assuming that the CFT lives on the boundary of some asymptotic AdS space,
for a region $A$ on that boundary,
there exists a corresponding minimal \tb{codimension-one} surface $\gamma_A$  such that 
(1)it
is homologous to $A$ in the asymptotic AdS bulk; i.e.,  its boundary coincides with
the boundary of $A$, or $\partial\gamma_A=\partial A$;
(2) 
its area is extremal (in our case minimal) among  all surfaces satisfying (1).
The union of $A$ and $\gamma_A$ encloses a volume denoted the \textit{entanglement wedge}
$W(A)$.
The Ryu-Takayanagi formula indicates that the entanglement entropy $S_A$
of the CFT states between $A$ and its complement $A^c$
is proportional to the area of $\gamma_A$, ignoring higher-order bulk contributions \cite{Ryu2006,Ryu2006a}:
\begin{equation}
S_A=\frac{\text{Area}(\gamma_A)}{4G_N} \;.
\end{equation}
This is illustrated in Fig.~\ref{FIG_RT_Rindler_sub_1}.

\subsection{Subregion Duality and Rindler Reconstruction}
Since AdS/CFT is a duality between the boundary and the bulk physics,
it is crucial to understand how much boundary information is needed to 
reconstruct a bulk state or operator, and how the state is reconstructed.
It is a subtle issue in the presence of temporal direction,
which we do not intend to discuss.
Fortunately, we only 
work on a spatial slice of the AdS$_3$ \tb{spacetime} 
like most of the tensor-network models,
when the laws of bulk reconstruction are significantly simplified:
The bulk state can be constructed from a boundary segment $A$
if and only if it is within the entanglement wedge
$W(A)$, as shown in Fig.~\ref{FIG_RT_Rindler}.

An educative example is to examine the tripartition $A,B,C$ of the boundary
and a  bulk operator $\mathcal{O}$ at the center of the hyperbolic disk ( Figs.~\ref{FIG_RT_Rindler_sub_2},~\ref{FIG_RT_Rindler_sub_3}.).
The entanglement wedge of any single one of regions $A$, $B$, or $C$
does not include the bulk site,
meaning  $\mathcal{O}$ cannot be reconstructed from these boundary states  .
However, the union of any two boundary segments has an entanglement wedge 
that covers $\mathcal{O}$, 
so given states on two of the three boundary segments,
$\mathcal{O}$ can be reconstructed.

This example indicates the highly nontrivial entanglement \tb{structure}
of the boundary states.
It is captured by the quantum error correction code \cite{Pastawski2015}.
and realized in the perfect tensor networks and random tensor-networks  \cite{Almheiri2015,Qi2018}.

\section{The Hyperbolic Fracton Model} \label{SEC_Hyper_Fracton}
Given the hints of holography discussed in Sec.~\ref{Sec_Hints_of_Holography},
it is natural to consider the fracton model discussed in
Sec.~\ref{SEC_EU_fracton}
transplanted 
to the hyperbolic lattice.
The hyperbolic lattice is a symmetric, uniform tiling of the hyperbolic disk,
which is a spatial slice of the AdS spacetime,
or a two-dimensional space of constant negative curvature, 
as shown in Fig.~\ref{FIG_ADS_Space}.
Most features of the fracton model are preserved 
on the hyperbolic lattice as we explain below.
We also note that the fracton model 
on a curved space has been discussed in
Refs.~\cite{Shirley2017,Slagle2018,Slagle2017-2}.

%
\subsection{The Hyperbolic Lattice and the Model}
\begin{figure}[t]
	\centering
	\captionsetup[subfigure]{justification=centering}
	\subfloat[Hyperbolic lattice\label{FIG_HY_Lattice_sub_2}]{\includegraphics[height=0.22\textwidth]{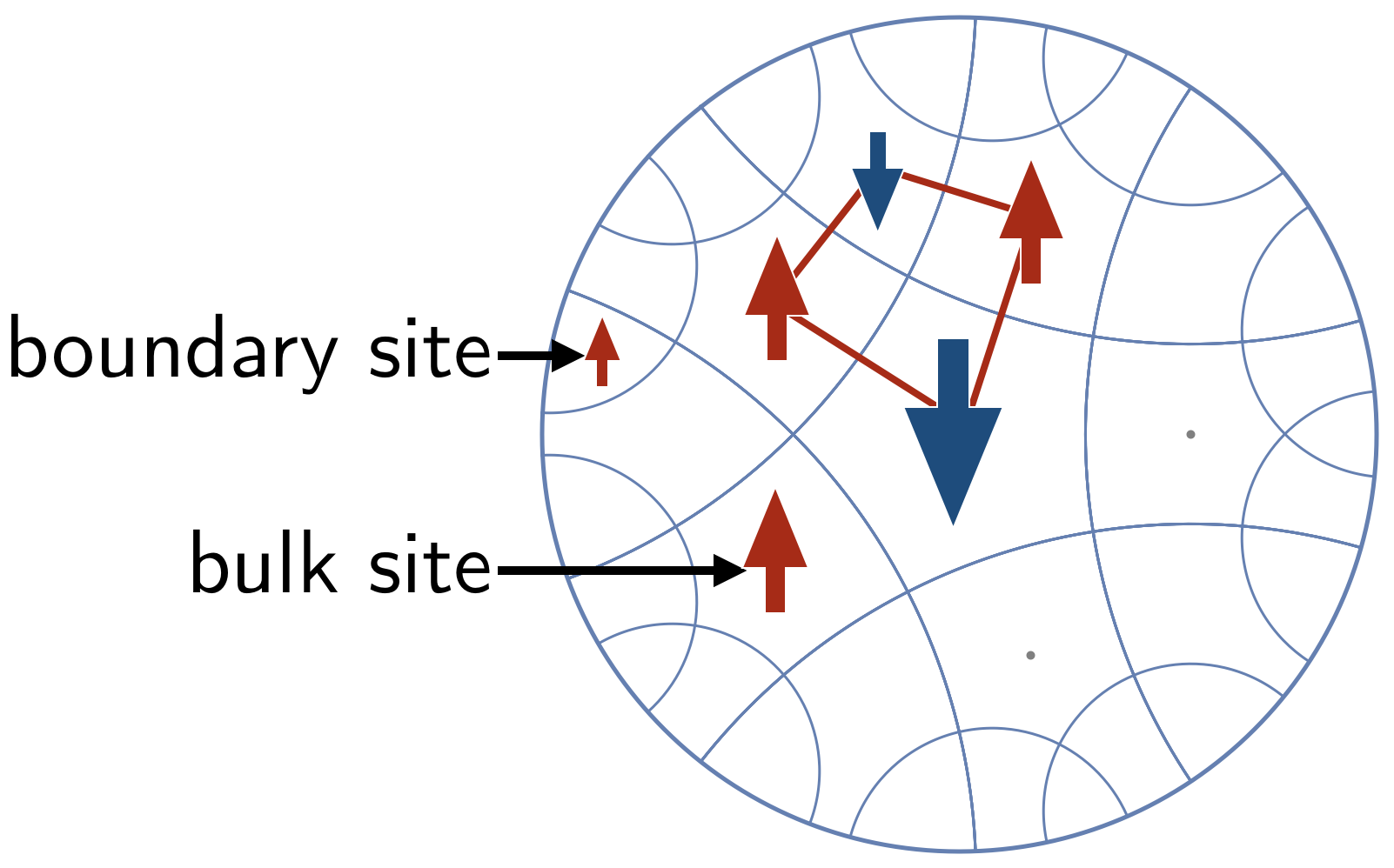}} \\
	\subfloat[Hyperbolic lattice\label{FIG_HY_Lattice_sub_3}]{\includegraphics[width=0.22\textwidth]{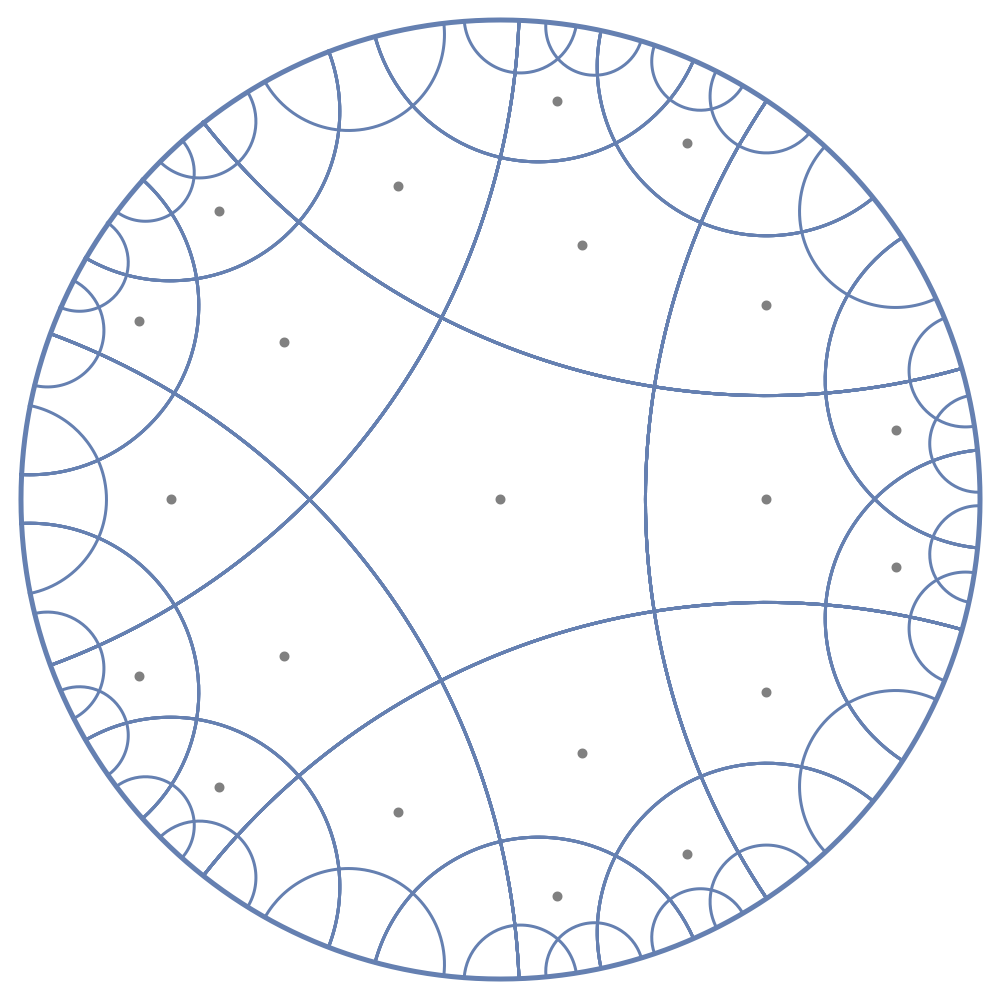}}
	\subfloat[Hyperbolic lattice\label{FIG_HY_Lattice_sub_4}]{\includegraphics[width=0.22\textwidth]{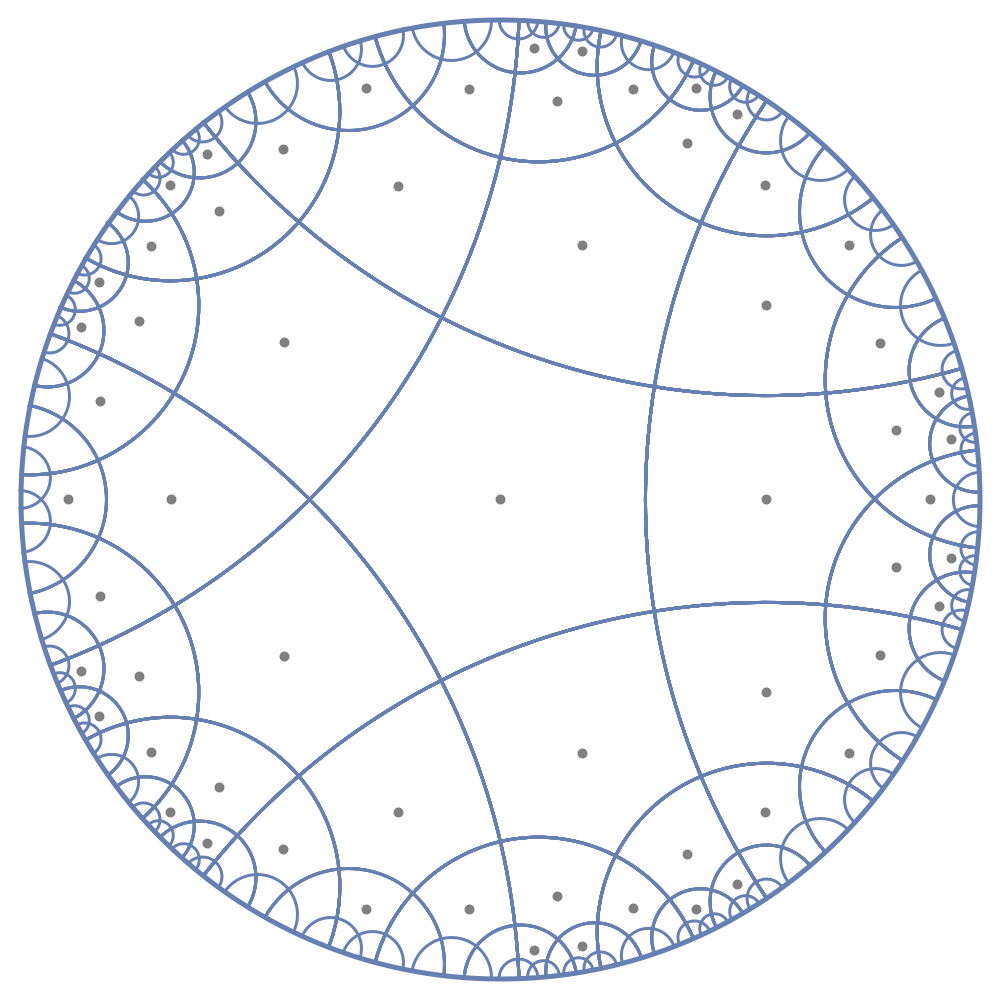}}
	\caption{
		\tb{
			Hyperbolic lattice for the fracton model.
			(a): The lattice as the (5,4) tessellation of the hyperbolic disk. 
			The spins sit at the center of the unit plaquettes (pentagons in the bulk or plaquettes on the boundary).
			The red square shows four pentagons that
			form the four-spin cluster interaction term $O_p$ [Eq.~\eqref{EQN_HY_Op}]. 
			(b,c): Lattices of different sizes.
			They can be obtained by adding more pentagon-edge geodesics
			far from the center.
		}
		\label{FIG_HYP_LAT}
	}
\end{figure}

%
%
We will use the \tb{$(5,4)$} tessellation of the hyperbolic disk ( Fig.~\ref{FIG_HYP_LAT}),
that is, tiling it with \tb{pentagons}, 
and each corner of a \tb{pentagon} is shared by  \tb{four pentagons} in total.
An Ising spin of value $\pm1$ is placed at \tb{the \textit{center} of each pentagon in the lattice}.
It is a natural generalization of the two-dimensional fracton model on the Euclidean lattice 
discussed in previous sections.

%
%
The $(5,4)$ tessellation has the \tb{four-spin cluster
	for four pentagons sharing the same corner}.
On the clusters we define the operator again
\begin{equation}\label{EQN_HY_Op}
\mathcal{O}_p=\prod_{i=1}^{4} S^z_i \;,
\end{equation}
where $i$ runs over \tb{four spins at the centers of the pentagons}, 
and $S^z_i=\pm1$.
The Hamiltonian is 
\begin{equation}\label{C4_eqn_hamiltonian}
\mathcal{H}=-\sum_{p}\mathcal{O}_p \;,
\end{equation}
and the values of $\mathcal{O}_p$ on different clusters
are independent of each other.
%
The hyperbolic lattice is illustrated in Fig.~\ref{FIG_HYP_LAT}.
%
%

%
%
When analyzing the fracton model on the Euclidean lattice,
we have used the operations of splitting the system by
straight lines very often,
\tb{since it is how the subsystem symmetries are obtained}.
They are essentially geodesics in Euclidean geometry,
made from the edges of
the square lattice.
\tb{By construction,}
these lines do not overlap with 
any spin site, 
so  every spin is unambiguously on one side of the line.

On the hyperbolic disk,
the geodesics
become  arcs on the disk that intersect the disk boundary
perpendicularly on both ends.
Thus the geodesics
defined by the $(5,4)$ tessellation,
i.e., those formed by the edges of the pentagons,
play an important role in our analysis.
They are referred to as {\it ``pentagon-edge geodesics''}.
All other conventional geodesics are simply refereed to as {\it ``geodesics''}.
\tb{
	The pentagon-edge geodesics are the blue 
	arcs in Fig.~\ref{FIG_HYP_LAT}.}
%

%
%
The hyperbolic lattice is infinite.
To study it in a controlled way,
we need to introduce a cutoff 
and unambiguously define the 
bulk and boundary sites.
This can be achieved 
by removing the infinitely many pentagon-edge geodesics far from the center,
as shown in Fig.~\ref{FIG_HYP_LAT}.
\tb{
	It is a common trick in AdS/CFT that yields 
	a finite-sized AdS space on the gravity side,
	and cuts off the ultraviolet modes
	of the CFT.
}

After \tb{the cutoff},
there will be finitely many pentagon-edge geodesics remaining,
whose number is denoted as $N_\text{g}$.
They will leave finitely many pentagons 
and their associated spins in the system,
which become the bulk.
On the boundary there will be $2N_\text{g}$ non-pentagon plaquettes, 
each partially bounded  by a segment of the disk boundary.
We place an Ising spin on each of them,
and define them to be the boundary degrees of freedom.
Hence $N_\text{g}$ can be thought of as a measure of the boundary size of the lattice.
In Fig.~\ref{FIG_HYP_LAT}, finite lattices of different sizes are illustrated.

\subsection{Ground States and Fracton Excitations}
Similarly to the Euclidean fracton model, 
the ground states and excitations can be 
explicitly constructed 
by simply replacing the straight lines with pentagon-edge geodesics.

%
%
\begin{figure}[t]
	\centering
	\captionsetup[subfigure]{justification=centering}
	\subfloat[Ground state\label{FIG_Hyper_Frac_sub_1}]
	{\includegraphics[width=0.22\textwidth]{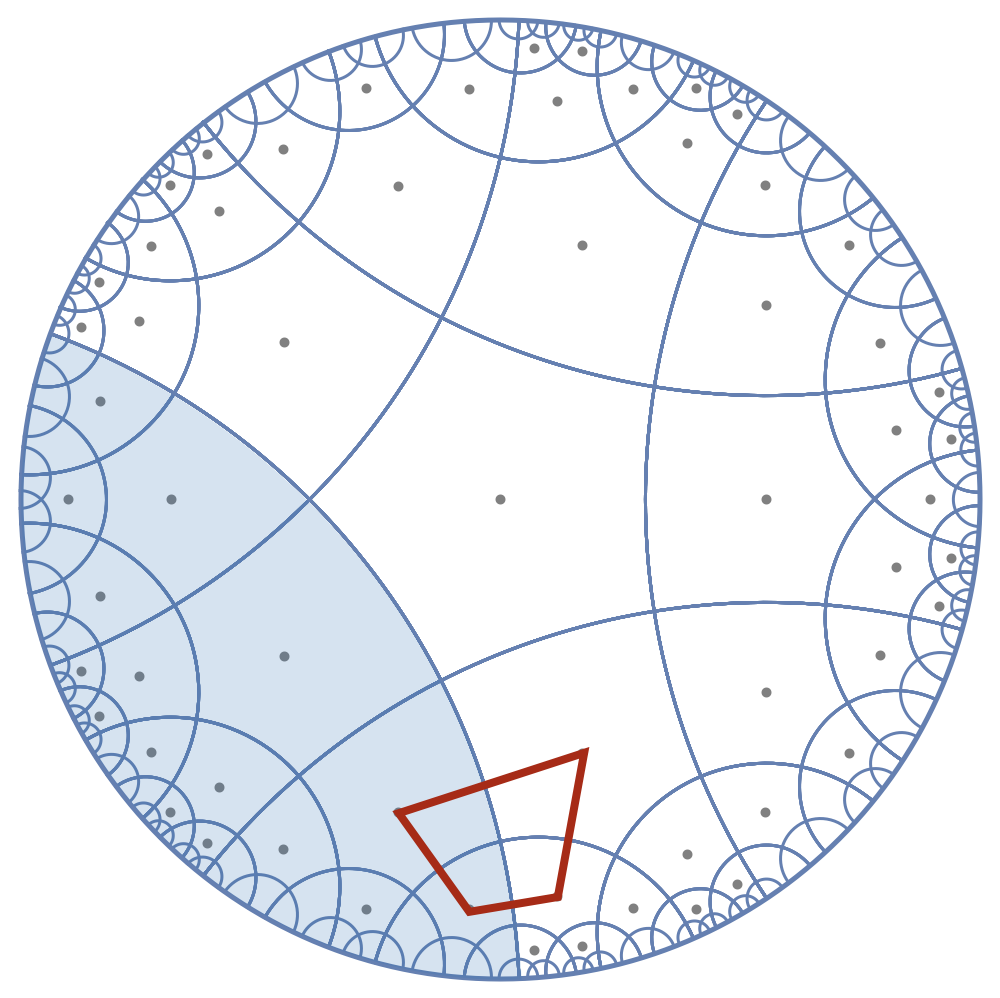}} 
	\subfloat[Another ground state\label{FIG_Hyper_Frac_sub_2}]
	{\includegraphics[width=0.22\textwidth]{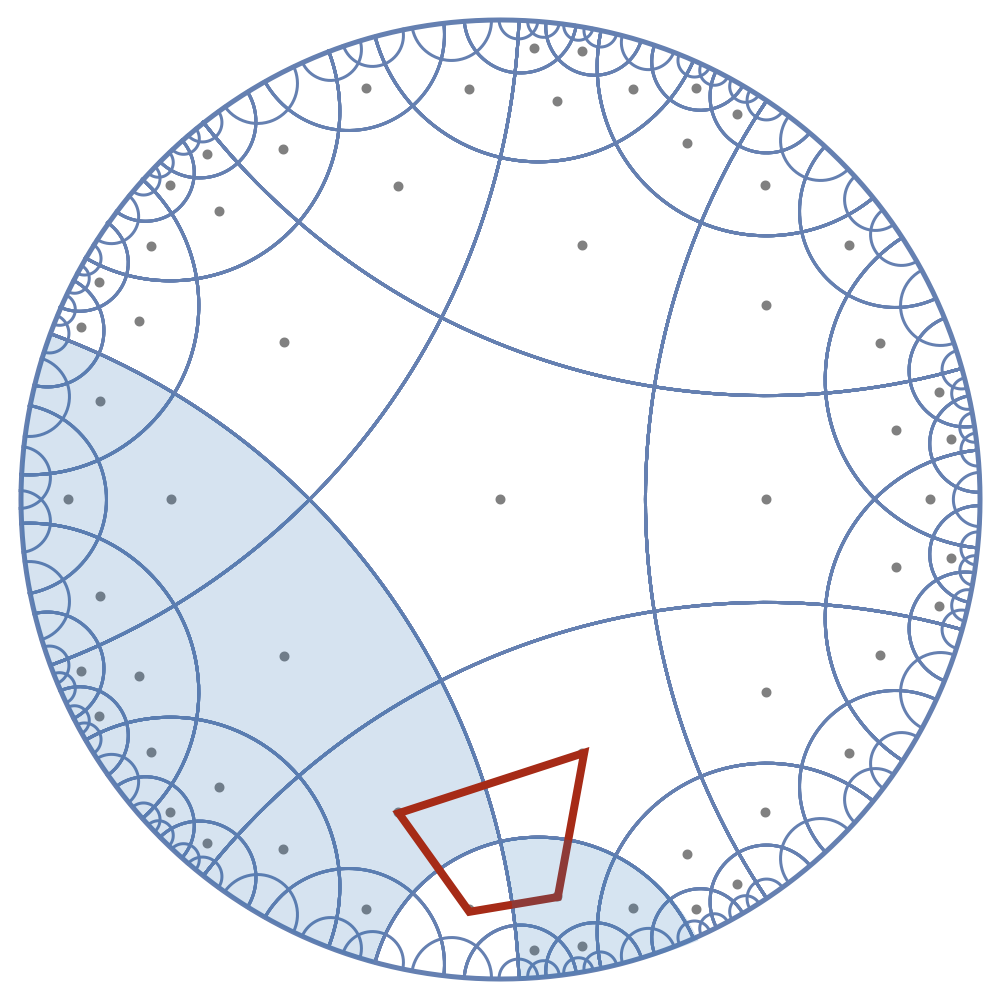}}\\
	\subfloat[Fracton excitation\label{FIG_Hyper_Frac_sub_3}]
	{\includegraphics[width=0.22\textwidth]{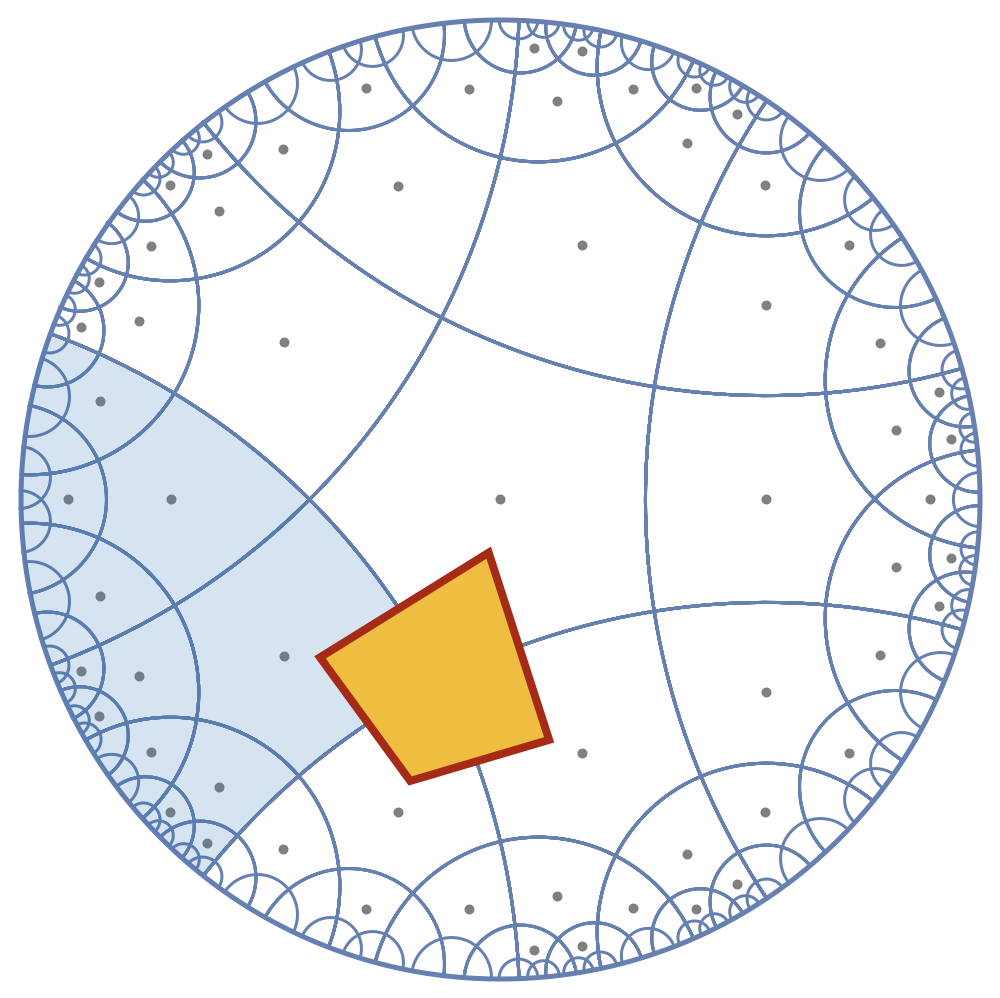}}
	\subfloat[Two fracton excitation\label{FIG_Hyper_Frac_sub_4}]
	{\includegraphics[width=0.22\textwidth]{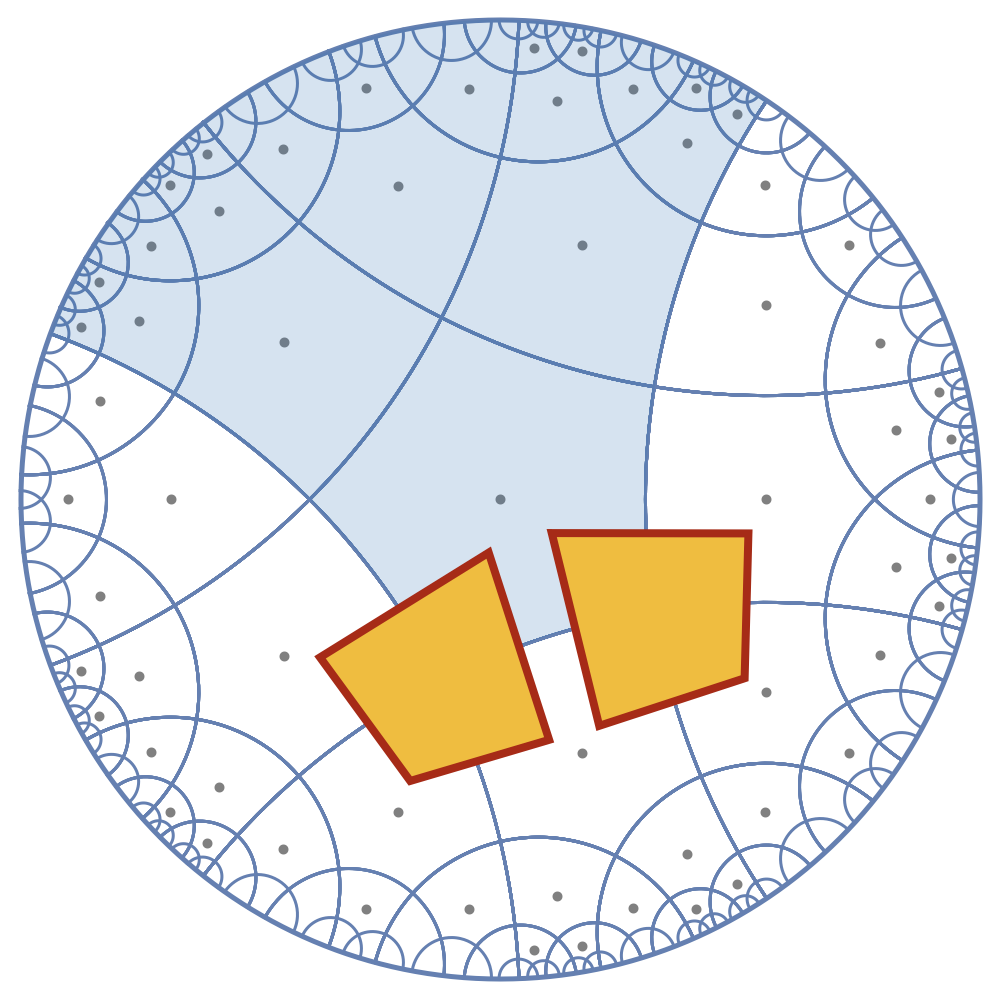}}\\
	\subfloat[Five-fracton bound state\label{FIG_Hyper_Frac_sub_5}]
	{\includegraphics[width=0.22\textwidth]{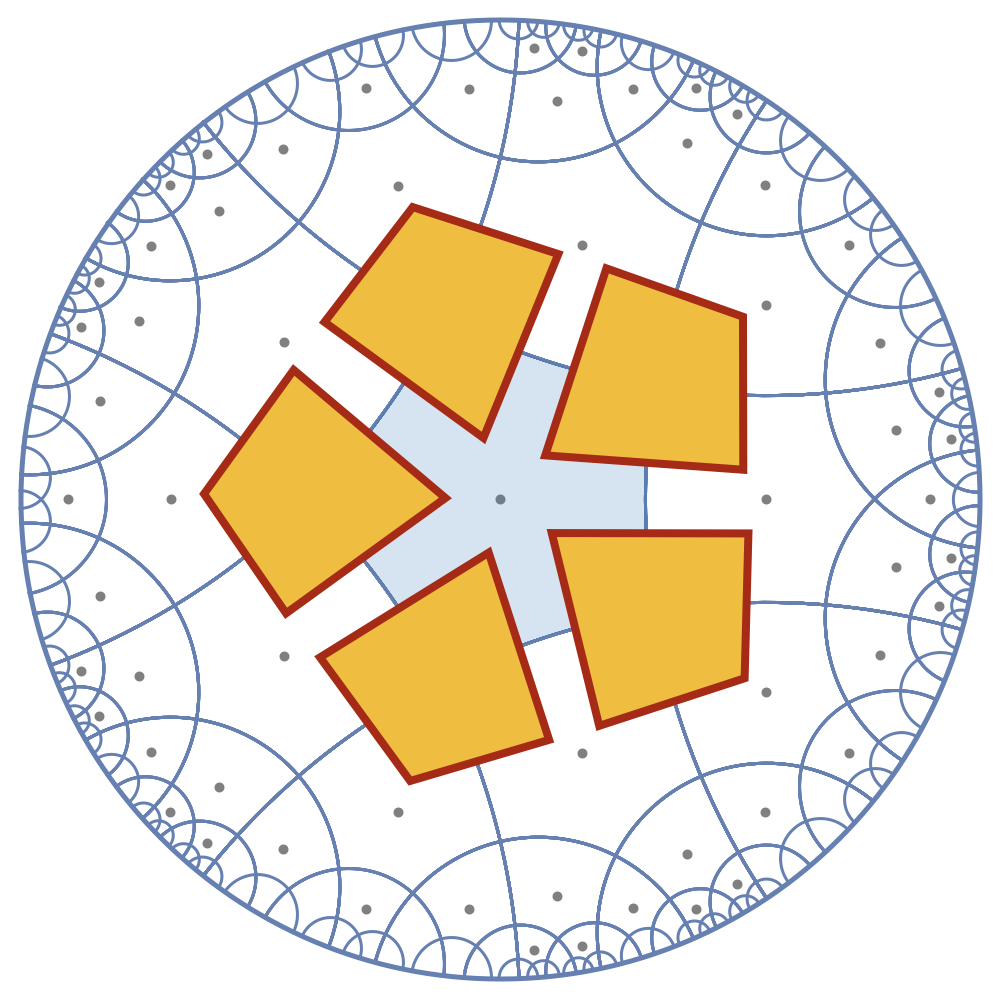}}
	\caption{The degenerate ground states and fracton excitations in the hyperbolic fracton model.
		(a),(b): Constructions of different ground states by flipping all spins on one side of a chosen pentagon-edge geodesic.
		The four-spin cluster highlighted in red has its operator [Eq.~\eqref{EQN_Op_Def}] value invariant.
		(c): A single-fracton excitation created by flipping a quadrant of the spins divided by two intersecting geodesics.
		It is a topological excitation since it involves flipping infinitely many spins.
		(d): Two-fracton excitation. 
		Unlike the case of the Euclidean lattice, they are not movable by local operations.
		(e): Five-fracton bound state created by a single spin flip.
		It is free to move on the lattice.
	}
\end{figure}

%
%
The ground state degeneracy and entropy for this model are respectively
\begin{eqnarray} 
\Omega & = & 2^{N_\text{g}+1} \;, \\
S & = & \kb\log \Omega = \kb\log2\times(N_\text{g}+1) \nonumber\\
& \approx & \frac{\kb\log2}{2}\times \text{(Boundary area)} \;,\label{EQN_GS_entropy}
\end{eqnarray}  
as one would expect.
Starting from the obvious ground state of all spins pointing up,
all the other ground states can be constructed
by repeating the procedure of 
selecting a pentagon-edge geodesic and flipping all the spins on one side of it.
Since a pentagon-edge geodesic always cuts the four-spin clusters in a two-left-two-right manner,
the value of any $\mathcal{O}_p$ remains invariant.
Therefore the system stays in the lowest energy state after the flipping operation.
Two such examples are illustrated in Fig.~\ref{FIG_Hyper_Frac_sub_1},\ref{FIG_Hyper_Frac_sub_2}.

%
%
A single fracton excitation is created by flipping the sign of one 
operator $\mathcal{O}_p$
while keeping the others invariant.
To do so, choose two pentagon-edge geodesics intersecting at the target,
which divide the lattice into four parts,
then flip a quadrant of the spins.
The target \tb{operator} has one spin flipped, 
while all the others have either zero, two, or four spins flipped.
Hence a single fracton excitation is created
as shown in Fig.~\ref{FIG_Hyper_Frac_sub_3}. 
It is topological 
in the thermodynamics limit $N_\text{g}\rightarrow \infty$,
\tb{in the sense that}
no local (i.e., finite number of) spin \tb{flipping} operation
can create a single fracton.
Like the case of the Euclidean lattice model,
it is localized in the system in the sense that
no local operation can move it
without creating more fractons and costing more energy.

%
%
Similar procedures can be employed to create
two-, three-, and four-fracton bound states,
which are all topological. 
The two-fracton bound state is illustrated in Fig.~\ref{FIG_Hyper_Frac_sub_4}. 
However, these excitations do not have 
enhanced mobility in submanifold
like the Euclidean case.
This is due to the different geometry of 
hyperbolic space:
roughly speaking, two parallel geodesics 
do not keep their distance constant,
so there is not a well defined ``$x-$direction'' 
for the bound states to propagate.

%
%
The first local excitation is the
five-fracton bound state,
created by a single spin flip in the bulk.
It can move freely on the lattice by local spin flipping
without costing more energy,
like the four-fracton bound state on the Euclidean lattice.
The five-fracton excitations are illustrated in Fig.~\ref{FIG_Hyper_Frac_sub_5}.

\section{Rindler Reconstruction  of the Hyperbolic fracton model} \label{SEC_Rindler}

\tb{Now we will start discussing the holographic properties realized in the hyperbolic fracton model.
}
The first key property of holography realized on this model is the {\it AdS-Rindler reconstruction}.
In our classical, static model, its simplified version becomes the following:
\begin{property}
	For a given spin configuration on a connected boundary segment,
	the bulk spins can be reconstructed
	if and only if the minimal convex wedge 
	of the boundary segment covers the bulk sites.
\end{property}

The minimal convex wedge is
basically the geodesic wedge slightly 
modified due to the discretization of the hyperbolic disk.
Its precise definition will be made clear soon.
This property holds for the bulk in the ground state,
and also for any excited state if the 
positions of fractons within the minimal convex wedge are given.
%

%
%
\begin{figure}[t]
	\centering
	\captionsetup[subfigure]{justification=centering}
	\subfloat[Before\label{FIG_Frac_Rindler_sub_1}]{\includegraphics[width=0.12\textwidth]{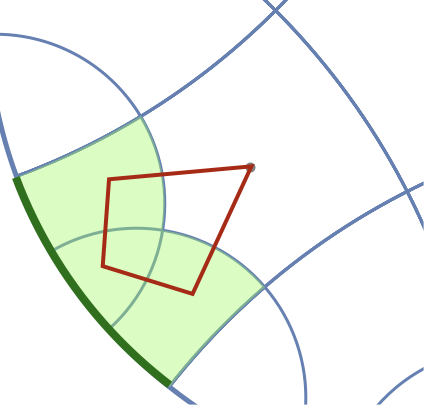}}\qquad
	\raisebox{0.05\textwidth}[0pt][0pt]{%
		\includegraphics[width=0.05\textwidth]{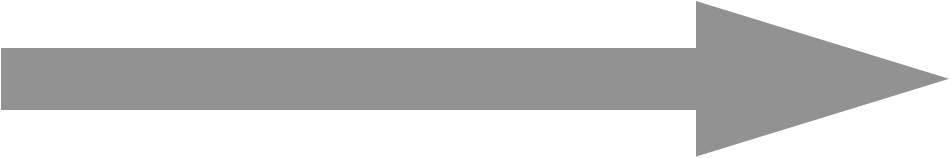}}\qquad
	\subfloat[After\label{FIG_Frac_Rindler_sub_2}]{\includegraphics[width=0.12\textwidth]{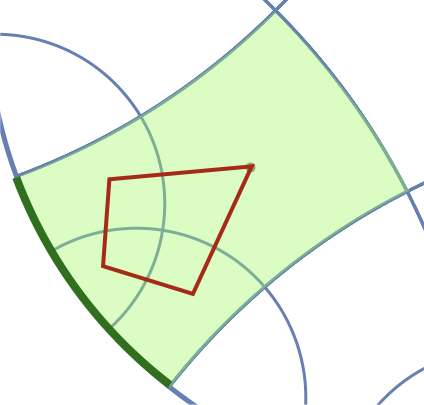}}\\
	\subfloat[Minimal convex wedge, \newline case one\label{FIG_Frac_Rindler_sub_3}]
	{\includegraphics[width=0.22\textwidth]{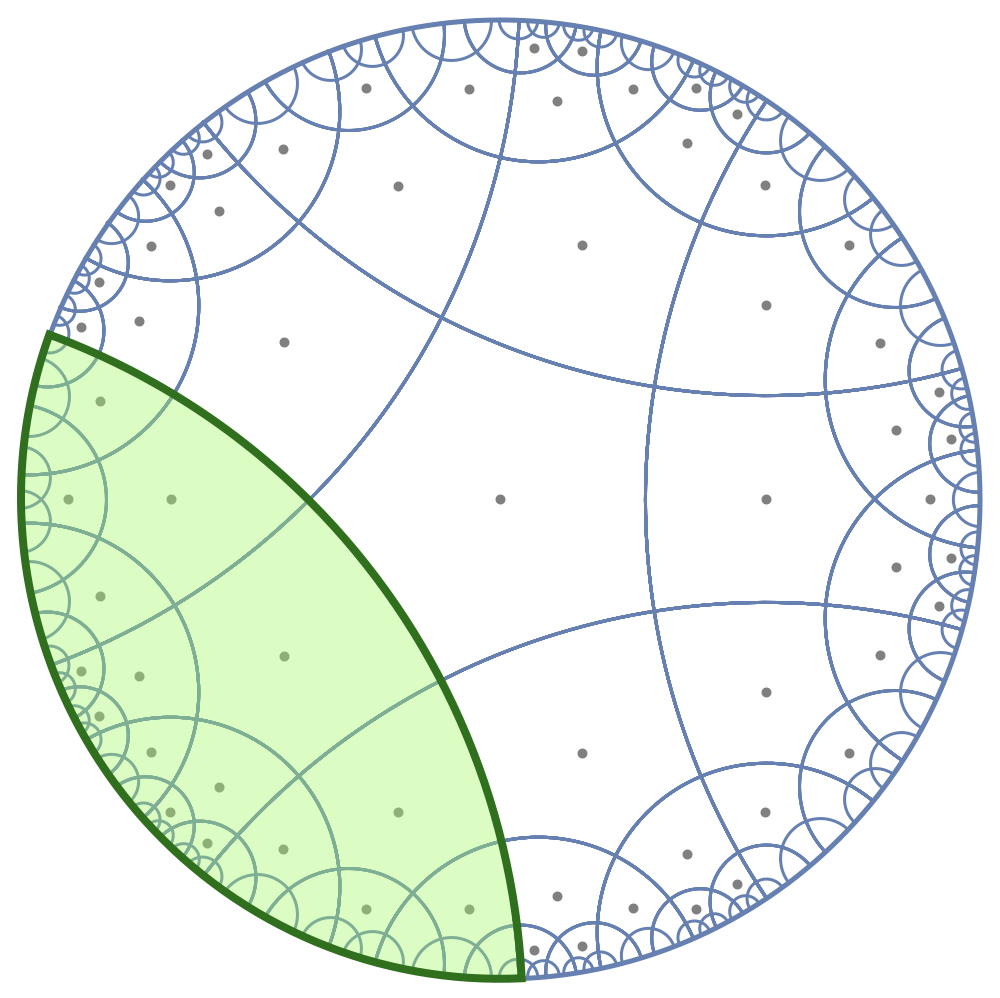}}
	\subfloat[Minimal convex wedge, \newline case two\label{FIG_Frac_Rindler_sub_4}]
	{\includegraphics[width=0.22\textwidth]{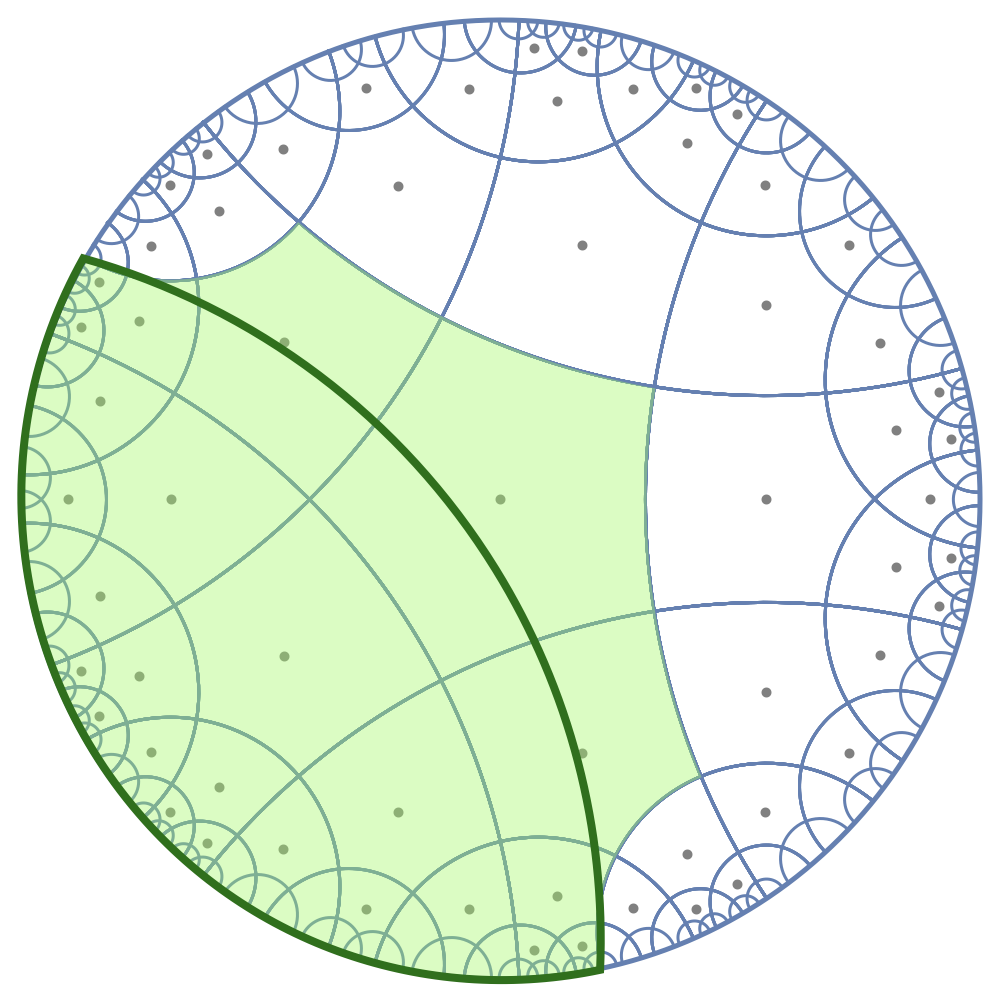}} \\
	\subfloat[Minimal convex wedge, \newline disconnected bounday\label{FIG_Frac_Rindler_sub_5}]
	{\includegraphics[width=0.22\textwidth]{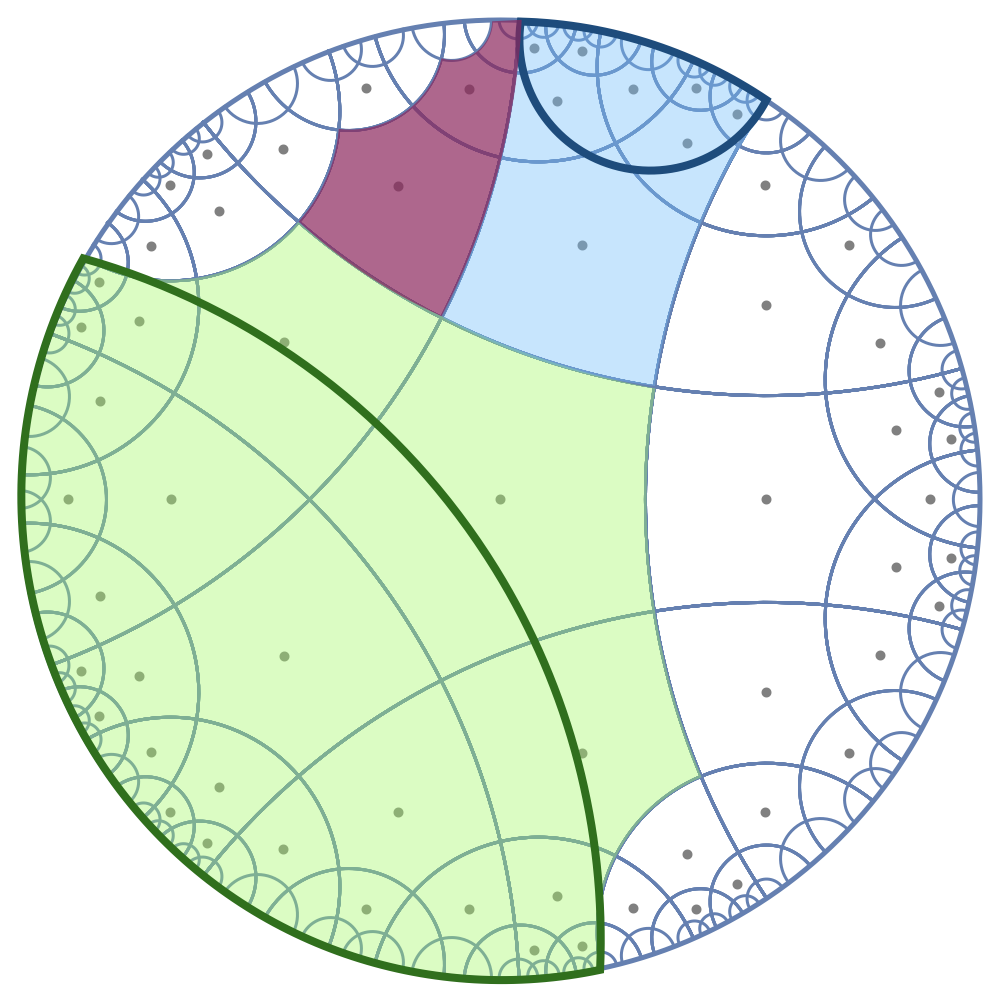}}
	\caption{Rindler reconstruction of the hyperbolic fracton model. 
		(a) Before and (b) after views  illustrate how the reconstruction works. 
		Given three sites on the boundary (green) and the value of the four-spin cluster (red square) operator [Eq.~\eqref{EQN_Op_Def}], 
		the fourth one on the same cluster can be reconstructed.
		(c): For a given boundary segment (boundary arc in dark green), the bulk that can be reconstructed is its minimal convex wedge, the region highlighted in green.
		In this example the minimal convex wedge ends exactly on a pentagon-edge geodesic.
		(d): Another example of a minimal convex wedge as the reconstructible bulk. 
		In this example its minimal convex chain is not a geodesic (arc in dark green).
		(e): An example of Rindler reconstruction for a disconnected boundary subregion.
		Each connected piece (in green or blue) individually has its own minimal convex wedge,
		but the collective minimal convex wedge is bigger than the sum of the two individual wedges.
		The extra segment is colored in magenta.
	}
\end{figure}
%
%
We start with the simpler case
whose entanglement wedge is covered by exactly a pentagon-edge geodesic,
as shown in Fig.~\ref{FIG_Frac_Rindler_sub_3}.
Examining the boundary spins, 
we notice that the plaquettes within the wedge
next to the boundary always contain  three boundary sites
and one bulk site.
Knowing that the four-spin cluster has to have $\mathcal{O}_p=1$ 
(or $-1$ if it is known to be a fracton), 
the  bulk site spin value is uniquely determined.
Thus we can reconstruct all these bulk spins neighboring the boundary spins.
This process is shown in Fig.~\ref{FIG_Frac_Rindler_sub_1}, \ref{FIG_Frac_Rindler_sub_2}. 

By repeating this procedure,
one can reconstruct the bulk spins inward layer by layer,
and exhaust all plaquettes within the entanglement wedge.
Such procedure comes to an end when the wedge boundary is reached.
Beyond the wedge, 
each \tb{four-spin cluster} contains at least two unknown spins at the same time;
thus, determining their values is impossible.
This is shown in Fig.~\ref{FIG_Frac_Rindler_sub_3}. 

A slight complication happens for a generic connected boundary segment,
whose entanglement wedge's boundary is not
a pentagon-edge geodesic, as shown in Fig.~\ref{FIG_Frac_Rindler_sub_4}.
In this case, the reconstructible bulk sites are within the {\it minimal convex wedge},
defined as follows:
\begin{definition}
	The {\bf minimal convex wedge} for a boundary segment is the
	bulk region delimited by a continuous chain of the pentagon's edges that satisfies the following conditions:
	(1) the chain is homologous to the boundary segment, i.e., shares the same ends;
	(2) it is a convex;
	(3) it contains the minimal number of pentagon edges.
	The chain is named a {\bf minimal convex chain}.
\end{definition}
This definition seems to be complicated,
but for a connected boundary segment,
it is simply the \tb{continuous geodesic wedge} extended by the 
pentagons partially overlapping with it:
\begin{property}
	The minimal convex wedge of a connected boundary segment
	consists of all the bulk sites whose pentagons have non zero 
	overlap with the \tb{geodesic wedge in the continuous case}.
\end{property}
It is a simple consequence of the hyperbolic disk discretization,
as the minimal bulk volume unit is a pentagon.

%
%
We also consider the case of a boundary segment consisting of two disconnected components.
In this case the entanglement wedge of the joint boundary segments can
be larger than the sum of the wedges for each individual component.
An example is shown in Fig.~\ref{FIG_Frac_Rindler_sub_5}.

We should point out that for a large subregion of the boundary,
the resulting entanglement wedge properly approximates its
continuous limit. 
However, for more complicated situations 
of boundary segments close to the phase transition,
or consisting of more components,
it becomes more complicated.
Such deviation from AdS/CFT is similar to the situation of holographic tensor networks
\tb{constructed by perfect tensors \cite{Pastawski2015}.
}
%


\section{ Mutual Information of the Hyperbolic Fracton Model}\label{SEC_RT_formula}
\noindent\tb{(Note by author: this entire section has been rewritten to make the physics clearer
	and more precise.)}\\

The second essential property of holography 
realized in the hyperbolic fracton model
is the Ryu-Takayanagi formula for entanglement entropy.

For a CFT with a gravitational dual in the AdS spacetime, 
there exists a geometric bulk description
for at least its static state at low energies.
For such states, the Ryu-Takayanagi (RT) formula
relates the entanglement entropy $S_A$ of a boundary segment $A$
with the area of the minimal covering surface $\gamma_A$ in the bulk,
\begin{equation}
S_A=\frac{\text{Area}(\gamma_A)}{4G_N} \;,
\end{equation}
where $G_N$ is Newton's constant,
and $\text{Area}(\gamma_A)$ refers to the length of the covering curve(s).
We shall show that its classical analog holds for the fracton model.

A few corrections need to be added to make the statement more accurate.
To begin with, a classical model has no quantum entanglement,
so instead of the entanglement entropy,
the quantity employed here is the
\textit{mutual information}.
The mutual information can be viewed as
the classical analog of the entanglement entropy.
Also, 
the minimal covering surface should be 
modified to be the boundary of the minimal convex wedge in the bulk,
which we named the \textit{minimal convex chain}
in the previous section.
Second, the mutual information 
may receive some corrections,
depending on the shape of the boundary subregion and its 
entanglement wedge. 
Here we will
discuss the corrections
that appear in relatively simple
boundary subregion configurations.
The mutual information grows linearly with the 
lattice size, 
but these corrections stay fixed. 

\subsection{Mutual Information as the Classical Analog of Entanglement Entropy}

The mutual information, as its name suggests,
measures how much information is shared between 
two subsystems.
It is defined as 
\begin{equation}
\label{Eqn_def_mutual_info}
I_\text{cl}(A;B)=S_\text{s}(A)+S_\text{s}(B)-S_\text{s}(A\cup B) \;,
\end{equation}
where $A,B$ are subsystems, and $S_\text{s}$ is the Shannon entropy.
$S_\text{s}(A\cup B)$ is the entropy for the union of two subsystems.
The subscript ``cl" is to remind us that it is a classical concept.

The mutual information is a proper classical analog of the entanglement entropy
between a bipartition of a quantum system. 
To see this, 
replace the classical Shannon entropy $S_\text{s}$ with 
von Neumann entropy $S_\text{v}$ for the corresponding subsystem's
reduced density matrices,
and note that $B = A^c$ is the complement of subregion $A$.
We have its quantum version 
\begin{equation}
I_\text{qu}(A;A^c)=S_\text{v}(A)+S_\text{v}(A^c)-S_\text{v}(A\cup  A^c) .
\end{equation}
For a pure state
\begin{eqnarray}
S_\text{v}(A\cup A^c) & = & 0  ,\\
S_\text{v}(A) & = & S_\text{v}(A^c)  ,
\end{eqnarray}
so we end up with exactly twice the entanglement entropy between $A$ and $A^c$,
\begin{equation}\label{EQN_MI_EE}
I_\text{qu}(A;A^c)=2S_\text{v}(A) = 2S_A \;,
\end{equation}
which indicates that its classical analog $I_\text{cl}$ is the correct choice, up to a factor of $2$.

\subsection{Mutual Information for Connected Subregions}
\begin{figure}[h]
	\centering
	\subfloat[]{
		\includegraphics[width=0.15\textwidth]{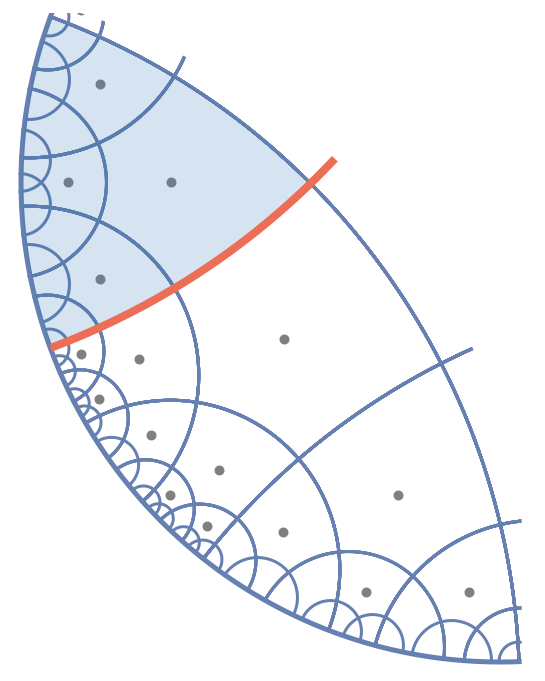}}
	\subfloat[]{
		\includegraphics[width=0.15\textwidth]{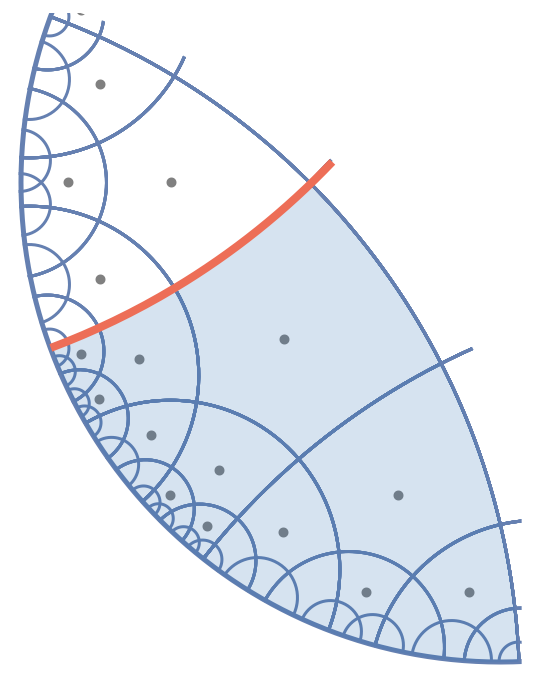}}
	\caption{Counting the entropy of a subsystem. 
		Every pentagon-edge geodesic crossing the system
		multiplies the degeneracy by two.
		These new states can be explicitly constructed as illustrated by
		(a) and (b), \tb{where the blue region indicates the spins being flipped.}
		\label{FIG_Entropy}}
\end{figure}

We start with the simple scenario when the subregion $A$ is connected.
We will show the following:
\begin{property}
	For both the vacuum and a given configuration of fractons,
	the mutual information for a bipartition of the boundary 
	into connected subregions obeys 
	the Ryu-Takayanagi formula
	\begin{equation} \label{EQN_RT_Mutal_Info}
	{I_\text{cl}(A;A^c)} \approx \kb\log 2|\gamma_A| \;.
	\end{equation}
	where $|\gamma_A|=\text{Area}(\gamma_A)$ is a short hand notation.
\end{property}
To calculate Eq.\eqref{Eqn_def_mutual_info},
we just need to compute the entropies for $A$, $B= A^c$, and the entire system
individually.
The entropy of the entire system is already given in Eq.\eqref{EQN_GS_entropy},
which is proportional to the number of pentagon-edge geodesics plus one.
The physics is that for each pentagon-edge geodesic the ground state is multiplied
by two, 
from the operation
of flipping spins on either side of the geodesic.
This is shown in Fig.\ref{FIG_Entropy}.

The same argument applies to counting the ground state degeneracy
of any connected subregion (for disconnected region it can be more complicated).
Due to Rindler reconstruction,
the entanglement wedge of boundary section $A$
has the same ground state degeneracy as
$A$ itself.
In either way of counting, the pentagon-edge geodesics of the system
are those intersecting $A$ with one or both ends.
Thus the degeneracy and entropy for the ground 
states of a subregion $A$ are
\begin{eqnarray} \label{eqn_sec4_gs_sub}
\Omega(A) & = & 2^{N_\text{g-A}+1} \;,\label{EQN_Sub_Degen} \\
S_\text{s}(A)  & = & \kb\log \Omega = \kb\log2\times(N_\text{g-A}+1) \;, \label{EQN_Sub_Entropy}
\end{eqnarray}
where $N_\text{g-A}$ is the number of pentagon-edge geodesics that cross 
the region.

Let us denote the minimal convex chain as $\gamma_A$.
Depending on the choice of subregion $A$, 
$\gamma_A$ can overlap with a pentagon-edge geodesics exactly,
or has some ``corners'', as shown in Fig.\ref{FIG_Frac_Rindler_sub_3},\ref{FIG_Frac_Rindler_sub_4}.

\begin{figure}[t]
	\centering
	\includegraphics[width=0.3\textwidth]{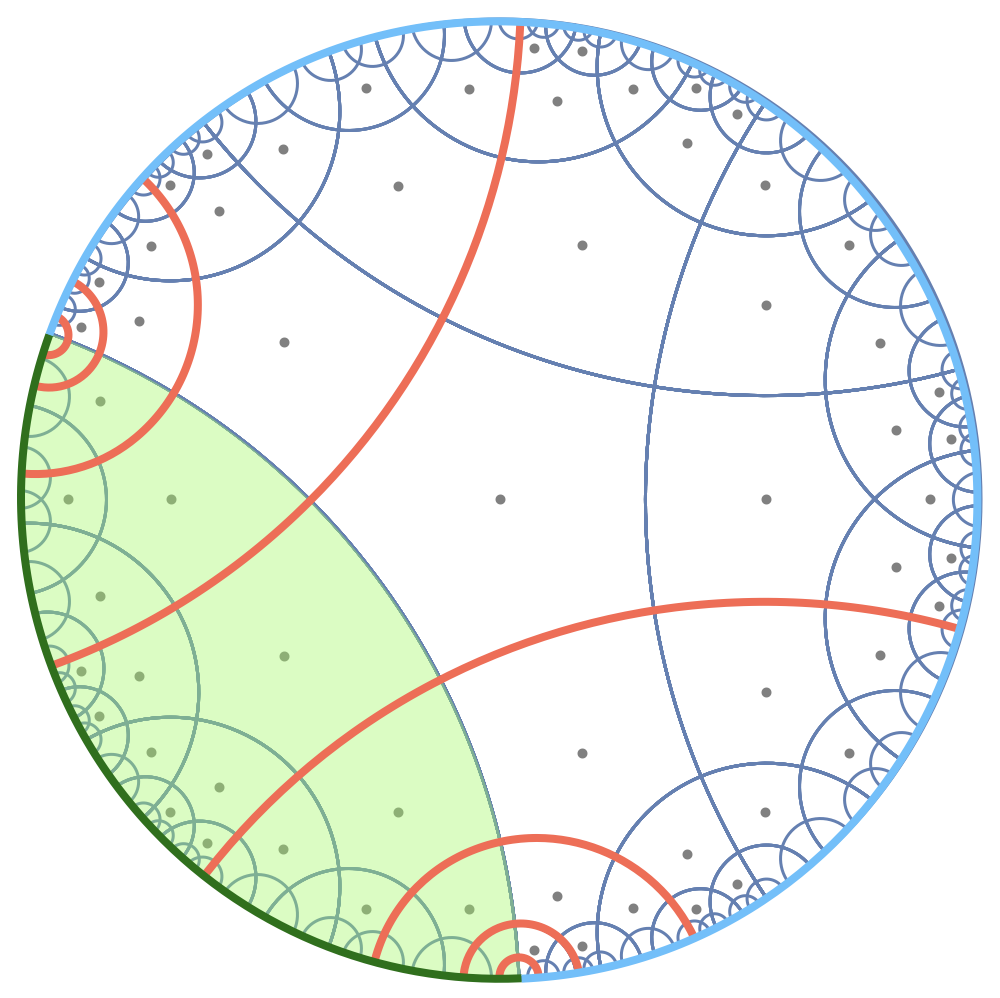}
	\caption{
		Mutual information as a classical analog of entanglement entropy obeys the Ryu-Takayanagi formula. 
		It is measured by the number of pentagon-edge geodesics
		that are shared by  $A$ (dark green arc) and $A^c$
		(light blue arc).
		The subregion-crossing  pentagon-edge geodesics are highlighted in orange,
		whose number is denoted $N_\text{g-$\gamma$}$.
		They are also the geodesics that intersect with the geodesic $\gamma_A$, which
		is the minimal curve that splits $A$ and $A^c$.
		Their relation $|\gamma_A|=N_\text{g-$\gamma$}+1$
		leads to the Ryu-Takayanagi formula for mutual information in Eq.~\eqref{EQN_RT_Mutal_Info}.
		\label{FIG_MI}
	}
\end{figure}

\noindent{\bf Case one: $\gamma_A$ is a pentagon-edge geodesic: } 
In the first case, we can divide the pentagon-edge geodesics,
whose total number is  $N_\text{g}$ ,
into four categories: 
\begin{enumerate}
	\item Those with both ends on $A$, whose number is denoted $N_\text{g-A}$;
	\item Those with both ends on $A^c$, whose number is denoted $N_\text{g-Ac}$;
	\item Those with one end on $A$ and the other on $A^c$, whose number is denoted $N_\text{g-$\gamma$}$;
	\item The geodesic $\gamma_A$. Its length is exactly $|\gamma_A|=N_\text{g-$\gamma$}+1$.
\end{enumerate}
These quantities satisfy the condition  
\begin{equation}
N_\text{g-A}+N_\text{g-Ac}+N_\text{g-$\gamma$}+1=N_\text{g} .
\end{equation}
For both the ground state or any given configuration of fracton excitations,
the entropy of states in region $A$ is 
\begin{equation} \label{C4_eqn_A_entropy} S_\text{s}(A)=(N_\text{g-A}+N_\text{g-$\gamma$}+1)\kb\log2 \;,
\end{equation}
as argued in Eq.~(\ref{EQN_Sub_Degen},\ref{EQN_Sub_Entropy}).
Similarly for region $A^c$,
\begin{equation} \label{C4_eqn_Ac_entropy} S_\text{s}(A^c)=(N_\text{g-Ac}+N_\text{g-$\gamma$}+1)\kb\log2 \;.
\end{equation}
Finally, the joint entropy of $A$ and $A^c$ is simply the entropy of the entire system, which is 
\begin{equation} \label{C4_eqn_join_entropy} S_\text{s}(A,A^c)=(N_\text{g}+1)\kb\log2 \;.
\end{equation}
Therefore the classical mutual information is 
\begin{equation}
I_\text{cl}(A;B)=
N_\text{g-$\gamma$}\kb\log2\approx \kb\log2 |\gamma_A|\;,
\end{equation}
in the limit of large $N_\text{g-$\gamma$}$. 
Here we consider the length of the edge of the pentagon to be $1$.
This calculation is illustrated in Fig. ~\ref{FIG_MI}.

Note that here, compared to Eqs.~(\ref{EQN_GS_entropy}, \ref{EQN_Fr_BH_Entropy}),  a factor of $\frac{1}{2}$ is missing.
But it is simply due to the fact that by definition $I_\text{cl}$ is twice the entanglement entropy [Eq.~\eqref{EQN_MI_EE}].\\

\noindent{\bf Case two: $\gamma_A$ is not a pentagon-edge geodesic: } 
Now let us consider more general situations when $\gamma_A$ is not a pentagon-edge geodesic.
The proof is
basically the same, but we just write it down for completeness.
We have the $N_\text{g}$ pentagon-edge geodesics now classified into three categories:
\begin{enumerate}
	\item Those with both ends on $A$, whose number is denoted $N_\text{g-A}$;
	\item Those with both ends on $A$, whose number is denoted $N_\text{g-A}$;
	\item Those with one end on $A$ and the other on $A^c$, whose number is denoted $N_\text{g-$\gamma$}$.
\end{enumerate}
Here a geodesic that starts and ends on $A$ is considered to be in the first category,
and vice versa for $A^c$.
These numbers obey the modified constraint 
\begin{equation}
N_\text{g-A}+N_\text{g-Ac}+N_\text{g-$\gamma$}=N_\text{g} \;.
\end{equation}
The different entropies remain the same as defined in
Eqs.~(\ref{C4_eqn_A_entropy}, \ref{C4_eqn_Ac_entropy}, \ref{C4_eqn_join_entropy}).
Therefore the classical mutual information becomes
\begin{equation}
I_\text{cl}(A;B)=(N_\text{g-$\gamma$}-1)\kb\log2,
\end{equation}
for large $N_\text{g-$\gamma$}$. 

Let us denote the number of corners of $\gamma_A$ as $N_\text{cor}$;
then 
\begin{equation}
I_\text{cl}(A;B)=(N_\text{g-$\gamma$}-1)\kb\log2\approx \kb\log2 (|\gamma_A|- N_\text{cor} ).
\end{equation}

Here $-N_\text{cor}$ is a correction to the RT formula,
which stays fixed as the lattice size grows.
It is, however, in some sense ``benign.''
The lattice discretized minimal convex chain $\gamma_A$ 
has some sharp corners.
As a consequence, 
its length becomes larger than the continuous covering geodesic.
The  $-N_\text{cor}$ reduces such deviation,
resulting in a mutual information closer to the continuous case.

\subsection{Mutual Information for Disconnected Subregions}

\begin{figure}[t]
	\centering
	\subfloat[]{
		\includegraphics[width=0.15\textwidth]{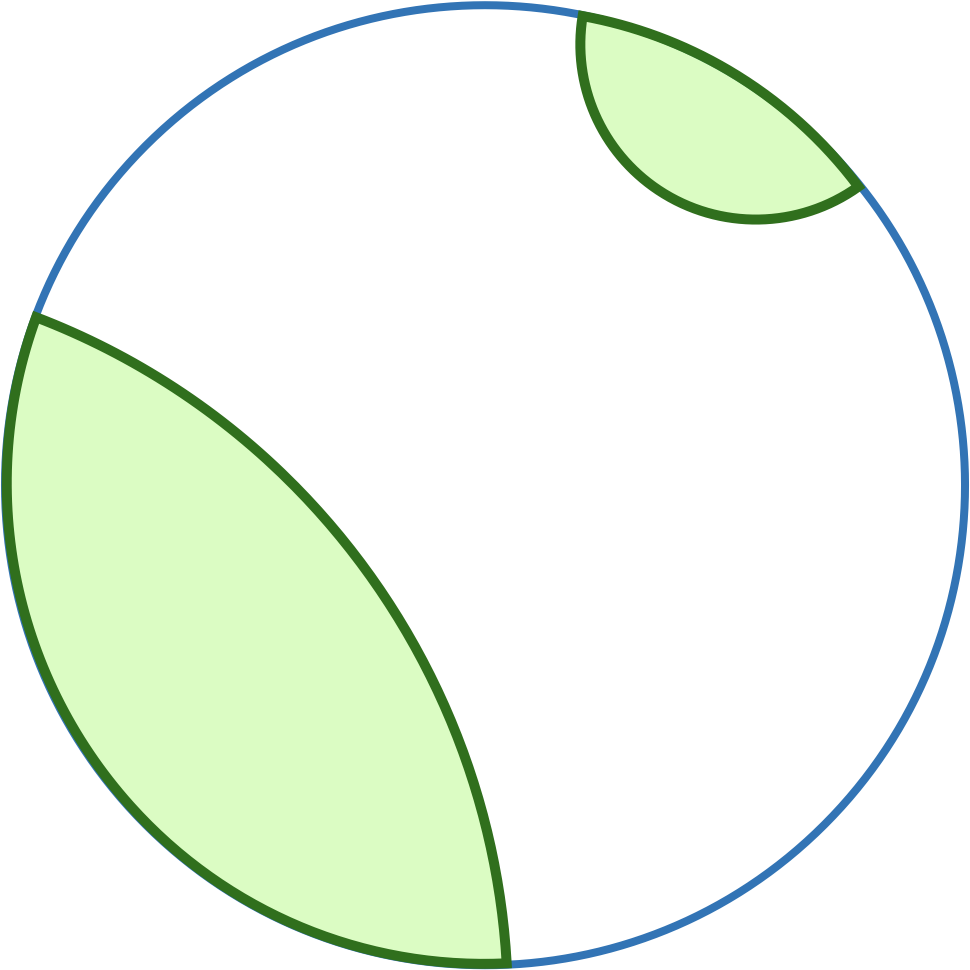}}
	\subfloat[]{
		\includegraphics[width=0.15\textwidth]{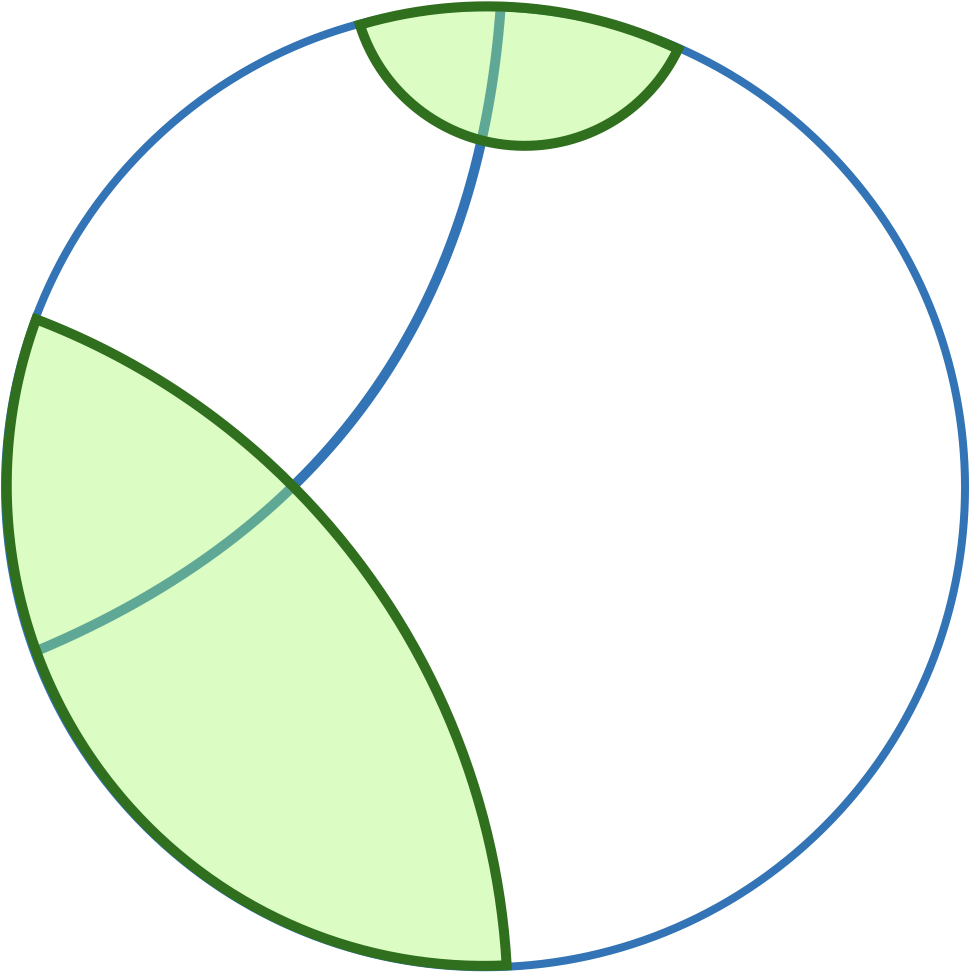}}
	\subfloat[\label{FIG_mutual_diss_3}]{
		\includegraphics[width=0.15\textwidth]{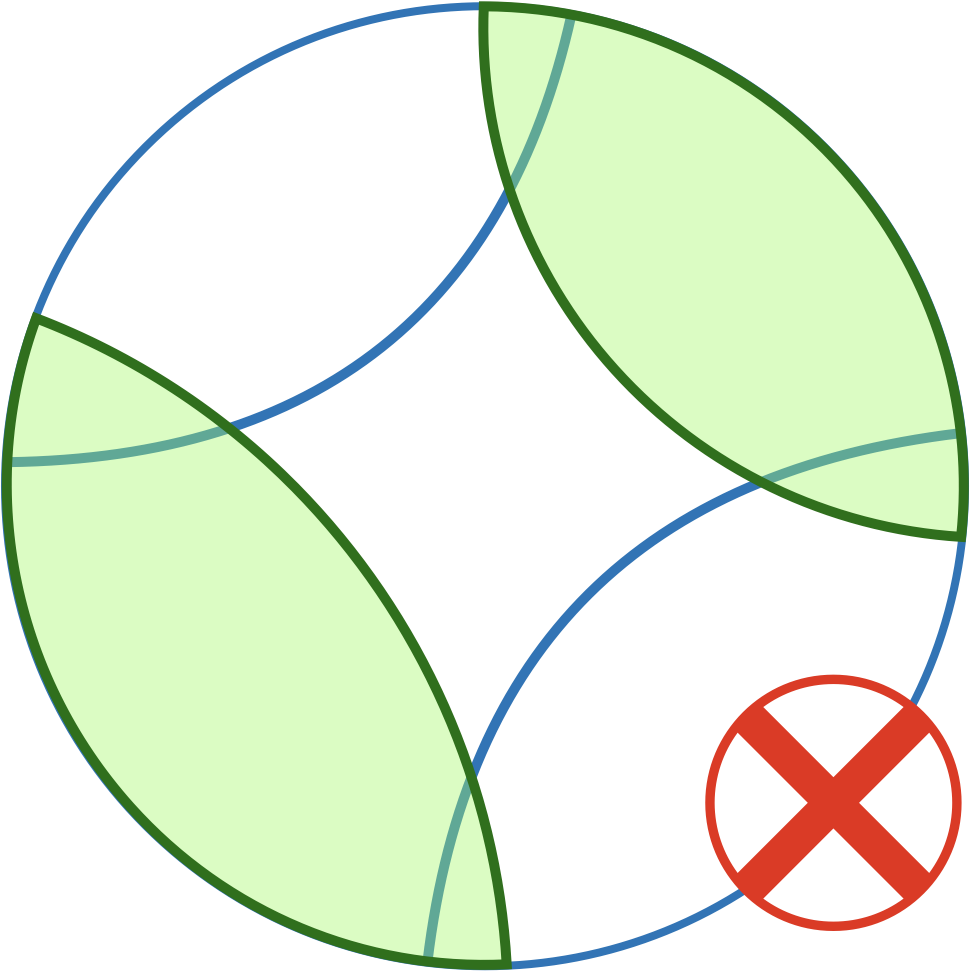}}
	\caption{Possible configurations for disconnected boundary subregion.
		Here we limit ourselves to the case when the entanglement wedges (green region)
		are covered by pentagon-edge geodesics exactly.
		Blue arcs are pentagon-edge geodesics.
		Situation (a) and (b) are possible, and 
		(b) will cause one bit of correction to the RT formula for the mutual information.
		The situation in (c) is impossible as the four pentagon-edge geodesics
		cannot form a square.
		\label{FIG_mutual_diss}}
\end{figure}

The situation becomes more complicated for a subregion with several
disconnected components.
Equation~\eqref{Eqn_def_mutual_info} can still be computed for each
subregion by identifying its entanglement wedge 
and computing its entropy. 
Here we analyze the possible 
correction to the simplest case of disconnected subregion $A$.

The simplest case is defined as follows:
for each component of $A$, its entanglement wedge is \textit{case one} 
discussed above; i.e., it is covered by a pentagon-edge geodesic.
The mutual information can be again computed
by counting the pentagon-edge geodesics.

One issue may lead to some corrections to the mutual information:
There are geodesics starting from one component of $A$
and ending in another,
instead of ending in $A^c$.
This is shown in Fig.\ref{FIG_mutual_diss}.
We have to consider the correction contributed by them.

First we note that between two components,
there can be at most one pentagon-edge geodesic.
That is, situations in Fig.\ref{FIG_mutual_diss_3} do not exist.
This is because there is a rectangle formed by
these pentagon-edge geodesics,
whose four angles are all $\pi/2$. 
Such rectangles cannot exist in the hyperbolic space.

So we only need to take care of the case with one pentagon-edge
geodesic between the two components.
Note that  it still goes through the entanglement wedge of $A^c$
and contributes one unit of entropy to $S_{A^c}$.
So it contributes one unit of 
mutual information,
but two units of the length of the minimal covering chain.
Therefore, the final correction is one unit:
\begin{equation}
I_\text{cl}(A;A_c)=(N_\text{g-$\gamma$}-1)\kb\log2\approx \kb\log2 (|\gamma_A|- N_\text{A-A} ),
\end{equation}
where $N_\text{A-A}$ denotes the number of geodesics starting from one component 
of $A$ and ending in another.

As the boundary subregion becomes more complicated,
more corrections will enter the mutual information.
In particular, for configurations close to the phase transition
of entanglement entropy,
the deviation can be big.
Similar issues with the holographic tensor-networks
are fixed by the random tensors \cite{Qi2018}.
It remains an open question
on how the modifications 
of the hyperbolic fracton model will 
amend such issue
and yield exact RT formula for arbitrary boundary bipartition.

\section{Naive Black Holes in the Hyperbolic fracton model}\label{SEC_BH}

\begin{figure}[h]
	\centering
	\includegraphics[width=0.3\textwidth]{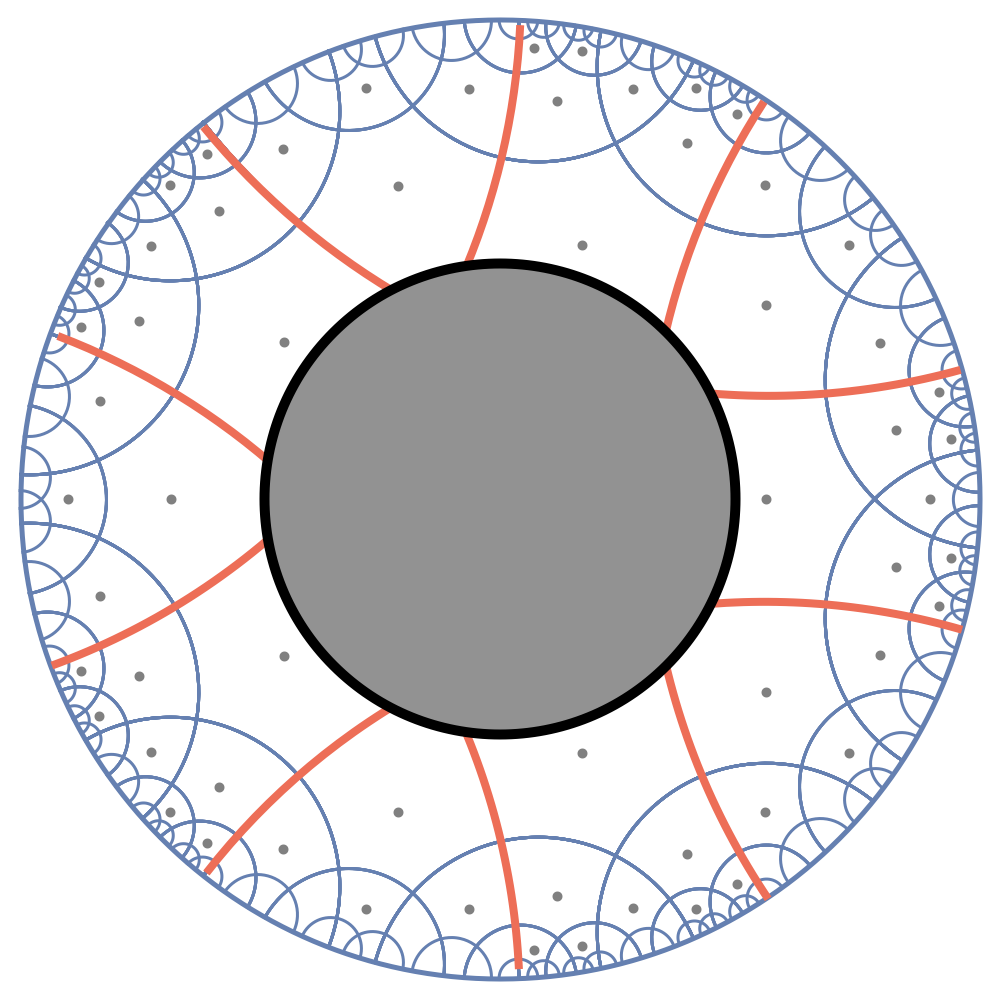}
	\caption{A  naive black hole in the hyperbolic fracton model.
		There is no geometrical change of the hyperbolic disk,
		but some bulk sites are hidden behind the horizon,
		and not accessible by observers.
		The horizon is the solid black line.
		The pentagon-edge geodesics crossing the black hole
		are highlighted in orange.
		\label{FIG_BH}
	}
\end{figure}

The black hole in this model has its entropy proportional to its horizon.
Here we consider a very naive black hole 
constructed by simply cutting out some 
bulk pentagons included in a closed convex,
but leaving the rest of the lattice unchanged.
The spins of the pentagon inside the black hole,
and all interactions associated with them,
are considered hidden behind the horizon.
An example is illustrated in Fig.~\ref{FIG_BH}.
\tb{
	Our approach is adapted from Ref.\cite{Pastawski2015},
	in which a black hole is constructed by removing some bulk tensors in the holographic tensor network. 
	Though there is no change of geometry outside the horizon, 
	this approach does show some resemblance 
	to  a black hole in an asymptotic AdS geometry,
	as we demonstrate below.
}
The horizon size of the black hole is approximately 
\begin{equation}
\text{Horizon area}=N_\text{BH} \;,
\end{equation}
where $N_\text{BH}$ are the semi-infinite pentagon-edge geodesics extended from the black hole, highlighted in orange in Fig.~\ref{FIG_BH}.
They used to be $N_\text{BH}/2$ complete geodesics.
The black hole entropy has several interpretations,
including the entropy for its microstates,
or its entanglement entropy with the outside.
Here we use the definition proposed by Witten \cite{Witten1998}, 
tailored for our model:
\begin{definition}
	The \textbf{black hole entropy} is the boundary or bulk ground state 
	Shannon entropy increase from introducing the black hole.
\end{definition}

This is a rather simple calculation:
since $N_\text{BH}/2$ pentagon-edge geodesics 
are cut into two pieces,
the system has effectively $N_\text{BH}/2$ more pentagon-edge geodesics 
for the topological spin-flipping operations to create new ground states.
Therefore we have the following:
\begin{property}
	The black hole entropy is
	\begin{equation}\label{EQN_Fr_BH_Entropy}
	S_\text{BH}= \frac{\kb\log 2}{2}   N_\text{BH}=\frac{\kb\log 2}{2} \times \text{(Horizon area)}\;,
	\end{equation}
	which has the proper scaling behavior.
\end{property}

The appearance of a black hole means the
boundary ground state degeneracy grows,
similarly to the Hilbert space enlargement discussed in Ref.~\cite{Pastawski2015}.
This is expected as only a very small portion of 
the boundary states correspond to the pure AdS geometry,
and most states corresponds to some black hole state in the bulk.

\section{Generalizations: Higher Dimension and Quantum Version}
\label{SEC_generalization}

Two  important questions naturally follow the major results of this work:
how to generalize the model to higher dimension,
and whether there is a quantum version of the model?
Both answers are positive, as we explain below.

\subsection{Three-Dimensional generalization}
The three-dimensional generalization of our model 
is a cubic Ising model with eight-spin interaction terms.

\begin{figure}[t]
	\centering
	\includegraphics[width=0.3\textwidth]{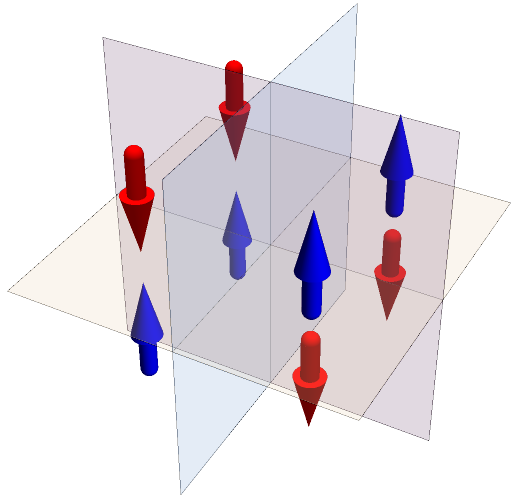}
	\caption{Building block of 3D classical fracton model [Eq.\eqref{eqn_3d_hamiltonian}]. 
		The spins sit at the centers of the cube, and eight cubes sharing the same corner
		are used to construct  the operator $\mathcal{O}$ in Eq.\eqref{eqn_3d_op}.
		The subsystem symmetry is flipping a line of spins   in the   $x-,\ y-$
		or $z-$ direction.
		\label{FIG_3Dmodel}
	}
\end{figure}

In this model,
each Ising spin sits at the \textit{center} of the cube of the lattice, 
as shown in Fig.\ref{FIG_3Dmodel}.
The operator $\mathcal{O}_c$ is
\begin{equation}\label{eqn_3d_op}
\mathcal{O}_c = \prod_{i=1}^{8} S_i^z		,
\end{equation}
where $i$ runs over 8 cubes sharing the same corner,
which forms the cube of the dual lattice.
The Hamiltonian is again 
\begin{equation}\label{eqn_3d_hamiltonian}
\mathcal{H}_\textsf{cl}=-\sum_{c}\mathcal{O}_c		,
\end{equation}
where $c$ runs over all eight spin operators.

This classical model has the subsystem symmetries of 
flipping all spins on a line   in  the $x-,\ y-$
or $z-$ direction.
An equivalent way to view them 
is to
have two perpendicularly intersecting
planes.
The two planes divide the lattice into four parts,
and flipping one quadrant of the spins leaves the energy 
invariant.
This way has a more straight forward 
adaptation to the AdS$_3$ lattice.

This model has a natural generalization to the AdS$_3$ space. 
We do not need to visualize the entire lattice, which is 
rather difficult.
Instead we can focus on the subsystem symmetries,
and it would be sufficient to demonstrate the 
holographic properties.

In its AdS$_3$ lattice, the original 2D planes become spherical surfaces
that intersect the boundary
of AdS$_3$ perpendicularly.
These intersecting 2D hypersurfaces
form cells for the spins to sit in. 
Every cluster of eight cells share the same corner since three spherical surfaces intersect
at the same point, 
which can be used to build the same Hamiltonian for each local 8-spin cluster.
\begin{figure}[t]
	\centering
	\includegraphics[width=0.3\textwidth]{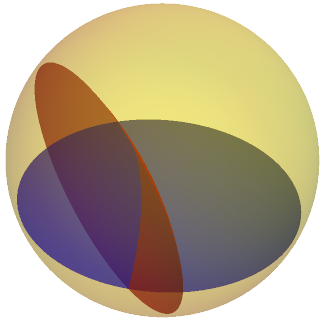}
	\caption{
		Subsystem symmetry of the fracton model [Eq.\eqref{eqn_3d_hamiltonian}] in AdS$_3$ space.
		The spherical surfaces (red and blue) in this representation are actually ``flat''
		in the AdS$_3$ space.
		The two surfaces split the entire AdS$_3$ lattice into four parts.
		Flipping spins in one part of the AdS$_3$ lattice does not change the system's
		energy, and is a subsystem symmetry.
		\label{Fig_AdS_3dmodel}
	}
\end{figure}
Each geodesic is now determined by two intersecting  spherical surfaces,
and they split the entire lattice into four parts as shown in Fig.\ref{Fig_AdS_3dmodel},
in analogy to each geodesic splitting the AdS$_2$ lattice into two parts. 
Flipping spins in 
one of the four parts keeps the energy of the system invariant,
which is the subsystem symmetry in AdS$_3$ space.
Again the number of independent subsystem symmetries 
is proportional to the number of geodesics, 
hence the boundary area. 

The Rindler reconstruction and RT formula for mutual information
holds as a consequence of the structure of the subsystem symmetries.

\subsection{Quantum Model with a Transverse Field}
Next let us make some remarks 
on the quantum version of the model.
The simplest case is to introduce a constant transverse field.
For a small transverse field,
we can assume that there will be a unique quantum ground state as the superposition 
of (almost) all classical ground states. 
The superposition does not necessarily have to have equal weight or phase.

The boundary state is then defined as 
a mixed state by tracing out all degrees of freedom in the bulk,
and such mixed state can be viewed as an ensemble of 
all classical ground states on the boundary with a certain probability distribution.
Assuming the probability distribution (or weight of the superposition) 
to be close to even among all classical states
for a small transverse field,
the entanglement entropy/mutual information 
will still obey the Ryu-Takayanagi formula up to some correction.
If a bulk spin is fixed by hand to be up or down
in the model, 
it can be reconstructed by looking at any element from the ensemble
on a region whose entanglement wedge covers the bulk site.
It is not too different from the classical model
in the sense that on the boundary one always works with a classical ensemble.

We have to point out that 
this is an interesting difference from the large-$N$ limit of gravity/CFT duality.
There, the bulk is semiclassical and the boundary is quantum,
which is the opposite of our construction.
Whether such difference has any profound meaning is to be studied
in the future.

\section{Comparison with the Holographic Tensor-Networks}
\label{SEC_comparison}
\noindent\tb{(Note by author: this entire section is newly added.)}\\

A key question emerging from this work is ,
what  features of gravity
can be captured by the fracton models,
and what cannot?
To pave the way to the answer,
it is useful to compare our model with holographic tensor networks
regarding their holographic properties.
These models are, after all, not exactly quantum gravity,
so some properties of AdS/CFT duality 
are still not captured.
Clarifying them can be helpful for future investigations and improvements.

Holographic tensor networks are a type of toy models of holography.
They are built by tensors with special properties, and
uniformly tiled on the discrete hyperbolic lattice.
Two representatives are the perfect tensors and the random tensors.
Essentially, these tensors saturate the upper bond
of entanglement 
between any bipartition of their legs.
This guarantees that the bulk information 
is not lost when ``pushed'' toward the boundary.
It is closely related to the
quantum-error-correcting properties of gravity,
which manifest  in the Rindler reconstruction 
and the RT formula for the entanglement entropy.

Let us focus on the holographic \textit{state} defined by the tensor networks,
i.e., simply a quantum state on the boundary without bulk inputs.
For the holographic state, we care about the boundary
state's entanglement properties, mainly the Renyi entropy
for connected or disconnected subregions.
The exact RT formula for any disconnected subregion
is realized in the random tensor network in its large-$N$ limit \cite{Qi2018}.
The hyperbolic fracton model, however,
suffers from various corrections as we explained in previous sections.

Both the tensor network model  and the hyperbolic fracton model
have a trivial $n-$dependence for the $n-$th Renyi entropy.
More fundamentally this is due to the fact that the entanglement spectrum
is always flat in such models.
In contrast, the CFT has a nontrivial $n-$dependence and a nonflat 
entanglement spectrum \cite{Dong2016NatCom,Faulkner2013JHEP}.
A related issue is that the boundary state defined
by the holographic code cannot be the ground state 
of a local Hamiltonian. 
Refinement of such undesirable properties 
will be an important progress.

\begin{figure}[t]
	\centering
	\includegraphics[width=0.3\textwidth]{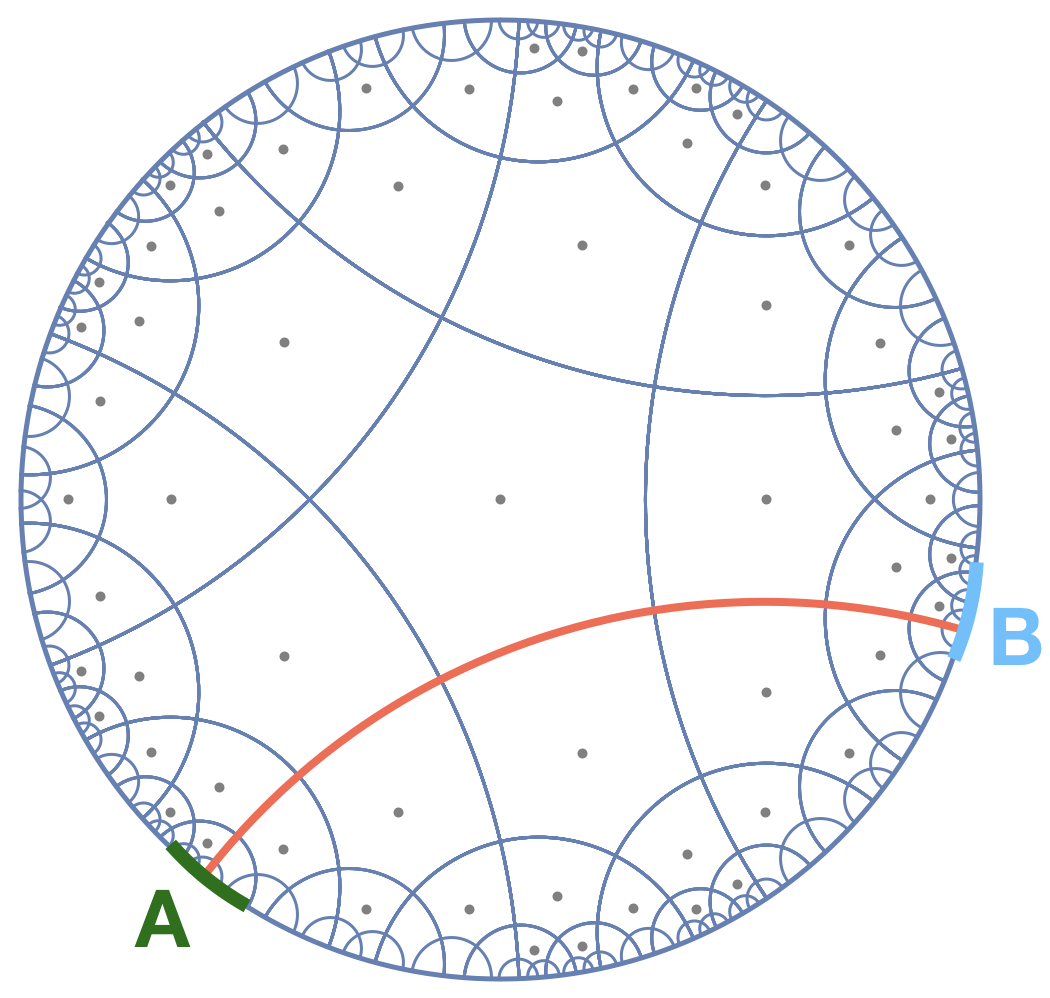}
	\caption{
		An example of subregions $A, B$ with non-vanishing mutual information.
		\label{FIG_mutual_AB}
	}
\end{figure}

Finally we point out an issue for the 
hyperbolic fracton model
that does not exist in the tenser-network models.
Let us consider two small boundary subregions denoted $A$ and $B$,
and examine their mutual information when 
$A$ and $B$ are far apart.
The two subregions should not have any mutual information
according to AdS/CFT, 
which is the case in the tensor-network models.
In the hyperbolic fracton model,
this is also true for most choices of $A$ and $B$.
However, 
there will be one bit of mutual information
when $A$ and $B$ cover  the two ends of the same 
pentagon-edge geodesic.
Such a choice is illustrated in Fig.~\ref{FIG_mutual_AB}.

Such issue has to do with the subsystem symmetry being ``rigid'';
that is, the pentagon-edge geodesics are fixed in the model.

The ground states of most gapped fracton models 
actually have a stabilizer map description
as discussed in Refs.~\cite{Haah2011,Schmitz2018,Ma2018a,Kubica2018,He2018PRB}.
Many of the holographic tensor networks 
are also built from ``perfect'' stabilizer tensors,
although the  construction is different.
The ``perfect'' stabilizer tensor may lend us some insight 
on how to modify the hyperbolic fracton model for improved
realization of AdS/CFT properties.

\section{Outlook}\label{SEC_summary_discussion}
Modern physics has witnessed increasing 
interactions between high energy theory, many-body physics,
and quantum information.
This work adds another example at this trisection,
by elaborating the holographic properties of a classical fracton model.
After an introduction of the fracton model accompanied by 
a discussion of various hints of its similarity with gravity,
we demonstrate that when defined on a hyperbolic disk,
it satisfies some key properties of
AdS/CFT,
including the Rindler reconstruction/subregion duality
and the RT formula for its mutual information.
A naively defined black hole in this model also
has the correct entropy.
\tb{Some generalizations and comparisons with 
	tensor-network toy models are also discussed.}

This work expands the scope of application of holography 
in condensed matter physics.
Not only can one study a strongly coupled/critical system 
as the CFT side of AdS/CFT;
there are also states of matter that exhibit meaningful physics on the AdS side.
In particular, it may be interesting to examine other fracton models
in AdS space, and classify them by their holographic properties.

\tb{
	A long-term ambition we initiate with this work
	is to concretely understand 
	what exactly are the similarities and differences
	between various fracton models
	and quantum gravity.
	In return
	it may help us study 
	how quantum gravity 
	or related many-body models can perform
	quantum error correction encoding,
	which is one of the most intriguing quantum information aspect questions of gravity.
}
We may be able to partially achieve this by quantitatively examining
the speculated web of connections in Fig.~\ref{FIG_Big_pic}.
Some works on fracton models \cite{He2018PRB} suggest that studying a quantum, lattice version of Higgsed linearized general relativity (or a higher-rank gauge theory) 
and constructing the tensor-network 
representation of its ground state are possible.
\tb{
	A reasonable approach could be to explore}
its connections to holographic tensor networks discussed 
in Ref.~\cite{Pastawski2015,Qi2018}.

Some questions remain open even for the classical model,
especially concerning the subregion duality and mutual information
for more complicated, disconnected boundary segments.
%
%

%
The higher-rank gauge theory is also interesting in its own right,
and it remains to be understood whether it is  holographic 
without being Higgsed into \tb{gapped} fracton models,
at both the classical  and quantum \tb{level}. 
A recent development has already shown that some versions 
of the theory can be consistently defined on a constant-curvature
manifold \cite{Gromov2017,Slagle2018}.

Another direction for future investigation
is to study other gapped fracton models protected
by different types of subsystem symmetries, 
or the fracton topological orders obtained by 
gauging these symmetries.
It is desirable to know what are
the necessary and sufficient conditions
for a model to be holographic,
and also construct some of them explicitly.

%
%

To conclude,
certain fracton models give rise to some interesting physics
that mimics general relativity.
In this work we point out the holographic aspect of this, and hope further investigation  could provide 
useful insight for both the condensed-matter and high-energy- theory communities.

\section*{Acknowledgment}
We thank Sugawara Hirotaka, Tadashi Takayanagi, Yasha Neiman, Nic Shannon, Sean Hartnoll, Keisuke Totsuka, Owen Benton, Geet Rakala, 
Arpan Bhattacharyya, Ludovic Jaubert, and Addison Richards
for helpful comments and discussions.
We thank Owen Benton, Nic Shannon, Yasha Neiman,
Sugawara Hirotaka, and Ludovic Jaubert for a careful reading of
the manuscript.
We appreciate the hospitality of Yukawa Institute for Theoretical Physics (YITP), Kyoto University,
where some of the discussion happened and part of the work was done.
H.Y. is supported by 
the Theory of Quantum Matter Unit at Okinawa Institute of Science and Technology
and
the Japan Society for the Promotion of Science (JSPS)
Research Fellowships for Young Scientists.

\bibliography{reference}

\begin{thebibliography}{65}%
\makeatletter
\providecommand \@ifxundefined [1]{%
 \@ifx{#1\undefined}
}%
\providecommand \@ifnum [1]{%
 \ifnum #1\expandafter \@firstoftwo
 \else \expandafter \@secondoftwo
 \fi
}%
\providecommand \@ifx [1]{%
 \ifx #1\expandafter \@firstoftwo
 \else \expandafter \@secondoftwo
 \fi
}%
\providecommand \natexlab [1]{#1}%
\providecommand \enquote  [1]{``#1''}%
\providecommand \bibnamefont  [1]{#1}%
\providecommand \bibfnamefont [1]{#1}%
\providecommand \citenamefont [1]{#1}%
\providecommand \href@noop [0]{\@secondoftwo}%
\providecommand \href [0]{\begingroup \@sanitize@url \@href}%
\providecommand \@href[1]{\@@startlink{#1}\@@href}%
\providecommand \@@href[1]{\endgroup#1\@@endlink}%
\providecommand \@sanitize@url [0]{\catcode `\\12\catcode `\$12\catcode
  `\&12\catcode `\#12\catcode `\^12\catcode `\_12\catcode `\%12\relax}%
\providecommand \@@startlink[1]{}%
\providecommand \@@endlink[0]{}%
\providecommand \url  [0]{\begingroup\@sanitize@url \@url }%
\providecommand \@url [1]{\endgroup\@href {#1}{\urlprefix }}%
\providecommand \urlprefix  [0]{URL }%
\providecommand \Eprint [0]{\href }%
\providecommand \doibase [0]{http://dx.doi.org/}%
\providecommand \selectlanguage [0]{\@gobble}%
\providecommand \bibinfo  [0]{\@secondoftwo}%
\providecommand \bibfield  [0]{\@secondoftwo}%
\providecommand \translation [1]{[#1]}%
\providecommand \BibitemOpen [0]{}%
\providecommand \bibitemStop [0]{}%
\providecommand \bibitemNoStop [0]{.\EOS\space}%
\providecommand \EOS [0]{\spacefactor3000\relax}%
\providecommand \BibitemShut  [1]{\csname bibitem#1\endcsname}%
\let\auto@bib@innerbib\@empty
\bibitem [{\citenamefont {Hooft}(1974)}]{Hooft1974}%
  \BibitemOpen
  \bibfield  {author} {\bibinfo {author} {\bibfnamefont {G.}~\bibnamefont
  {Hooft}},\ }\href {\doibase 10.1016/0550-3213(74)90154-0} {\bibfield
  {journal} {\bibinfo  {journal} {Nucl. Phys. B}\ }\textbf {\bibinfo {volume}
  {72}},\ \bibinfo {pages} {461} (\bibinfo {year} {1974})}\BibitemShut
  {NoStop}%
\bibitem [{\citenamefont {Susskind}(1995)}]{Susskind1995}%
  \BibitemOpen
  \bibfield  {author} {\bibinfo {author} {\bibfnamefont {L.}~\bibnamefont
  {Susskind}},\ }\href {\doibase 10.1063/1.531249} {\bibfield  {journal}
  {\bibinfo  {journal} {J. Math. Phys.}\ }\textbf {\bibinfo {volume} {36}},\
  \bibinfo {pages} {6377} (\bibinfo {year} {1995})}\BibitemShut {NoStop}%
\bibitem [{\citenamefont {Maldacena}(1999)}]{Maldacena1999}%
  \BibitemOpen
  \bibfield  {author} {\bibinfo {author} {\bibfnamefont {J.}~\bibnamefont
  {Maldacena}},\ }\href {\doibase 10.1023/A:1026654312961} {\bibfield
  {journal} {\bibinfo  {journal} {Int. J. Theor. Phys.}\ }\textbf {\bibinfo
  {volume} {38}},\ \bibinfo {pages} {1113} (\bibinfo {year}
  {1999})}\BibitemShut {NoStop}%
\bibitem [{\citenamefont {Witten}(1998)}]{Witten1998}%
  \BibitemOpen
  \bibfield  {author} {\bibinfo {author} {\bibfnamefont {E.}~\bibnamefont
  {Witten}},\ }\href {\doibase 10.4310/ATMP.1998.v2.n2.a2} {\bibfield
  {journal} {\bibinfo  {journal} {Adv. Theor. Math. Phys.}\ }\textbf {\bibinfo
  {volume} {2}},\ \bibinfo {pages} {253} (\bibinfo {year} {1998})}\BibitemShut
  {NoStop}%
\bibitem [{\citenamefont {{Gubser}}\ \emph {et~al.}(1998)\citenamefont
  {{Gubser}}, \citenamefont {{Klebanov}},\ and\ \citenamefont
  {{Polyakov}}}]{gubser1998gauge}%
  \BibitemOpen
  \bibfield  {author} {\bibinfo {author} {\bibfnamefont {S.~S.}\ \bibnamefont
  {{Gubser}}}, \bibinfo {author} {\bibfnamefont {I.~R.}\ \bibnamefont
  {{Klebanov}}}, \ and\ \bibinfo {author} {\bibfnamefont {A.~M.}\ \bibnamefont
  {{Polyakov}}},\ }\href {\doibase 10.1016/S0370-2693(98)00377-3} {\bibfield
  {journal} {\bibinfo  {journal} {Phys. Lett. B}\ }\textbf {\bibinfo {volume}
  {428}},\ \bibinfo {pages} {105} (\bibinfo {year} {1998})}\BibitemShut
  {NoStop}%
\bibitem [{\citenamefont {Aharony}\ \emph {et~al.}(2000)\citenamefont
  {Aharony}, \citenamefont {Gubser}, \citenamefont {Maldacena}, \citenamefont
  {Ooguri},\ and\ \citenamefont {Oz}}]{Aharony2000}%
  \BibitemOpen
  \bibfield  {author} {\bibinfo {author} {\bibfnamefont {O.}~\bibnamefont
  {Aharony}}, \bibinfo {author} {\bibfnamefont {S.~S.}\ \bibnamefont {Gubser}},
  \bibinfo {author} {\bibfnamefont {J.}~\bibnamefont {Maldacena}}, \bibinfo
  {author} {\bibfnamefont {H.}~\bibnamefont {Ooguri}}, \ and\ \bibinfo {author}
  {\bibfnamefont {Y.}~\bibnamefont {Oz}},\ }\href {\doibase
  10.1016/S0370-1573(99)00083-6} {\bibfield  {journal} {\bibinfo  {journal}
  {Phys. Rep.}\ }\textbf {\bibinfo {volume} {323}},\ \bibinfo {pages} {183}
  (\bibinfo {year} {2000})}\BibitemShut {NoStop}%
\bibitem [{\citenamefont {Aharony}\ \emph {et~al.}(2008)\citenamefont
  {Aharony}, \citenamefont {Bergman}, \citenamefont {Jafferis},\ and\
  \citenamefont {Maldacena}}]{Aharony2008}%
  \BibitemOpen
  \bibfield  {author} {\bibinfo {author} {\bibfnamefont {O.}~\bibnamefont
  {Aharony}}, \bibinfo {author} {\bibfnamefont {O.}~\bibnamefont {Bergman}},
  \bibinfo {author} {\bibfnamefont {D.~L.}\ \bibnamefont {Jafferis}}, \ and\
  \bibinfo {author} {\bibfnamefont {J.}~\bibnamefont {Maldacena}},\ }\href
  {\doibase 10.1088/1126-6708/2008/10/091} {\bibfield  {journal} {\bibinfo
  {journal} {JHEP}\ }\textbf {\bibinfo {volume} {2008}},\ \bibinfo {pages}
  {091} (\bibinfo {year} {2008})}\BibitemShut {NoStop}%
\bibitem [{\citenamefont {Klebanov}\ and\ \citenamefont
  {Polyakov}(2002)}]{Klebanov2002}%
  \BibitemOpen
  \bibfield  {author} {\bibinfo {author} {\bibfnamefont {I.}~\bibnamefont
  {Klebanov}}\ and\ \bibinfo {author} {\bibfnamefont {A.}~\bibnamefont
  {Polyakov}},\ }\href {\doibase 10.1016/S0370-2693(02)02980-5} {\bibfield
  {journal} {\bibinfo  {journal} {Phys. Lett. B}\ }\textbf {\bibinfo {volume}
  {550}},\ \bibinfo {pages} {213} (\bibinfo {year} {2002})}\BibitemShut
  {NoStop}%
\bibitem [{\citenamefont {Hawking}(2005)}]{Hawking2005}%
  \BibitemOpen
  \bibfield  {author} {\bibinfo {author} {\bibfnamefont {S.~W.}\ \bibnamefont
  {Hawking}},\ }\href {\doibase 10.1103/PhysRevD.72.084013} {\bibfield
  {journal} {\bibinfo  {journal} {Phys. Rev. D}\ }\textbf {\bibinfo {volume}
  {72}},\ \bibinfo {pages} {084013} (\bibinfo {year} {2005})}\BibitemShut
  {NoStop}%
\bibitem [{\citenamefont {Guica}\ \emph {et~al.}(2009)\citenamefont {Guica},
  \citenamefont {Hartman}, \citenamefont {Song},\ and\ \citenamefont
  {Strominger}}]{Guica2009}%
  \BibitemOpen
  \bibfield  {author} {\bibinfo {author} {\bibfnamefont {M.}~\bibnamefont
  {Guica}}, \bibinfo {author} {\bibfnamefont {T.}~\bibnamefont {Hartman}},
  \bibinfo {author} {\bibfnamefont {W.}~\bibnamefont {Song}}, \ and\ \bibinfo
  {author} {\bibfnamefont {A.}~\bibnamefont {Strominger}},\ }\href {\doibase
  10.1103/PhysRevD.80.124008} {\bibfield  {journal} {\bibinfo  {journal} {Phys.
  Rev. D}\ }\textbf {\bibinfo {volume} {80}},\ \bibinfo {pages} {124008}
  (\bibinfo {year} {2009})}\BibitemShut {NoStop}%
\bibitem [{\citenamefont {Hartnoll}\ \emph {et~al.}(2016)\citenamefont
  {Hartnoll}, \citenamefont {Lucas},\ and\ \citenamefont {Sachdev}}]{Hartnoll}%
  \BibitemOpen
  \bibfield  {author} {\bibinfo {author} {\bibfnamefont {S.}~\bibnamefont
  {Hartnoll}}, \bibinfo {author} {\bibfnamefont {A.}~\bibnamefont {Lucas}}, \
  and\ \bibinfo {author} {\bibfnamefont {S.}~\bibnamefont {Sachdev}},\
  }\href@noop {} {\emph {\bibinfo {title} {Holographic quantum matter}}}\
  (\bibinfo  {publisher} {The MIT Press, Cambridge},\ \bibinfo {year} {2016})\
  \Eprint {http://arxiv.org/abs/1612.07324} {arXiv:1612.07324} \BibitemShut
  {NoStop}%
\bibitem [{\citenamefont {Ryu}\ and\ \citenamefont
  {Takayanagi}(2006{\natexlab{a}})}]{Ryu2006}%
  \BibitemOpen
  \bibfield  {author} {\bibinfo {author} {\bibfnamefont {S.}~\bibnamefont
  {Ryu}}\ and\ \bibinfo {author} {\bibfnamefont {T.}~\bibnamefont
  {Takayanagi}},\ }\href {\doibase 10.1088/1126-6708/2006/08/045} {\bibfield
  {journal} {\bibinfo  {journal} {JHEP}\ }\textbf {\bibinfo {volume} {2006}},\
  \bibinfo {pages} {045} (\bibinfo {year} {2006}{\natexlab{a}})}\BibitemShut
  {NoStop}%
\bibitem [{\citenamefont {Ryu}\ and\ \citenamefont
  {Takayanagi}(2006{\natexlab{b}})}]{Ryu2006a}%
  \BibitemOpen
  \bibfield  {author} {\bibinfo {author} {\bibfnamefont {S.}~\bibnamefont
  {Ryu}}\ and\ \bibinfo {author} {\bibfnamefont {T.}~\bibnamefont
  {Takayanagi}},\ }\href {\doibase 10.1103/PhysRevLett.96.181602} {\bibfield
  {journal} {\bibinfo  {journal} {Phys. Rev. Lett.}\ }\textbf {\bibinfo
  {volume} {96}},\ \bibinfo {pages} {181602} (\bibinfo {year}
  {2006}{\natexlab{b}})}\BibitemShut {NoStop}%
\bibitem [{\citenamefont {Bousso}(2002)}]{Bousso2002}%
  \BibitemOpen
  \bibfield  {author} {\bibinfo {author} {\bibfnamefont {R.}~\bibnamefont
  {Bousso}},\ }\href {\doibase 10.1103/RevModPhys.74.825} {\bibfield  {journal}
  {\bibinfo  {journal} {Rev. Mod. Phys.}\ }\textbf {\bibinfo {volume} {74}},\
  \bibinfo {pages} {825} (\bibinfo {year} {2002})}\BibitemShut {NoStop}%
\bibitem [{\citenamefont {Swingle}(2012)}]{Swingle2012}%
  \BibitemOpen
  \bibfield  {author} {\bibinfo {author} {\bibfnamefont {B.}~\bibnamefont
  {Swingle}},\ }\href {\doibase 10.1103/PhysRevD.86.065007} {\bibfield
  {journal} {\bibinfo  {journal} {Phys. Rev. D}\ }\textbf {\bibinfo {volume}
  {86}},\ \bibinfo {pages} {065007} (\bibinfo {year} {2012})}\BibitemShut
  {NoStop}%
\bibitem [{\citenamefont {Pastawski}\ \emph {et~al.}(2015)\citenamefont
  {Pastawski}, \citenamefont {Yoshida}, \citenamefont {Harlow},\ and\
  \citenamefont {Preskill}}]{Pastawski2015}%
  \BibitemOpen
  \bibfield  {author} {\bibinfo {author} {\bibfnamefont {F.}~\bibnamefont
  {Pastawski}}, \bibinfo {author} {\bibfnamefont {B.}~\bibnamefont {Yoshida}},
  \bibinfo {author} {\bibfnamefont {D.}~\bibnamefont {Harlow}}, \ and\ \bibinfo
  {author} {\bibfnamefont {J.}~\bibnamefont {Preskill}},\ }\href {\doibase
  10.1007/JHEP06(2015)149} {\bibfield  {journal} {\bibinfo  {journal} {JHEP}\
  }\textbf {\bibinfo {volume} {2015}},\ \bibinfo {pages} {149} (\bibinfo {year}
  {2015})}\BibitemShut {NoStop}%
\bibitem [{\citenamefont {Almheiri}\ \emph {et~al.}(2015)\citenamefont
  {Almheiri}, \citenamefont {Dong},\ and\ \citenamefont
  {Harlow}}]{Almheiri2015}%
  \BibitemOpen
  \bibfield  {author} {\bibinfo {author} {\bibfnamefont {A.}~\bibnamefont
  {Almheiri}}, \bibinfo {author} {\bibfnamefont {X.}~\bibnamefont {Dong}}, \
  and\ \bibinfo {author} {\bibfnamefont {D.}~\bibnamefont {Harlow}},\ }\href
  {\doibase 10.1007/JHEP04(2015)163} {\bibfield  {journal} {\bibinfo  {journal}
  {JHEP}\ }\textbf {\bibinfo {volume} {2015}},\ \bibinfo {pages} {163}
  (\bibinfo {year} {2015})}\BibitemShut {NoStop}%
\bibitem [{\citenamefont {Hayden}\ \emph {et~al.}(2016)\citenamefont {Hayden},
  \citenamefont {Nezami}, \citenamefont {Qi}, \citenamefont {Thomas},
  \citenamefont {Walter},\ and\ \citenamefont {Yang}}]{Hayden2016}%
  \BibitemOpen
  \bibfield  {author} {\bibinfo {author} {\bibfnamefont {P.}~\bibnamefont
  {Hayden}}, \bibinfo {author} {\bibfnamefont {S.}~\bibnamefont {Nezami}},
  \bibinfo {author} {\bibfnamefont {X.-L.}\ \bibnamefont {Qi}}, \bibinfo
  {author} {\bibfnamefont {N.}~\bibnamefont {Thomas}}, \bibinfo {author}
  {\bibfnamefont {M.}~\bibnamefont {Walter}}, \ and\ \bibinfo {author}
  {\bibfnamefont {Z.}~\bibnamefont {Yang}},\ }\href {\doibase
  10.1007/JHEP11(2016)009} {\bibfield  {journal} {\bibinfo  {journal} {JHEP}\
  }\textbf {\bibinfo {volume} {2016}},\ \bibinfo {pages} {9} (\bibinfo {year}
  {2016})}\BibitemShut {NoStop}%
\bibitem [{\citenamefont {{Qi}}\ and\ \citenamefont {{Yang}}(2018)}]{Qi2018}%
  \BibitemOpen
  \bibfield  {author} {\bibinfo {author} {\bibfnamefont {X.-L.}\ \bibnamefont
  {{Qi}}}\ and\ \bibinfo {author} {\bibfnamefont {Z.}~\bibnamefont {{Yang}}},\
  }\href@noop {} {\  (\bibinfo {year} {2018})},\ \Eprint
  {http://arxiv.org/abs/1801.05289} {arXiv:1801.05289 [hep-th]} \BibitemShut
  {NoStop}%
\bibitem [{\citenamefont {{Qi}}(2013)}]{Qi2013}%
  \BibitemOpen
  \bibfield  {author} {\bibinfo {author} {\bibfnamefont {X.-L.}\ \bibnamefont
  {{Qi}}},\ }\href@noop {} {\  (\bibinfo {year} {2013})},\ \Eprint
  {http://arxiv.org/abs/1309.6282} {arXiv:1309.6282 [hep-th]} \BibitemShut
  {NoStop}%
\bibitem [{\citenamefont {Gu}\ \emph {et~al.}(2016)\citenamefont {Gu},
  \citenamefont {Lee}, \citenamefont {Wen}, \citenamefont {Cho}, \citenamefont
  {Ryu},\ and\ \citenamefont {Qi}}]{Gu2016}%
  \BibitemOpen
  \bibfield  {author} {\bibinfo {author} {\bibfnamefont {Y.}~\bibnamefont
  {Gu}}, \bibinfo {author} {\bibfnamefont {C.~H.}\ \bibnamefont {Lee}},
  \bibinfo {author} {\bibfnamefont {X.}~\bibnamefont {Wen}}, \bibinfo {author}
  {\bibfnamefont {G.~Y.}\ \bibnamefont {Cho}}, \bibinfo {author} {\bibfnamefont
  {S.}~\bibnamefont {Ryu}}, \ and\ \bibinfo {author} {\bibfnamefont {X.-L.}\
  \bibnamefont {Qi}},\ }\href {\doibase 10.1103/PhysRevB.94.125107} {\bibfield
  {journal} {\bibinfo  {journal} {Phys. Rev. B}\ }\textbf {\bibinfo {volume}
  {94}},\ \bibinfo {pages} {125107} (\bibinfo {year} {2016})}\BibitemShut
  {NoStop}%
\bibitem [{\citenamefont {Lee}\ and\ \citenamefont {Qi}(2016)}]{Lee2016}%
  \BibitemOpen
  \bibfield  {author} {\bibinfo {author} {\bibfnamefont {C.~H.}\ \bibnamefont
  {Lee}}\ and\ \bibinfo {author} {\bibfnamefont {X.-L.}\ \bibnamefont {Qi}},\
  }\href {\doibase 10.1103/PhysRevB.93.035112} {\bibfield  {journal} {\bibinfo
  {journal} {Phys. Rev. B}\ }\textbf {\bibinfo {volume} {93}},\ \bibinfo
  {pages} {035112} (\bibinfo {year} {2016})}\BibitemShut {NoStop}%
\bibitem [{\citenamefont {Haah}(2011)}]{Haah2011}%
  \BibitemOpen
  \bibfield  {author} {\bibinfo {author} {\bibfnamefont {J.}~\bibnamefont
  {Haah}},\ }\href {\doibase 10.1103/PhysRevA.83.042330} {\bibfield  {journal}
  {\bibinfo  {journal} {Phys. Rev. A}\ }\textbf {\bibinfo {volume} {83}},\
  \bibinfo {pages} {042330} (\bibinfo {year} {2011})}\BibitemShut {NoStop}%
\bibitem [{\citenamefont {Vijay}\ \emph {et~al.}(2015)\citenamefont {Vijay},
  \citenamefont {Haah},\ and\ \citenamefont {Fu}}]{Vijay2015}%
  \BibitemOpen
  \bibfield  {author} {\bibinfo {author} {\bibfnamefont {S.}~\bibnamefont
  {Vijay}}, \bibinfo {author} {\bibfnamefont {J.}~\bibnamefont {Haah}}, \ and\
  \bibinfo {author} {\bibfnamefont {L.}~\bibnamefont {Fu}},\ }\href {\doibase
  10.1103/PhysRevB.92.235136} {\bibfield  {journal} {\bibinfo  {journal} {Phys.
  Rev. B}\ }\textbf {\bibinfo {volume} {92}},\ \bibinfo {pages} {235136}
  (\bibinfo {year} {2015})}\BibitemShut {NoStop}%
\bibitem [{\citenamefont {Vijay}\ \emph {et~al.}(2016)\citenamefont {Vijay},
  \citenamefont {Haah},\ and\ \citenamefont {Fu}}]{Vijay2016}%
  \BibitemOpen
  \bibfield  {author} {\bibinfo {author} {\bibfnamefont {S.}~\bibnamefont
  {Vijay}}, \bibinfo {author} {\bibfnamefont {J.}~\bibnamefont {Haah}}, \ and\
  \bibinfo {author} {\bibfnamefont {L.}~\bibnamefont {Fu}},\ }\href {\doibase
  10.1103/PhysRevB.94.235157} {\bibfield  {journal} {\bibinfo  {journal} {Phys.
  Rev. B}\ }\textbf {\bibinfo {volume} {94}},\ \bibinfo {pages} {235157}
  (\bibinfo {year} {2016})}\BibitemShut {NoStop}%
\bibitem [{\citenamefont {Hal{\'{a}}sz}\ \emph {et~al.}(2017)\citenamefont
  {Hal{\'{a}}sz}, \citenamefont {Hsieh},\ and\ \citenamefont
  {Balents}}]{Halasz2017}%
  \BibitemOpen
  \bibfield  {author} {\bibinfo {author} {\bibfnamefont {G.~B.}\ \bibnamefont
  {Hal{\'{a}}sz}}, \bibinfo {author} {\bibfnamefont {T.~H.}\ \bibnamefont
  {Hsieh}}, \ and\ \bibinfo {author} {\bibfnamefont {L.}~\bibnamefont
  {Balents}},\ }\href {\doibase 10.1103/PhysRevLett.119.257202} {\bibfield
  {journal} {\bibinfo  {journal} {Phys. Rev. Lett.}\ }\textbf {\bibinfo
  {volume} {119}},\ \bibinfo {pages} {257202} (\bibinfo {year}
  {2017})}\BibitemShut {NoStop}%
\bibitem [{\citenamefont {Bulmash}\ and\ \citenamefont
  {Barkeshli}(2018)}]{Bulmash2018}%
  \BibitemOpen
  \bibfield  {author} {\bibinfo {author} {\bibfnamefont {D.}~\bibnamefont
  {Bulmash}}\ and\ \bibinfo {author} {\bibfnamefont {M.}~\bibnamefont
  {Barkeshli}},\ }\href {\doibase 10.1103/PhysRevB.97.235112} {\bibfield
  {journal} {\bibinfo  {journal} {Phys. Rev. B}\ }\textbf {\bibinfo {volume}
  {97}},\ \bibinfo {pages} {235112} (\bibinfo {year} {2018})}\BibitemShut
  {NoStop}%
\bibitem [{\citenamefont {Shirley}\ \emph {et~al.}(2018)\citenamefont
  {Shirley}, \citenamefont {Slagle}, \citenamefont {Wang},\ and\ \citenamefont
  {Chen}}]{Shirley2017}%
  \BibitemOpen
  \bibfield  {author} {\bibinfo {author} {\bibfnamefont {W.}~\bibnamefont
  {Shirley}}, \bibinfo {author} {\bibfnamefont {K.}~\bibnamefont {Slagle}},
  \bibinfo {author} {\bibfnamefont {Z.}~\bibnamefont {Wang}}, \ and\ \bibinfo
  {author} {\bibfnamefont {X.}~\bibnamefont {Chen}},\ }\href {\doibase
  10.1103/PhysRevX.8.031051} {\bibfield  {journal} {\bibinfo  {journal} {Phys.
  Rev. X}\ }\textbf {\bibinfo {volume} {8}},\ \bibinfo {pages} {031051}
  (\bibinfo {year} {2018})}\BibitemShut {NoStop}%
\bibitem [{\citenamefont {Shirley}\ \emph
  {et~al.}(2019{\natexlab{a}})\citenamefont {Shirley}, \citenamefont {Slagle},\
  and\ \citenamefont {Chen}}]{Shirley2018}%
  \BibitemOpen
  \bibfield  {author} {\bibinfo {author} {\bibfnamefont {W.}~\bibnamefont
  {Shirley}}, \bibinfo {author} {\bibfnamefont {K.}~\bibnamefont {Slagle}}, \
  and\ \bibinfo {author} {\bibfnamefont {X.}~\bibnamefont {Chen}},\ }\href
  {\doibase 10.21468/SciPostPhys.6.1.015} {\bibfield  {journal} {\bibinfo
  {journal} {SciPost Phys.}\ }\textbf {\bibinfo {volume} {6}},\ \bibinfo
  {pages} {15} (\bibinfo {year} {2019}{\natexlab{a}})}\BibitemShut {NoStop}%
\bibitem [{\citenamefont {{Ma}}\ \emph {et~al.}(2018)\citenamefont {{Ma}},
  \citenamefont {{Hermele}},\ and\ \citenamefont {{Chen}}}]{Ma2018}%
  \BibitemOpen
  \bibfield  {author} {\bibinfo {author} {\bibfnamefont {H.}~\bibnamefont
  {{Ma}}}, \bibinfo {author} {\bibfnamefont {M.}~\bibnamefont {{Hermele}}}, \
  and\ \bibinfo {author} {\bibfnamefont {X.}~\bibnamefont {{Chen}}},\ }\href
  {\doibase 10.1103/PhysRevB.98.035111} {\bibfield  {journal} {\bibinfo
  {journal} {Phys. Rev. B}\ }\textbf {\bibinfo {volume} {98}},\ \bibinfo {eid}
  {035111} (\bibinfo {year} {2018})}\BibitemShut {NoStop}%
\bibitem [{\citenamefont {Schmitz}\ \emph {et~al.}(2018)\citenamefont
  {Schmitz}, \citenamefont {Ma}, \citenamefont {Nandkishore},\ and\
  \citenamefont {Parameswaran}}]{Schmitz2018}%
  \BibitemOpen
  \bibfield  {author} {\bibinfo {author} {\bibfnamefont {A.~T.}\ \bibnamefont
  {Schmitz}}, \bibinfo {author} {\bibfnamefont {H.}~\bibnamefont {Ma}},
  \bibinfo {author} {\bibfnamefont {R.~M.}\ \bibnamefont {Nandkishore}}, \ and\
  \bibinfo {author} {\bibfnamefont {S.~A.}\ \bibnamefont {Parameswaran}},\
  }\href {\doibase 10.1103/PhysRevB.97.134426} {\bibfield  {journal} {\bibinfo
  {journal} {Phys. Rev. B}\ }\textbf {\bibinfo {volume} {97}},\ \bibinfo
  {pages} {134426} (\bibinfo {year} {2018})}\BibitemShut {NoStop}%
\bibitem [{\citenamefont {Ma}\ \emph {et~al.}(2017)\citenamefont {Ma},
  \citenamefont {Lake}, \citenamefont {Chen},\ and\ \citenamefont
  {Hermele}}]{Ma2017}%
  \BibitemOpen
  \bibfield  {author} {\bibinfo {author} {\bibfnamefont {H.}~\bibnamefont
  {Ma}}, \bibinfo {author} {\bibfnamefont {E.}~\bibnamefont {Lake}}, \bibinfo
  {author} {\bibfnamefont {X.}~\bibnamefont {Chen}}, \ and\ \bibinfo {author}
  {\bibfnamefont {M.}~\bibnamefont {Hermele}},\ }\href {\doibase
  10.1103/PhysRevB.95.245126} {\bibfield  {journal} {\bibinfo  {journal} {Phys.
  Rev. B}\ }\textbf {\bibinfo {volume} {95}},\ \bibinfo {pages} {245126}
  (\bibinfo {year} {2017})}\BibitemShut {NoStop}%
\bibitem [{\citenamefont {Ma}\ \emph {et~al.}(2018)\citenamefont {Ma},
  \citenamefont {Schmitz}, \citenamefont {Parameswaran}, \citenamefont
  {Hermele},\ and\ \citenamefont {Nandkishore}}]{Ma2018a}%
  \BibitemOpen
  \bibfield  {author} {\bibinfo {author} {\bibfnamefont {H.}~\bibnamefont
  {Ma}}, \bibinfo {author} {\bibfnamefont {A.~T.}\ \bibnamefont {Schmitz}},
  \bibinfo {author} {\bibfnamefont {S.~A.}\ \bibnamefont {Parameswaran}},
  \bibinfo {author} {\bibfnamefont {M.}~\bibnamefont {Hermele}}, \ and\
  \bibinfo {author} {\bibfnamefont {R.~M.}\ \bibnamefont {Nandkishore}},\
  }\href {\doibase 10.1103/PhysRevB.97.125101} {\bibfield  {journal} {\bibinfo
  {journal} {Phys. Rev. B}\ }\textbf {\bibinfo {volume} {97}},\ \bibinfo
  {pages} {125101} (\bibinfo {year} {2018})}\BibitemShut {NoStop}%
\bibitem [{\citenamefont {Nandkishore}\ and\ \citenamefont
  {Hermele}(2019)}]{Nandkishore2018}%
  \BibitemOpen
  \bibfield  {author} {\bibinfo {author} {\bibfnamefont {R.~M.}\ \bibnamefont
  {Nandkishore}}\ and\ \bibinfo {author} {\bibfnamefont {M.}~\bibnamefont
  {Hermele}},\ }\href {\doibase 10.1146/annurev-conmatphys-031218-013604}
  {\bibfield  {journal} {\bibinfo  {journal} {Annual Review of Condensed Matter
  Physics}\ }\textbf {\bibinfo {volume} {10}},\ \bibinfo {pages} {295}
  (\bibinfo {year} {2019})},\ \Eprint
  {http://arxiv.org/abs/https://doi.org/10.1146/annurev-conmatphys-031218-013604}
  {https://doi.org/10.1146/annurev-conmatphys-031218-013604} \BibitemShut
  {NoStop}%
\bibitem [{\citenamefont {Pretko}\ and\ \citenamefont
  {Radzihovsky}(2018)}]{Pretko2018}%
  \BibitemOpen
  \bibfield  {author} {\bibinfo {author} {\bibfnamefont {M.}~\bibnamefont
  {Pretko}}\ and\ \bibinfo {author} {\bibfnamefont {L.}~\bibnamefont
  {Radzihovsky}},\ }\href {\doibase 10.1103/PhysRevLett.120.195301} {\bibfield
  {journal} {\bibinfo  {journal} {Phys. Rev. Lett.}\ }\textbf {\bibinfo
  {volume} {120}},\ \bibinfo {pages} {195301} (\bibinfo {year}
  {2018})}\BibitemShut {NoStop}%
\bibitem [{\citenamefont {Pretko}(2017{\natexlab{a}})}]{Pretko2017}%
  \BibitemOpen
  \bibfield  {author} {\bibinfo {author} {\bibfnamefont {M.}~\bibnamefont
  {Pretko}},\ }\href {\doibase 10.1103/PhysRevD.96.024051} {\bibfield
  {journal} {\bibinfo  {journal} {Phys. Rev. D}\ }\textbf {\bibinfo {volume}
  {96}},\ \bibinfo {pages} {024051} (\bibinfo {year}
  {2017}{\natexlab{a}})}\BibitemShut {NoStop}%
\bibitem [{\citenamefont {Pretko}(2017{\natexlab{b}})}]{Pretko2017a}%
  \BibitemOpen
  \bibfield  {author} {\bibinfo {author} {\bibfnamefont {M.}~\bibnamefont
  {Pretko}},\ }\href {\doibase 10.1103/PhysRevB.96.035119} {\bibfield
  {journal} {\bibinfo  {journal} {Phys. Rev. B}\ }\textbf {\bibinfo {volume}
  {96}},\ \bibinfo {pages} {035119} (\bibinfo {year}
  {2017}{\natexlab{b}})}\BibitemShut {NoStop}%
\bibitem [{\citenamefont {Pretko}(2017{\natexlab{c}})}]{Pretko2017b}%
  \BibitemOpen
  \bibfield  {author} {\bibinfo {author} {\bibfnamefont {M.}~\bibnamefont
  {Pretko}},\ }\href {\doibase 10.1103/PhysRevB.95.115139} {\bibfield
  {journal} {\bibinfo  {journal} {Phys. Rev. B}\ }\textbf {\bibinfo {volume}
  {95}},\ \bibinfo {pages} {115139} (\bibinfo {year}
  {2017}{\natexlab{c}})}\BibitemShut {NoStop}%
\bibitem [{\citenamefont {Shirley}\ \emph
  {et~al.}(2019{\natexlab{b}})\citenamefont {Shirley}, \citenamefont {Slagle},\
  and\ \citenamefont {Chen}}]{Shirly2018-2}%
  \BibitemOpen
  \bibfield  {author} {\bibinfo {author} {\bibfnamefont {W.}~\bibnamefont
  {Shirley}}, \bibinfo {author} {\bibfnamefont {K.}~\bibnamefont {Slagle}}, \
  and\ \bibinfo {author} {\bibfnamefont {X.}~\bibnamefont {Chen}},\ }\href
  {\doibase 10.21468/SciPostPhys.6.4.041} {\bibfield  {journal} {\bibinfo
  {journal} {SciPost Phys.}\ }\textbf {\bibinfo {volume} {6}},\ \bibinfo
  {pages} {41} (\bibinfo {year} {2019}{\natexlab{b}})}\BibitemShut {NoStop}%
\bibitem [{\citenamefont {Devakul}\ \emph {et~al.}(2018)\citenamefont
  {Devakul}, \citenamefont {Williamson},\ and\ \citenamefont
  {You}}]{DevakulPhysRevB2018}%
  \BibitemOpen
  \bibfield  {author} {\bibinfo {author} {\bibfnamefont {T.}~\bibnamefont
  {Devakul}}, \bibinfo {author} {\bibfnamefont {D.~J.}\ \bibnamefont
  {Williamson}}, \ and\ \bibinfo {author} {\bibfnamefont {Y.}~\bibnamefont
  {You}},\ }\href {\doibase 10.1103/PhysRevB.98.235121} {\bibfield  {journal}
  {\bibinfo  {journal} {Phys. Rev. B}\ }\textbf {\bibinfo {volume} {98}},\
  \bibinfo {pages} {235121} (\bibinfo {year} {2018})}\BibitemShut {NoStop}%
\bibitem [{\citenamefont {You}\ \emph {et~al.}(2018)\citenamefont {You},
  \citenamefont {Devakul}, \citenamefont {Burnell},\ and\ \citenamefont
  {Sondhi}}]{YouPhysRevB2018}%
  \BibitemOpen
  \bibfield  {author} {\bibinfo {author} {\bibfnamefont {Y.}~\bibnamefont
  {You}}, \bibinfo {author} {\bibfnamefont {T.}~\bibnamefont {Devakul}},
  \bibinfo {author} {\bibfnamefont {F.~J.}\ \bibnamefont {Burnell}}, \ and\
  \bibinfo {author} {\bibfnamefont {S.~L.}\ \bibnamefont {Sondhi}},\ }\href
  {\doibase 10.1103/PhysRevB.98.035112} {\bibfield  {journal} {\bibinfo
  {journal} {Phys. Rev. B}\ }\textbf {\bibinfo {volume} {98}},\ \bibinfo
  {pages} {035112} (\bibinfo {year} {2018})}\BibitemShut {NoStop}%
\bibitem [{\citenamefont {Baxter}(2007)}]{baxter2016exactly}%
  \BibitemOpen
  \bibfield  {author} {\bibinfo {author} {\bibfnamefont {R.~J.}\ \bibnamefont
  {Baxter}},\ }\href@noop {} {\emph {\bibinfo {title} {Exactly solved models in
  statistical mechanics}}}\ (\bibinfo  {publisher} {Dove Publications, New
  York},\ \bibinfo {year} {2007})\BibitemShut {NoStop}%
\bibitem [{\citenamefont {Jack}\ \emph {et~al.}(2005)\citenamefont {Jack},
  \citenamefont {Berthier},\ and\ \citenamefont {Garrahan}}]{Jack05}%
  \BibitemOpen
  \bibfield  {author} {\bibinfo {author} {\bibfnamefont {R.~L.}\ \bibnamefont
  {Jack}}, \bibinfo {author} {\bibfnamefont {L.}~\bibnamefont {Berthier}}, \
  and\ \bibinfo {author} {\bibfnamefont {J.~P.}\ \bibnamefont {Garrahan}},\
  }\href {\doibase 10.1103/PhysRevE.72.016103} {\bibfield  {journal} {\bibinfo
  {journal} {Phys. Rev. E}\ }\textbf {\bibinfo {volume} {72}},\ \bibinfo
  {pages} {016103} (\bibinfo {year} {2005})}\BibitemShut {NoStop}%
\bibitem [{\citenamefont {Garrahan}\ and\ \citenamefont
  {Newman}(2000)}]{GarrahanPhysRevE}%
  \BibitemOpen
  \bibfield  {author} {\bibinfo {author} {\bibfnamefont {J.~P.}\ \bibnamefont
  {Garrahan}}\ and\ \bibinfo {author} {\bibfnamefont {M.~E.~J.}\ \bibnamefont
  {Newman}},\ }\href {\doibase 10.1103/PhysRevE.62.7670} {\bibfield  {journal}
  {\bibinfo  {journal} {Phys. Rev. E}\ }\textbf {\bibinfo {volume} {62}},\
  \bibinfo {pages} {7670} (\bibinfo {year} {2000})}\BibitemShut {NoStop}%
\bibitem [{\citenamefont {{Savvidy}}\ and\ \citenamefont
  {{Wegner}}(1994)}]{Savvidy1994}%
  \BibitemOpen
  \bibfield  {author} {\bibinfo {author} {\bibfnamefont {G.~K.}\ \bibnamefont
  {{Savvidy}}}\ and\ \bibinfo {author} {\bibfnamefont {F.~J.}\ \bibnamefont
  {{Wegner}}},\ }\href {\doibase 10.1016/0550-3213(94)90003-5} {\bibfield
  {journal} {\bibinfo  {journal} {Nuclear Physics B}\ }\textbf {\bibinfo
  {volume} {413}},\ \bibinfo {pages} {605} (\bibinfo {year} {1994})},\ \Eprint
  {http://arxiv.org/abs/hep-th/9308094} {hep-th/9308094} \BibitemShut {NoStop}%
\bibitem [{\citenamefont {{Savvidy}}\ and\ \citenamefont
  {{Savvidy}}(1996)}]{Savvidy1996}%
  \BibitemOpen
  \bibfield  {author} {\bibinfo {author} {\bibfnamefont {G.~K.}\ \bibnamefont
  {{Savvidy}}}\ and\ \bibinfo {author} {\bibfnamefont {K.~G.}\ \bibnamefont
  {{Savvidy}}},\ }\href {\doibase 10.1142/S0217732396001399} {\bibfield
  {journal} {\bibinfo  {journal} {Modern Physics Letters A}\ }\textbf {\bibinfo
  {volume} {11}},\ \bibinfo {pages} {1379} (\bibinfo {year} {1996})},\ \Eprint
  {http://arxiv.org/abs/hep-th/9506184} {hep-th/9506184} \BibitemShut {NoStop}%
\bibitem [{\citenamefont {{Pietig}}\ and\ \citenamefont
  {{Wegner}}(1998)}]{Pietig1998}%
  \BibitemOpen
  \bibfield  {author} {\bibinfo {author} {\bibfnamefont {R.}~\bibnamefont
  {{Pietig}}}\ and\ \bibinfo {author} {\bibfnamefont {F.~J.}\ \bibnamefont
  {{Wegner}}},\ }\href {\doibase 10.1016/S0550-3213(98)00342-3} {\bibfield
  {journal} {\bibinfo  {journal} {Nuclear Physics B}\ }\textbf {\bibinfo
  {volume} {525}},\ \bibinfo {pages} {549} (\bibinfo {year} {1998})},\ \Eprint
  {http://arxiv.org/abs/hep-lat/9712002} {hep-lat/9712002} \BibitemShut
  {NoStop}%
\bibitem [{\citenamefont {{'t Hooft}}(1993)}]{Hooft1993}%
  \BibitemOpen
  \bibfield  {author} {\bibinfo {author} {\bibfnamefont {G.}~\bibnamefont {{'t
  Hooft}}},\ }\href@noop {} {\  (\bibinfo {year} {1993})},\ \Eprint
  {http://arxiv.org/abs/gr-qc/9310026} {arXiv:gr-qc/9310026 [gr-qc]}
  \BibitemShut {NoStop}%
\bibitem [{\citenamefont {{Leclerc}}(2006)}]{leclerc2006faddeev}%
  \BibitemOpen
  \bibfield  {author} {\bibinfo {author} {\bibfnamefont {M.}~\bibnamefont
  {{Leclerc}}},\ }\href@noop {} {\  (\bibinfo {year} {2006})},\ \Eprint
  {http://arxiv.org/abs/gr-qc/0612125} {gr-qc/0612125} \BibitemShut {NoStop}%
\bibitem [{\citenamefont {{Rasmussen}}\ \emph {et~al.}(2016)\citenamefont
  {{Rasmussen}}, \citenamefont {{You}},\ and\ \citenamefont
  {{Xu}}}]{rasmussen2016stable}%
  \BibitemOpen
  \bibfield  {author} {\bibinfo {author} {\bibfnamefont {A.}~\bibnamefont
  {{Rasmussen}}}, \bibinfo {author} {\bibfnamefont {Y.-Z.}\ \bibnamefont
  {{You}}}, \ and\ \bibinfo {author} {\bibfnamefont {C.}~\bibnamefont {{Xu}}},\
  }\href@noop {} {\  (\bibinfo {year} {2016})},\ \Eprint
  {http://arxiv.org/abs/1601.08235} {arXiv:1601.08235 [cond-mat.str-el]}
  \BibitemShut {NoStop}%
\bibitem [{\citenamefont {Rasmussen}\ and\ \citenamefont
  {Jermyn}(2018)}]{RasmussenPhysRevB2018}%
  \BibitemOpen
  \bibfield  {author} {\bibinfo {author} {\bibfnamefont {A.}~\bibnamefont
  {Rasmussen}}\ and\ \bibinfo {author} {\bibfnamefont {A.~S.}\ \bibnamefont
  {Jermyn}},\ }\href {\doibase 10.1103/PhysRevB.97.165141} {\bibfield
  {journal} {\bibinfo  {journal} {Phys. Rev. B}\ }\textbf {\bibinfo {volume}
  {97}},\ \bibinfo {pages} {165141} (\bibinfo {year} {2018})}\BibitemShut
  {NoStop}%
\bibitem [{\citenamefont {Slagle}\ \emph {et~al.}(2019)\citenamefont {Slagle},
  \citenamefont {Aasen},\ and\ \citenamefont {Williamson}}]{Slagle2018arXiv}%
  \BibitemOpen
  \bibfield  {author} {\bibinfo {author} {\bibfnamefont {K.}~\bibnamefont
  {Slagle}}, \bibinfo {author} {\bibfnamefont {D.}~\bibnamefont {Aasen}}, \
  and\ \bibinfo {author} {\bibfnamefont {D.}~\bibnamefont {Williamson}},\
  }\href {\doibase 10.21468/SciPostPhys.6.4.043} {\bibfield  {journal}
  {\bibinfo  {journal} {SciPost Phys.}\ }\textbf {\bibinfo {volume} {6}},\
  \bibinfo {pages} {43} (\bibinfo {year} {2019})}\BibitemShut {NoStop}%
\bibitem [{\citenamefont {{Polchinski}}(2010)}]{Polchinski2010}%
  \BibitemOpen
  \bibfield  {author} {\bibinfo {author} {\bibfnamefont {J.}~\bibnamefont
  {{Polchinski}}},\ }\href@noop {} {\  (\bibinfo {year} {2010})},\ \Eprint
  {http://arxiv.org/abs/1010.6134} {arXiv:1010.6134 [hep-th]} \BibitemShut
  {NoStop}%
\bibitem [{\citenamefont {{Nastase}}(2007)}]{nastase2007introduction}%
  \BibitemOpen
  \bibfield  {author} {\bibinfo {author} {\bibfnamefont {H.}~\bibnamefont
  {{Nastase}}},\ }\href@noop {} {\  (\bibinfo {year} {2007})},\ \Eprint
  {http://arxiv.org/abs/0712.0689} {arXiv:0712.0689 [hep-th]} \BibitemShut
  {NoStop}%
\bibitem [{\citenamefont {{Klebanov}}(2001)}]{klebanov2001tasi}%
  \BibitemOpen
  \bibfield  {author} {\bibinfo {author} {\bibfnamefont {I.~R.}\ \bibnamefont
  {{Klebanov}}},\ }in\ \href@noop {} {\emph {\bibinfo {booktitle} {Strings,
  Branes, and Gravity TASI 99}}},\ \bibinfo {editor} {edited by\ \bibinfo
  {editor} {\bibfnamefont {J.~A.}\ \bibnamefont {{Harvey}}}, \bibinfo {editor}
  {\bibfnamefont {S.}~\bibnamefont {{Kachru}}}, \ and\ \bibinfo {editor}
  {\bibfnamefont {E.}~\bibnamefont {{Silverstein}}}}\ (\bibinfo  {publisher}
  {World Scientific, Singapore},\ \bibinfo {year} {2001})\ pp.\ \bibinfo
  {pages} {615--650}\BibitemShut {NoStop}%
\bibitem [{\citenamefont {{Maldacena}}(2003)}]{maldacena2003tasi}%
  \BibitemOpen
  \bibfield  {author} {\bibinfo {author} {\bibfnamefont {J.~M.}\ \bibnamefont
  {{Maldacena}}},\ }\href@noop {} {\  (\bibinfo {year} {2003})},\ \Eprint
  {http://arxiv.org/abs/hep-th/0309246} {arXiv:hep-th/0309246} \BibitemShut
  {NoStop}%
\bibitem [{\citenamefont {Bekenstein}(1973)}]{Bekenstein1973}%
  \BibitemOpen
  \bibfield  {author} {\bibinfo {author} {\bibfnamefont {J.~D.}\ \bibnamefont
  {Bekenstein}},\ }\href {\doibase 10.1103/PhysRevD.7.2333} {\bibfield
  {journal} {\bibinfo  {journal} {Phys. Rev. D}\ }\textbf {\bibinfo {volume}
  {7}},\ \bibinfo {pages} {2333} (\bibinfo {year} {1973})}\BibitemShut
  {NoStop}%
\bibitem [{\citenamefont {Hawking}(1975)}]{Hawking}%
  \BibitemOpen
  \bibfield  {author} {\bibinfo {author} {\bibfnamefont {S.~W.}\ \bibnamefont
  {Hawking}},\ }\href {\doibase 10.1007/bf02345020} {\bibfield  {journal}
  {\bibinfo  {journal} {Commun. Math. Phys.}\ }\textbf {\bibinfo {volume}
  {43}},\ \bibinfo {pages} {199} (\bibinfo {year} {1975})}\BibitemShut
  {NoStop}%
\bibitem [{\citenamefont {{Slagle}}\ \emph {et~al.}(2018)\citenamefont
  {{Slagle}}, \citenamefont {{Prem}},\ and\ \citenamefont
  {{Pretko}}}]{Slagle2018}%
  \BibitemOpen
  \bibfield  {author} {\bibinfo {author} {\bibfnamefont {K.}~\bibnamefont
  {{Slagle}}}, \bibinfo {author} {\bibfnamefont {A.}~\bibnamefont {{Prem}}}, \
  and\ \bibinfo {author} {\bibfnamefont {M.}~\bibnamefont {{Pretko}}},\
  }\href@noop {} {\  (\bibinfo {year} {2018})},\ \Eprint
  {http://arxiv.org/abs/1807.00827} {arXiv:1807.00827 [cond-mat.str-el]}
  \BibitemShut {NoStop}%
\bibitem [{\citenamefont {Slagle}\ and\ \citenamefont
  {Kim}(2018)}]{Slagle2017-2}%
  \BibitemOpen
  \bibfield  {author} {\bibinfo {author} {\bibfnamefont {K.}~\bibnamefont
  {Slagle}}\ and\ \bibinfo {author} {\bibfnamefont {Y.~B.}\ \bibnamefont
  {Kim}},\ }\href {\doibase 10.1103/PhysRevB.97.165106} {\bibfield  {journal}
  {\bibinfo  {journal} {Phys. Rev. B}\ }\textbf {\bibinfo {volume} {97}},\
  \bibinfo {pages} {165106} (\bibinfo {year} {2018})}\BibitemShut {NoStop}%
\bibitem [{\citenamefont {Dong}(2016)}]{Dong2016NatCom}%
  \BibitemOpen
  \bibfield  {author} {\bibinfo {author} {\bibfnamefont {X.}~\bibnamefont
  {Dong}},\ }\href {https://doi.org/10.1038/ncomms12472} {\bibfield  {journal}
  {\bibinfo  {journal} {Nature Communications}\ }\textbf {\bibinfo {volume}
  {7}},\ \bibinfo {pages} {12472} (\bibinfo {year} {2016})}\BibitemShut
  {NoStop}%
\bibitem [{\citenamefont {{Faulkner}}\ \emph {et~al.}(2013)\citenamefont
  {{Faulkner}}, \citenamefont {{Lewkowycz}},\ and\ \citenamefont
  {{Maldacena}}}]{Faulkner2013JHEP}%
  \BibitemOpen
  \bibfield  {author} {\bibinfo {author} {\bibfnamefont {T.}~\bibnamefont
  {{Faulkner}}}, \bibinfo {author} {\bibfnamefont {A.}~\bibnamefont
  {{Lewkowycz}}}, \ and\ \bibinfo {author} {\bibfnamefont {J.}~\bibnamefont
  {{Maldacena}}},\ }\href {\doibase 10.1007/JHEP11(2013)074} {\bibfield
  {journal} {\bibinfo  {journal} {Journal of High Energy Physics}\ }\textbf
  {\bibinfo {volume} {2013}},\ \bibinfo {eid} {74} (\bibinfo {year} {2013})},\
  \Eprint {http://arxiv.org/abs/1307.2892} {arXiv:1307.2892 [hep-th]}
  \BibitemShut {NoStop}%
\bibitem [{\citenamefont {{Kubica}}\ and\ \citenamefont
  {{Yoshida}}(2018)}]{Kubica2018}%
  \BibitemOpen
  \bibfield  {author} {\bibinfo {author} {\bibfnamefont {A.}~\bibnamefont
  {{Kubica}}}\ and\ \bibinfo {author} {\bibfnamefont {B.}~\bibnamefont
  {{Yoshida}}},\ }\href@noop {} {\  (\bibinfo {year} {2018})},\ \Eprint
  {http://arxiv.org/abs/1805.01836} {arXiv:1805.01836 [quant-ph]} \BibitemShut
  {NoStop}%
\bibitem [{\citenamefont {He}\ \emph {et~al.}(2018)\citenamefont {He},
  \citenamefont {Zheng}, \citenamefont {Bernevig},\ and\ \citenamefont
  {Regnault}}]{He2018PRB}%
  \BibitemOpen
  \bibfield  {author} {\bibinfo {author} {\bibfnamefont {H.}~\bibnamefont
  {He}}, \bibinfo {author} {\bibfnamefont {Y.}~\bibnamefont {Zheng}}, \bibinfo
  {author} {\bibfnamefont {B.~A.}\ \bibnamefont {Bernevig}}, \ and\ \bibinfo
  {author} {\bibfnamefont {N.}~\bibnamefont {Regnault}},\ }\href {\doibase
  10.1103/PhysRevB.97.125102} {\bibfield  {journal} {\bibinfo  {journal} {Phys.
  Rev. B}\ }\textbf {\bibinfo {volume} {97}},\ \bibinfo {pages} {125102}
  (\bibinfo {year} {2018})}\BibitemShut {NoStop}%
\bibitem [{\citenamefont {Gromov}(2019)}]{Gromov2017}%
  \BibitemOpen
  \bibfield  {author} {\bibinfo {author} {\bibfnamefont {A.}~\bibnamefont
  {Gromov}},\ }\href {\doibase 10.1103/PhysRevLett.122.076403} {\bibfield
  {journal} {\bibinfo  {journal} {Phys. Rev. Lett.}\ }\textbf {\bibinfo
  {volume} {122}},\ \bibinfo {pages} {076403} (\bibinfo {year}
  {2019})}\BibitemShut {NoStop}%
\end{thebibliography}%

%
%
%
%
%
%
%
%

\end{document}